\documentclass[fleqn,10pt,draft]{article} %%% -> preprint works !
\usepackage{amstex}
\usepackage{righttag}
\usepackage{ctagsplt}
\usepackage{intlim}
\def\oo{\infty}

\def\NC{{\it Nuovo Cimento }}
\def\NuPh{{\it Nucl. Phys. }}
\def\PL{{\it Phys. Lett. }}
\def\PR{{\it Phys. Rev. }}

\def\NK{{N_{k}}} % numero loops
\def\NP{{N_{p}}} % numero impulsi esterni
\def\NE{{N_{e}}} % numero linee esterne 
\def\ND{{N_{d}}} % numero denominatori
\def\NPS{{N_{sp}}} % numero prodotti scalari indipendenti
\def\NPP{{N_{pr}}} % numero prodotti scalari indipendenti
\def\MPRD{M_{p}} % indice numeratore
\def\MDEN{M_{d}} % indice denominatore
\def\group#1#2#3{\left[#1;{\frac[0pt]{{#2}}{{#3}}}\right]}
\def\groupl#1#2#3{\left[#1;{\frac[0pt]{{0\dots#2}}{{0\dots#3}}}\right]}

\def\genpk{{(p\cdot k)}}
\def\indpk{{(p\cdot k \text{ irred.})}}
\def\UOMOG{U^{HO}}
\def\VOMOG{V^{HO}}
\def\IOMOG{I^{HO}}
\def\UNOMOG{U^{NH}}
\def\VNOMOG{V^{NH}}
\def\INOMOG{I^{NH}}
\def\vOMOG{v_{HO}}
\def\vNOMOG{v_{NH}}
\def\mup{\mu_{+}}
\def\mum{\mu_{-}}

\def\dk#1{[d^Dk_{#1}]}

\def\e{\epsilon}
\def\fe{\phantom{\epsilon}}
\def\Y{y}
\def\K{K}
\def\R{R}
\def\C{\eta}
\def\RHO{\boldsymbol{\rho}}
\def\PI{\boldsymbol{\pi}}
\def\sng{{si}}
\def\bW#1{\bold{W}_{\boldsymbol{#1}}}
\def\Gammae{\Gamma_\e}
\def\SYS{{\tt SYS}}
\def\IDENTIFY{\par (S. Laporta, High-precision calculation of 
 multi-loop Feynman integrals by difference equations)}
\def\factor{1.0}
\def\ssss{{}}                               
\def\baselinestretch{\factor}
\newcommand\mytoday{\number\day\space \ifcase\month\or
  January\or February\or March\or April\or May\or June\or
    July\or August\or September\or October\or November\or December\fi
      \space\number\year}
\newtheorem{alg}{Algorithm}
\def\eqref#1{Eq.(\ref{#1})}
\def\eqrefb#1#2{Eqs.(\ref{#1})-(\ref{#2})}
\def\itref#1{(\ref{#1})}

\begin{document}
\title{\vspace{2cm}
High-precision calculation of 
 multi-loop Feynman integrals by difference equations}
\author{S. Laporta\thanks{{E-mail: \tt laporta{\char"40}bo.infn.it}} \\ 
 \hfil \\ {\small \it Dipartimento di Fisica, Universit\`a di Bologna, }
 \hfil \\ {\small \it Via Irnerio 46, I-40126 Bologna, Italy} 
 }
\date{}
\maketitle
%\vspace{-8.5cm} \hspace{13.2cm} {\mytoday} \vspace{+8.5cm}
\vspace{-8.5cm} \hspace{13.2cm} {April 2000} \vspace{+8.5cm}
\vspace{1cm}
\begin{abstract} 
We describe a new method of calculation of generic multi-loop master 
integrals based on the numerical solution of systems of
difference equations
in one variable.
We show algorithms for the construction of the systems using
integration-by-parts 
identities and methods of solutions by means of expansions in factorial series 
and  Laplace's transformation.
We also describe new algorithms for the identification of master integrals 
and the reduction of generic Feynman integrals to 
master integrals,
and procedures for generating and solving systems of differential
equations in masses and momenta for master integrals.
We apply our method to the calculation of the master integrals of 
massive vacuum and self-energy diagrams up to three loops
and of massive vertex and box diagrams up to two loops.
Implementation in a computer program of our approach is described.
Important features of the implementation are: 
the ability to deal with hundreds of master integrals and
the ability to obtain very high precision results 
expanded at will in the number of dimensions.
\end{abstract}

\pagenumbering{roman}
\setcounter{page}{0}
\vfill\eject 
\pagenumbering{arabic}
\setcounter{page}{1}
\ssss
\section{Introduction}
Nowadays the technique probably more popularly used
for calculating the contribution of
Feynman diagrams is that
based on the integration-by-parts in $D$ dimensions~\cite{Tkachov,Tkachov2}.
After some algebraic operations, 
like contracting Lorentz indices and calculating fermion traces,
the contribution of the diagram is expressed as a combination 
of several Feynman integrals with 
different powers of numerators and denominators.

This expression composed of many integrals  
is reduced to a combination of 
a limited number of `master integrals'
using recurrence relations obtained by combining the  
identities obtained by integration-by-parts.
Then, the master integrals are calculated numerically or analytically,
with some other method.

As this technique is applied to more and more complicated diagrams,
some difficulties appear.

First, 
a working general algorithm 
for identifying master integrals and for obtaining such recurrence relations
is not known at present\footnote{
An algorithm which in principle may solve this problem
has been recently proposed in \cite{Tar99}, 
but up to now no practical application was shown.}.
Up to now, for each diagram laborious handwork 
was needed for obtaining such recurrence relations\footnote{See section
\ref{ssystem}.}.

Second, the number of master integrals grows rapidly 
with the number of loops and legs of the diagrams.
Considering for example the reduction to master integrals
of the contribution to the $g$-$2$ of the electron in QED,
it is known that the one-, two-, and three-loop contributions
are reduced by integration-by-parts
to respectively, 1, 3 and 17 master integrals 
(see \cite{3-loop,pol} for the analytical calculation of the three-loop
contribution).
At the four-loop level,
to which integration-by-parts has been still not applied,
one expects several hundreds of master integrals.

Third, the calculation of a single 
multi-loop master integral is a difficult problem 
for which a general method 
applicable to any diagram, with any values of masses and momenta,
able to provide high-precision values and   
suitable for automatic calculation,
is not known at present.
Up to now a variety of different methods has been used, 
according to the topology and the values of masses and momenta
of the particular diagram considered.  

Therefore, to face problems like four-loop $g$-$2$, 
it arises the need of a completely automated approach to the calculation
of Feynman integrals, and of finding out suitable new methods, algorithms and
techniques of calculations which allow one to solve or avoid the above difficulties.

In this paper we describe the methods and techniques to be used in such
automated approach.
In particular we present:
\begin{enumerate}
\item\label{met1}
A new method for determining the master integrals
and for reducing generic Feynman integrals to master integrals,
applicable to arbitrary Feynman diagrams.
\item\label{met2}
A new method for calculating master integrals
based on the numerical solution
of the recurrence relations provided by the integration-by-parts method,
seen as linear difference equations in one index,
applicable to arbitrary Feynman diagrams.
\item\label{met3}
A new method of generation and solution of systems of differential equations
in masses and momenta for master integrals,
applicable to arbitrary Feynman diagrams.
\end{enumerate}

The part more innovative and important of this work 
is the method of calculation
of master integrals based on \emph{difference equations}.
As this mathematical topic appears (surprisingly) to be 
practically 
absent 
from the literature,
in order to improve the intelligibility of the paper 
we will give an extensive discussion on this argument, 
including methods of solutions, techniques of calculations and examples of
applications. Moreover, study of boundary conditions of difference equations
will lead us to a detailed discussion of 
asymptotic behaviour of Feynman integrals 
\emph{for large powers of one denominator}, another topic 
which has received very scarce attention in the literature.

Another important point is the automation of the approach.
All methods, algorithms and techniques developed in this work
have been implemented in a comprehensive program called $\SYS$, 
written on purpose, which, among other things, contains a simplified 
algebraic manipulator.
The aim of this program is to calculate the value of 
a generic Feynman integral in completely automatic way, 
by reducing it to master integrals and by calculating the master integrals.
The program turned out to be able 
to deal with diagrams with hundreds of master integrals
and to obtain very high precision values 
(for example 100-200 digits) expanded at will in $\e=(4-D)/2$, 
with all coefficients in numerical form, divergent terms included.
Limitations of the current implementation will be described.

A crucial point is the test of the approach.
By using $\SYS$ we have analyzed  vacuum, self-energy, vertex and box diagrams 
up to three or two loops
and we have calculated the values of master integrals for some values
of masses and momenta, 
comparing them with already known results, when possible.  
We will show 
the results obtained.
In particular we have calculated all the previously unknown 
single-scale massive three-loop self-energy master integrals.
These numerical calculations, involving the
calculation of hundreds of master integrals at the same time, 
allowed us to prove the reliability of the approach in real cases and 
to accumulate a considerable experience of calculation.

The plan of the paper is as follows:
In section~\ref{idebewin} we illustrate an algorithm
for the construction and solution of systems of integration-by-parts
identities which allows one to reduce
a generic Feynman integral to a combination of master integrals;
this algorithm is the basis for the algorithms developed
in the following sections.
In section~\ref{difequfey} we show how to obtain systems of
difference equations in one variable satisfied by the master integrals.
In section~\ref{secsolfact} we show how to find solutions of difference
equations using expansions in factorial series.
In section~\ref{detcon} we discuss the methods to find
the values of the arbitrary constants appearing in the solutions of
the difference equations.
In section~\ref{sumfact} we describe the techniques used to sum the 
factorial series expansions.
In section~\ref{example1loop} we describe in detail the 
application of our methods to the one-loop self-energy diagram.
In section~\ref{Laplacesec} we illustrate an alternative method of solution
of difference equations based on the Laplace's transformation.
In section~\ref{resu0} we apply our methods to various diagrams
from one to three loops.
In section~\ref{differentialequ} we show how to deal with
integrals with some particular values of masses and momenta,
by solving systems of differential equations in masses and momenta.
In section~\ref{calcprog} we show
some technical information on the computer program \SYS~ 
used in the calculations.
In section~\ref{Conclusions} we give our conclusions.

\section{Systems of identities between Feynman integrals}\label{idebewin}
After the introduction of the notation used for integrals and identities,
in sections~\ref{ssystem}-\ref{algsolsys} 
we describe the new method used for solving the systems of 
integration-by-parts identities and for obtaining reduction
to master integrals. The description is much detailed  
because the algorithm here shown is an essential part of this work,
being the basis of all similar algorithms of solution of systems of identities 
which will be described in sections 
\ref{constsysdif}, \ref{trasys} and \ref{consys}.
In section~\ref{identificamaster} we show how the algorithm can be used 
for determining the master integrals.
In sections~\ref{secab}-\ref{exasecab}
we show how to use the algorithm for reducing integrals,
and we describe its more interesting feature: 
the absence of intermediate 
integrals with increasing exponents of the denominators 
in the intermediate steps of the reduction to master integrals.

\subsection{Generalities}\label{Generalities}
Let us introduce some notations used throughout the paper.
We consider a generic Feynman diagram with $\NK$ loops, $\NE$ external and
$\ND$ internal lines.
The loop momenta are $k_i$, $i=1,\ldots,\NK$;
the independent external momenta are $p_i$, $i=1,\ldots,\NP$,
where $\NP=\NE-1$ (if $\NE>0$) for total momentum conservation.
The denominators of the propagators are $D_i=q_i^2+m_i^2 $,
where $q_i$ is the momentum flowing in the internal line $i$
and $m_i$ is the mass of the internal line;
each $q_i$ is a linear combination of the momenta $\{p_j\}$ and $\{k_j\}$.
The usual prescription $q_i^2+m_i^2=q_i^2+m_i^2-i0$ is implied if necessary.
A generic Feynman integral regularized in $D$-dimensional
euclidean momentum space 
has the form
\begin{equation}\label{genericterm}
\int {\dk1\; \dk2 \dots \dk{\NK}} \  V_{\gamma\delta} \ ,
\end{equation}
where $\dk{i}=d^D k_i/\pi^{D/2}$ and $V_{\gamma\delta}$ is
the generic integrand
\begin{equation}\label{genericv}
V_{\gamma\delta}=\dfrac
{\prod_{i=1}^\NP \prod_{j=1}^\NK (p_i\cdot k_j)^{\delta_{ij1}}
 \prod_{i=1}^\NK \prod_{j=i}^\NK (k_i\cdot k_j)^{\delta_{ij2}}}
{\prod_{i=1}^\ND  D_i^{\gamma_i}} \ ,
\quad \gamma_i\ge 0,\  \delta_{ijl}\ge 0\ ,\quad 
\end{equation}
$\gamma=\{\gamma_1,\ldots,\gamma_\ND\}$ and $\delta=\{\delta_{ijl}\}$.
The numerator of \eqref{genericv} is a product of powers of
all the possible scalar products involving the loop momenta $k$;
the total number $\NPS$ of such scalar products is
\begin{equation}\label{npsdef}
\NPS= \NP\NK + {\NK(\NK+1)}/{2} 
\ .
\end{equation}

\subsection{Algebraic and integration-by-parts identities}\label{introduceide}
Let us consider the generic integrand \itref{genericv}.
For each denominator $D_j$  we write
the identity
\begin{equation}\label{simpalg}
\frac{{\genpk}_j}{D_j}=\frac{1}{C_j}
 \left(1-\frac{D_j-C_j \;{\genpk}_j}{D_j}\right) \ ,\qquad  j=1,\ldots,\ND\ ,  
\end{equation}
where ${\genpk}_j$ indicates one scalar product involving loop momenta which
appears in the expression of $D_j$, and $C_j$ is its coefficient
in the expression.
The scalar products ${\genpk}_j$ must be chosen all different.
These  algebraic identities are applied in sequence to $V_{\gamma\delta}$
and to the terms subsequently generated, more times if it is necessary.
As a result, the original integrand $V_{\gamma\delta}$ is transformed into 
a sum of new terms,
each one not containing ${\genpk}_j$ and $D_j$ simultaneously,
with the general form 
\begin{equation}\label{genericvind}
V'_{ni\alpha\beta}=\dfrac
{\prod_{j=1}^{\NPS-n} {\indpk}_j^{\beta_{j}} }
{\prod_{j=1}^n  D_{i_j}^{\alpha_j}}\ ,    \quad n\le\ND\ , \quad
\alpha_j,\beta_j \ge 0\ ,
\end{equation}
where
the subscript $ni\alpha\beta$ shows the dependence on the number $n$ of
denominators, their particular combination $i=\{i_1,\ldots,i_n\}$,
and the exponents $\alpha=\{\alpha_1,\ldots,\alpha_n\}$
and $\beta=\{\beta_1,\ldots,\beta_{\NPS-n}\}$.
The symbols ${\indpk}_j$, $j=1,\ldots,\NPS-n$ indicate the
$\NPS - n$ `irreducible' scalar products which cannot be simplified
further on by \eqref{simpalg} with one of the denominators
$D_{i_1}$, $\ldots$, $D_{i_n}$ of
$V'_{ni\alpha\beta}$;
if $n=\NPS$ the numerator of \eqref{genericvind} is unity.
We stress that one different set of irreducible scalar products corresponds
to each different combination of denominators.

Integrating by parts in $D$ dimensions~\cite{Tkachov,Tkachov2}
one can write the identities
\begin{equation}\label{intbyparts}
\begin{split}
&\int {\dk1 \dots \dk{\NK}}
\frac{\partial}{\partial (k_j)_\mu} 
\left( (p_l)_\mu 
V'_{ni\alpha\beta}
 \right) =0 \;, \quad  j=1,\dots,\NK, \ 
                       l=1,\dots,\NP, \\
&\int {\dk1 \dots \dk{\NK}}
\frac{\partial}{\partial (k_j)_\mu} 
\left( (k_l)_\mu 
V'_{ni\alpha\beta}
\right) =0 \;, \quad  j,l=1,\dots,\NK\ , \\
\end{split}
\end{equation}
where $V'_{ni\alpha\beta}$ is defined in \eqref{genericvind}.
For each different $V'_{ni\alpha\beta}$
\eqref{intbyparts} gives $\NK(\NP+\NK)$ identities.
The ratio  $V'_{ni\alpha\beta}$ 
contains only irreducible scalar products (relative to the particular
combination of denominators);
the calculation of the derivative and the contraction
of the index $\mu$
form terms also containing reducible scalar products,
which must be transformed into irreducible scalar products
using the algebraic identities \itref{simpalg}.
As a final result, the identity will contain
a linear combination of integrals of two
kinds: integrals containing all the $n$ denominators
$\{D_{i_1}$, $D_{i_2}$, $\ldots$, $D_{i_n}\}$
appearing in $V'_{ni\alpha\beta}$
and integrals with one denominator
missing as effect of algebraic identities.

\subsection{A new method for solving the system of identities}\label{ssystem}
Each identity obtained from \eqref{intbyparts} is a linear combination
of integrals like 
\begin{equation}\label{generictermvp}
\int {\dk1 \dots \dk{\NK}} \  V'_{ni\alpha\beta} \ ,
\end{equation}
with polynomials of degree zero or one 
in the number of dimensions $D$ as coefficients,
and where the integrand has the general form of \eqref{genericvind}.
A generic set of these identities
forms a homogeneous linear system of equations,
with the integrals as unknowns. 
As it is well-known, the system is under-determined, and some integrals exist,
the ``master integrals'', whose values cannot be determined from the system.

In the literature\footnote{
For a review see \cite{Harlander}. 
In the program {\tt BUBBLES}\cite{Ritbergen}
reduction of systems of identities is performed by semi-automatic means.
A different approach to solving integration-by-parts identities was
developed in \cite{Baikov}. 
}
it is usual
to look for the general solution of
the \emph{infinite} system \itref{intbyparts} in the form of 
combination of identities which,
lowering and raising exponents,
transform integrals of the form \itref{genericterm} or \itref{generictermvp}
 into linear combinations of
carefully chosen master integrals.
Up to now, the methods used to find these identities 
have been based on laborious human analysis;
as the number of these identities (and the difficulties encountered) 
grow rapidly with the number of denominators and loops, 
it becomes very difficult to analyze in this way diagrams 
beyond a certain limit.

On the contrary, our approach is different, as it consists in the solution
of systems made up of a \emph{finite} number of identities.
The identities \itref{intbyparts} are generated explicitly using 
suitable $V'_{ni\alpha\beta}$,  
with parameters $n$, $i$, $\alpha$, $\beta$ taken from 
a large (but finite) set of values carefully chosen.
The set of the generated identities forms a linear system with
integrals as unknowns,
which is solved using the well-known Gauss elimination method;
the solution gives the expressions of the integrals
as linear combinations
of the master integrals with coefficients rational in $D$.
The advantage of this approach is that it is simple,
applicable without modifications to diagrams with any topology,
and suitable for completely automatic calculations.\footnote{
A similar method, in very preliminary form, 
was developed by the author in \cite{3-loop} 
to reduce to master integrals the triple-cross vertex diagrams contribution
to the $g-2$ of the electron. In that work a system of about 100000
identities was built and solved, and this required some months of computer
time; with our new method the same calculation may be performed in a
fraction of hour. 
Some similar technique has been recently used in \cite{prep1},
to solve small systems of identities in order
to obtain differential equations for two-loop massless box diagrams.}
It does not consist only in a mere ``mechanical'' approach because,
as we will see in section~\ref{secab}, the solution of the whole system
allows us to discover
very useful identities of a kind \emph{a priori} not expected;
these identities allow one to avoid the appearing of intermediate 
integrals with increasing exponents of the denominators 
in the intermediate steps of the reduction to master integrals.

Let us now consider more in detail the solution of the system.
The identities are generated and inserted in the system one at a time.
Let $\sum_j c_j W_j=0$ be an identity obtained from \eqref{intbyparts},
where $W_j$ are integrals and $c_j$ the coefficients.
The identities already existing in the system,
expressing some of the integrals $W_j$ in terms of other integrals 
are substituted in the new identity, which becomes $\sum_j c_j' W_j'=0$.
One particular integral $W_l'$ is chosen between the integrals $\{W_j'\}$,
and the identity itself is rewritten as $W_l'=\sum_{j\not =l} c_j'' W_j'$
in order to express the particular integral 
in terms of the remaining integrals.
Then the new identity is added to the system
and the chosen integral $W_l'$ is substituted in the rest of the system.

The choice of the integral $W_l'$ is carried out following
an ordering\footnote{
The importance of a total ordering for the solution of systems
of differential equations derived from integration-by-parts identities 
was remarked in \cite{Tar99}.}
of all the integrands $V'_{ni\alpha\beta}$
as functions of the parameters:
the number $n$ of denominators, the combination $i$ of denominators,
and the exponents.
The ordering defines the priority of extraction between integrals.
Priorities are arranged so that integrals
with a higher number of denominators are extracted first,
and expressed as linear combinations of the integrals
with a lower number of denominators.
Remaining details of ordering used
are shown in the algorithm of the next section.
The form of the master integrals depends on the choice of the ordering:
see section~\ref{identificamaster}.

Now we consider the choice of
the order of the generation and processing of the identities.
The final solution of the system is independent of the order of processing,
but the computing time is not.
Each addition of a new identity to the system
implies
a substitution of an integral in all identities of the system which contains
it; therefore the order must be carefully chosen
in order to minimize the number of substitutions required.
A bad choice may cause the computing time to blow up.

A good choice of the ordering
of the ratios $V'_{ni\alpha\beta}$ appearing in \eqref{intbyparts}
is the inverse of the above considered ordering of integrands used
for the extraction of integrals.
The identities corresponding to ratios $V'_{ni\alpha\beta}$
with the lowest priority of extraction (as integrands), therefore
with the lowest number of denominators,
are generated and processed first; then the identities
with a higher number of denominators are processed.
The last processed identities
will have the highest number of denominators.
This choice is efficient as long as the number $n$ of denominators of the
ratio $V'_{ni\alpha\beta}$ used in \eqref{intbyparts} is equal
to the number $n'$ of denominators of the integral $W_l'$ extracted from
the identity generated. In some cases
(relatively rare, but of great importance, see section~\ref{secab})
one finds $n'<n$;
the identities obtained with ratios $V'_{n'i'\alpha'\beta'}$, 
containing integrals with $n'$ denominators,
have already been inserted into the system, so that the ``late'' integral $W_l'$ 
must be substituted in a large number of identities, and that may
require a considerable amount of time.

It is useful to split the system of identities \itref{intbyparts}
into several \emph{subsystems}.
Each subsystem is made up of all the identities
obtained by inserting in \eqref{intbyparts} terms $V'$
containing one particular combination $i$ of denominators
$\{D_{i_1}$, $D_{i_2}$, \ldots, $D_{i_n}\}$.
The integrals with a number of denominators $n$
less than the number of loops $\NK$ of the diagram are null
in the framework of the dimensional regularization,
and must be not considered, so that 
the total number $S$ of subsystems is 
\begin{equation}\label{defins}
S=\sum_{n=\NK}^\ND \binom{\ND}{n} < 2^\ND\ .
\end{equation}
Let us now define the non-negative quantities $\MPRD$, $\MDEN$,
\begin{equation}\label{groupdef}
\sum_{j=1}^{\NPS-n} \beta_j=\MPRD \ , \quad 
\sum_{j=1}^{n} \left(\alpha_j-1\right)=\MDEN \ , 
\end{equation}
which are  the total sum of the powers of
the scalar products in the
numerator and the sum of the powers of the denominators minus
the number of denominators
of the generic integrand $V'_{ni\alpha\beta}$ of \eqref{genericvind},
respectively.
Using this definition we can split the infinite set of the integrands
$V'_{ni\alpha\beta}$
into finite sets of integrands with equal $n$, $\MPRD$ and $\MDEN$ 
that we indicate with the symbol $\group{n}{\MPRD}{\MDEN}$.
The set of the integrands with $0\le\MPRD\le a $ and $0\le\MDEN\le b$
will be indicated with the symbol  $\groupl{n}{a}{b}$.
In the following sections we will say in short 
that an integral \itref{generictermvp} or an identity
\itref{intbyparts} belongs to a specified set,
understanding that it is the corresponding $V'_{ni\alpha\beta}$ 
which belongs to that set.

\subsection{The algorithm of solution of the system}\label{algsolsys}
The algorithm of solution is the following:
\begin{alg}\label{algsys}\label{algsys3}
\begin{enumerate}
\item Let $n=\NK$. \label{quin1}
\item Let $i_1=1$, $i_2=2$, \ldots, $i_n=n$\label{quin15}. 
\item Consider the combinations of $n$ different denominators, chosen
      in the set $\{D_1$, $\ldots$, $D_\ND\}$;
      let $D_{i_1},D_{i_2},\ldots,D_{i_n}$ be one of these combinations.
      \label{quiN}
\item Choose two integer non-negative constants $a_i$ and $b_i$,
      $ i = \{ i_1, \ldots, i_n \} $. \label{quiab}
\item Let $\MDEN=0$. \label{quiMden} 
\item Let $\MPRD=0$. \label{quiMprd}

\item \label{cf6a}
Consider the ratios of the kind of \eqref{genericvind}, \label{quiset}
      containing the $n$ denominators \par
$D_{i_1},D_{i_2},\ldots,D_{i_n}$ raised to some power;
let $W$ be one of these ratios:\label{quissset}
\begin{equation*}
W(n,i,\alpha,\beta)=\dfrac
{\prod_{j=1}^{\NPS-n} {\indpk}_j^{\beta_{j}} }
{D_{i_1}^{\alpha_1} D_{i_2}^{\alpha_2} \cdots D_{i_n}^{\alpha_n}},
\end{equation*}
where the non-negative exponents $\alpha_j$, $\beta_j$ are such that 
\begin{equation*}
\sum_{j=1}^{\NPS-n} \beta_j=\MPRD \ , \quad 
\sum_{j=1}^{n} \left(\alpha_j-1\right)=\MDEN \ ,
\end{equation*}
i.e., they are such that $W$ belongs to the set $\group{n}{\MPRD}{\MDEN}$.

\item Set $V_{ni\alpha\beta}'=W$ in \eqref{intbyparts} and generate all the 
      integration-by-parts identities.\label{quiW}\label{quiins1}
\item \label{quiins2}
      Let $\int \dk1\ldots\dk\NK \ \sum_j c_j W_j =0 $ be one of the  identities
      of \eqref{intbyparts}, where $W_j$ is a generic integrand with
      $n$ or $n-1$ denominators, 
      \label{quiIde}
      then:
\begin{enumerate}
\item Substitute all the integrals already known in the left-hand side of the
      new identity;
      \par
      let $\int \dk1\ldots\dk\NK \sum_j c_j' W_j'=0$ be the result.
      If the new identity is a linear combination of other identities
      of the system, go to step~\ref{nullo}.
\item If the new identity is linearly independent,
      choose an integrand $W_l'$ from the identity,
\begin{equation*}
W_l'(n',i',\alpha',\beta')=\dfrac
{\prod_{j=1}^{\NPS-n'} {\indpk}_j^{\beta'_{j}} }
{D_{i_1'}^{\alpha_1'} D_{i_2'}^{\alpha_2'} \cdots D_{i_n'}^{\alpha_n'}} \ ,
\end{equation*}
belonging to $\group{n'}{\MPRD'}{\MDEN'}$,
$\NK \le n' \le n$,
following the order of priority:
\begin{enumerate} \label{ordering}
\item the greatest number of the denominators $n'$; 
\item the greatest $\MDEN'$; \label{ord2}
\item the greatest $\MPRD'$;  \label{ord3}
\item the greatest $i_1'$, the greatest $i_2'$, \ldots,  the greatest $i_n'$;
\item the greatest $\alpha_1'$, the greatest $\alpha_2'$, \ldots,
      the greatest $\alpha_n'$;
\item the greatest $\beta_1'$,  the greatest $\beta_2'$,  \ldots,
      the greatest $\beta_{\NPS-n'}$.
     \label{cf6b}
\end{enumerate}
\item Substitute and add the following identity to the system:
\begin{equation*}
\int \dk1\ldots\dk\NK \; W_l'= - \int \dk1\ldots\dk\NK \sum_{j\not=l} 
( c_j'/ c_l' ) W_j'   \ .
\end{equation*}
\end{enumerate}
\item Generate a new integration-by-parts identity among the $\NK(\NP+\NK)$ 
      possible identities of \eqref{intbyparts} 
      and go to step~\ref{quiIde}, otherwise continue.\label{nullo}
\item Choose a new integrand $W$ with different exponents $\alpha$ and $\beta$,
      belonging to the set $\group{n}{\MPRD}{\MDEN}$, among the
      $\binom{\NPS-n+\MPRD-1}{\MPRD}\binom{n+\MDEN-1}{\MDEN}$ elements of
      the set
      and go to step~\ref{quiW}, otherwise continue.
\item $\MPRD=\MPRD+1$; if $\MPRD\le a_i$ go to step~\ref{quiset}.
\item $\MDEN=\MDEN+1$; if $\MDEN\le b_i$ go to step~\ref{quiMprd}.
\item Choose a new combination of indices $i_1<i_2<\ldots<i_n$
      among the $\binom{\ND}{n}$ possible combinations of
      numbers $\{1$, $2$, $\ldots$, $\ND\}$  and go to step~\ref{quiN},
      otherwise continue.
\item $n=n+1$; if $n\le \ND$ go to step~\ref{quin15}, otherwise end. \label{quin2}
\end{enumerate}
\end{alg}
In this algorithm two arbitrary integer constants, $a_i$ and $b_i$,
must be chosen for each different combination of denominators 
$\{D_{i_1}, \ldots, D_{i_n} \}$.
These constants are cutoffs and define which
identities must be included in the system and which must be excluded;
$a_i$ limits the sum of the powers of scalar products in the
numerator and $b_i$ limits the sum of the exponents of the denominators. 

With a suitable choice of the parameters $a_i$ and $b_i$
(see section~\ref{secab}),
this algorithm allows one to reduce any given integral $I$
to a sum of master integrals $B_l$
\begin{equation}\label{sumc3}
I=\sum_{l=1}^L r_l B_l\ ,
\end{equation}
where $r_l$ are rational functions of $D$, masses and scalar products
of external momenta.

\subsection{Determination of master integrals}
\label{identificamaster}

Our algorithm of solution of the system of identities provides
a general method for determining the master integrals:
it suffices, for each different combination of $n$ denominators
$\{D_{i_1}$, $D_{i_2}$, \ldots, $D_{i_n}\}$
to build the subsystem with suitable $a_i$ and $b_i$,
to solve it using the algorithm,
and to find the integrals belonging to $\groupl{n}{a_i}{b_i}$
which are not reduced to terms with $n-1$
denominators.
These should be master integrals.
Of course, since we are dealing with a system
made up with a finite number of identities,
there is the possibility
that a master integral is erroneously identified or not found
because some essential identities
have not been included in the system
as they exceed the limits of generation.
Explicit solutions of subsystems with increasing values of $a_i$ and $b_i$
show that empirical `thresholds'
$\bar a_i$ and $\bar b_i$ exist,
depending on the combination of denominators,
such that all the subsystems built with $a_i\ge \bar a_i$ and $b_i\ge \bar b_i$
give a stable and correct identification of the master integrals;
for the diagrams so far examined (see section~\ref{resu0})
one finds $0\le \bar a_i \le 3$ 
(it is equal to the maximum number of scalar products
appearing in  the master integrals) and $\bar b_i=0$.
Corresponding number of identities varies from some tens to some
hundreds.

The particular order of selection of the integrals used in step~\ref{ordering}
of the algorithm favours
for each different combination of denominators
the choice of master integrals with all
the denominators $D_j$ raised to the first power
and with the numerator containing increasing numbers of scalar products:
first the scalar integral, with 1 as
numerator, then integrals with one scalar product, and so on.

As result of the procedure just described 
we find the set of $L$ master integrals $B_l$ with the form
\begin{equation}
B_{l}= 
\int {\dk1  \dots \dk{\NK}} \  
\dfrac
{\prod_{j=1}^{\NPS-n} {\indpk}_j^{\beta_{jl}} }
{D_{i_1} D_{i_2} \cdots D_{i_n}}  \ ,
\end{equation}
where the combination of indices $i$ and exponents $\beta$
depends on the index $l$.
The ordering $B_{1}$, $B_{2}$, \ldots, $B_L$ 
follows the ordering
of integrals provided by the algorithm.

\subsection{Choice of constants $a_i$ and $b_i$: the ``golden rule''}\label{secab}

The minimal values of $a_i$ and $b_i$ necessary to 
reduce one particular integral strongly depend on  
the structure of the integral and are not easy to determine.
Therefore we limit ourselves to give some general rules, based on the
experience acquired by using the algorithm, rather than rigorous results.
Actually, these rules turned out to be very effective.

Let us consider the system of identities 
obtained by inserting in \eqref{intbyparts} all the terms $V'_{ni\alpha\beta}$ 
belonging to the set $G_{ab}$
\begin{equation}\label{G2}
G_{ab} =\bigcup_{n=\NK}^{\ND}\groupl{n}{a}{b} = \groupl{\NK \ldots
 \ND}{a}{b}  \ 
\end{equation}
containing all the possible $S$ subsystems of identities (see \eqref{defins}),
and choosing $a_i=a$ and $b_i=b$ for all the combinations
of denominators.
The number of elements $V'_{ni\alpha\beta}$
forming the set $G_{ab}$ is given by 
\begin{equation}\label{ng2bis}
N_{ele}\left(G_{ab}\right)= 
\sum_{n=\NK}^\ND
\binom{\ND}{n} 
\binom{\NPS-n+a}{a}
\binom{n+b}{b} 
\ , 
\end{equation}
while the number of identities of the system is
\begin{equation}\label{ng2}
N_{ide}\left(G_{ab}\right)= 
\NK\left(\NP+\NK\right) N_{ele}(G_{ab}) \ . 
\end{equation}
Explicit calculations suggest the following empirical ``golden rule'':
the solution of the system provides identities which
reduce any integral belonging to
the set
$G_{ab}$
to master integrals
\begin{equation}\label{expnabtot2}
\text{Integrals}\  
G_{ab}
@>>{\text{Identities }G_{ab}}>
\text{Master Integrals} \ .
\end{equation}

In order to better understand this rule,
we have analyzed in some test cases the reduction of single integrals
belonging to $\groupl{\ND}{a}{b}$,
by modifying the steps~\ref{quin1} and~\ref{quin2} of the algorithm
so that the various subsystems of identities $\groupl{n}{a}{b}$
are generated starting from $\ND$ denominators and decreasing $n$
up to $\NK$ denominators
(note that the solution of the whole system or parts of it
with such `inverted' order is extremely time consuming).
For each value of $n$, intermediate expressions contain master integrals
with a number of denominators from $n$ to $\ND$, and
integrals belonging to the set $\groupl{n-1}{a}{b+1}$,
that is, integrals with one exponent of denominators increased by one
compared with the original integral \emph{whichever is $n$}.
This is very different from what one obtains 
by performing the reduction one denominator at a time with
recurrence relations which lower one index and raise another one,
where the exponents of denominators of the integrals are observed to increase 
whenever the number of denominators decreases, 
causing the number of intermediate integrals to blow up.
Our algorithm avoids this blowing-up.

The quite unexpected behaviour of our algorithm seems to be due to
the fact that solving the whole system we consider all the identities
generated, not only those which reduce single integrals,
but also those many ``additional'' identities
apparently useless which reduce \emph{combinations} of integrals
with the same number of denominators.
Their effect is to reduce systematically 
all the generated combinations of integrals 
with increased exponents of the denominators. 
The reason for this behaviour is not known; however, 
its systematic presence suggests that (hopefully) it may have some 
simple explanation.
The presence of these ``additional'' identities
is important, particularly if $\ND$ is large,
because without them the number of identities to consider 
(and the number of intermediate integrals)
would be orders of magnitude larger.

The ``golden rule'' \itref{expnabtot2}
turns out to be valid if $a\ge a_0$ and $b\ge b_0$, where $a_0$ and $b_0$ are
some empirical `thresholds' depending on the structure of denominators of
the diagram; 
for the diagrams examined in section~\ref{resu0} 
we found $1 \le a_0 \le 3$ 
(values almost always equal to the maximum number of scalar products
appearing in the master integrals)
and $b_0=0,1$.
Values of the thresholds are unfortunately not known \emph{a priori};
if the chosen values of $a$ and $b$ are below the threshold,
the result of the solution of the system is that
some integrals belonging to $G_{ab}$ are not
completely reduced to master integrals, and
some non-master integrals with a few denominators still survive
in the final expressions.
In this case one may increase $a$ or $b$, or alternatively 
one may suitably enlarge only the subsystems of identities 
corresponding to residual integrals.

We found some very rare exceptions to the 
``golden rule'' \itref{expnabtot2}, when some denominators have zero mass;
in these cases, whatever the values of $a$ and $b$ are,
there are always few integrals belonging to $\group{\ND}{0}{b}$
(note, changing with the value of $b$)  
which are not completely reduced to master integrals.
In the example of the next section we will  show just one of these exceptions.

Now, let us consider an important application: 
the  reduction to master integrals of one combination of many integrals, 
for example, 
expressing the contribution of a diagram to some physical quantity.
Let us suppose that all the integrals of this combination belong to a set
$G_{\bar a \bar b}$ with some (minimal) $\bar a$
and $\bar b$;
if the values of $\bar a$ and $\bar b$ are over thresholds, 
according \eqref{expnabtot2}
the solution of the system will provide identities which reduce to master
integrals all the integrals of the set.

\subsection{An example}\label{exasecab}
Let us consider the  self-energy diagram with 5 denominators
$D_1=(p-k_1)^2+1$, $D_2=(p-k_1-k_2)^2+1$, $D_3=(p-k_2)^2+1$, $D_4=k_1^2$,
$D_5=k_2^2$ and $p^2=-1$.
Following the notation of section~\ref{Generalities}, this diagram has 
$\NP=1$, $\NK=2$, $\ND=5$ and  $\NPS=5$.
For instance, we want to transform the integral
\begin{equation}
J= \int 
\frac
{[dk]} 
{D_1^2 D_2 D_3 D_4 D_5} \ , 
\end{equation}
where $[dk]=\dk1 \; \dk2$,
into a combination of master integrals.
The pairs between scalar products and denominators used in the
algebraic identities are
$(p\cdot k_1,D_1)$, $(k_1\cdot k_2,D_2)$, $(p\cdot k_2, D_3)$,
$(k_1\cdot k_1,D_4)$ and $(k_2\cdot k_2,D_5)$.
We must identify the master integrals;
therefore,
following section~\ref{identificamaster},
we build the system of identities
with $a_i=1$ and $b_i=0$ for all the combinations of denominators,
and we look for the integrals which are not reduced.
The master integrals turn out be:
$B_{123}=\int [dk] / D_1 D_2 D_3 $,
$B_{345}=\int [dk] / D_3 D_4 D_5 $,
$B_{12} =\int [dk] / D_1 D_2 $,
$B_{13} =\int [dk] / D_1 D_3 $ and 
$B_{23} =\int [dk] / D_2 D_3 $.

We consider the whole system made up of the identities of the
set $G_{11}$, choosing $a=b=1$;
according to \eqref{ng2} it contains 1776 identities.
Solving it with the algorithm~\ref{algsys} 
(by using the program \SYS~ described in section \ref{calcprog})
we find 1122 independent identities.
Examining them, we find
the identities which reduce to master integrals all the 291 integrals
$\groupl{2\ldots 5}{1}{1}$, integral $J$ included,
with the exception of the two integrals
$\int {[dk] }/D_1 D_2 D_3 D_4^2 D_5$ and
$\int {[dk] }/D_1 D_2 D_3 D_4 D_5^2$,
whose reduction still contains the integrals
$\int {[dk] }/D_2 D_4^2 D_5^2$ and
$\int {[dk] }/D_3 D_4^2 D_5^2$;
in order to complete the reduction of these two integrals,
we may add  the two sets of 30 identities $\groupl{3}{1}{2}$ with
these combinations of three denominators to the system.
The remaining 831 identities reduce to master integrals complicated 
combinations of integrals $\group{2\ldots 5}{0\ldots 2}{2}$.

We want to illustrate how the mechanism of ``additional'' identities works.
Let us consider the partial reduction of the integral $J$
from 5 to 4 denominators.
We consider the subsystem of the identities $\group{5}{0}{0 \ldots 1}$,
which is formed by 36 identities; solving it,
we find:
\begin{enumerate}
\item \label{p5001} 6 identities  reducing the integrals 
                    $\group{5}{0}{0\ldots 1}$ to 4 denominators;
\item \label{p5002} 9 identities  reducing combinations of integrals 
                    $\group{5}{0}{2}$ to 4 denominators; 
\item \label{p4012}20 identities  between integrals $\group{4}{0}{1}$
                           and $\group{4}{0}{2}$
                           with different combinations of denominators;
\end{enumerate}
The identities of the groups~\ref{p5002} and~\ref{p4012} are the
``additional'' ones.
Among the identities of the group~\ref{p5001} we find 
\begin{equation}\label{J}
J =
\int {[dk] } \left(
\frac{1}{4}W_{1} +\frac{1}{2}W_{2} \right) \ ,
\end{equation}
\begin{equation*}
\begin{split}
W_{1}&=
 \frac{3}{D_2^2 D_3 D_4 D_5}
-\frac{3}{D_1 D_2^2 D_3 D_4}
+\frac{2}{D_2 D_3 D_4^2 D_5}
-\frac{1}{D_1 D_2 D_3^2 D_4} +\\
&+\frac{1}{D_1 D_2 D_3^2 D_5}
+\frac{1}{D_2 D_3 D_4 D_5^2}
-\frac{1}{D_1 D_3^2 D_4 D_5}
-\frac{1}{D_1 D_3 D_4 D_5^2}
    \ ,\\
W_{2}&=
  \frac{1}{D_1 D_3^2 D_4^2 D_5}   
 -\frac{1}{D_1 D_2 D_4^2 D_5^2}
 -\frac{1}{D_1 D_2^2 D_4^2 D_5} 
 +\frac{1}{D_1 D_3 D_4^2 D_5^2}   
 -\frac{1}{D_2 D_3 D_4^2 D_5^2}   
 	\ .\\	   
\end{split}
\end{equation*}
We see that the original integral, which contains a square denominator 
(set $\group{5}{0}{1}$),
is transformed into a combination of integrals with 4 denominators,
with one or two square denominators (sets $\group{4}{0}{1}$ and
$\group{4}{0}{2}$).

Considering the reduction from 4 denominators to master integrals,
if we peruse the solution of the subsystems $\groupl{2\ldots 4}{1}{1}$
we find 8 relations which transform 
each one of the terms of $W_{1}$ into master integrals; 
on the contrary we do not find relations which transform
\emph{single} terms of $W_{2}$ into master integrals. 
But if we also consider the previous identities of the group~\ref{p4012}, 
we find one relation which transforms the \emph{whole} combination $W_{2}$.
The presence of this relation makes it unnecessary 
to search for identities which singly transform the integrals of $W_{2}$.

\section{Difference equations for master integrals}\label{difequfey}
By using the algorithm shown in the above sections, 
a generic Feynman integral can be reduced to master integrals.
Now we consider their calculation.

We consider the generic integral \itref{generictermvp}
as a function of the exponents $U(\alpha,\beta)$;
the integration-by-parts identities \itref{intbyparts}
form a system of recurrence relations satisfied by the function $U$.
The key observation
is that such recurrence relations are \emph{linear difference equations}\footnote{
Other authors are also encountering other kinds of difference equations 
in the evaluation of diagrams 
\cite{Vermaseren,Bastei}
and solving them, 
by using techniques
different from the standard methods of \cite{Milne} used in this paper.
}
satisfied by the function $U(\alpha,\beta)$.
This observation has two important implications:
\begin{enumerate}
\item
Theory of linear difference equations\cite{Milne} is a well established
(but not very well-known) mathematical topic 
for which various useful mathematical tools exist.
\item
Difference equations can be solved numerically, and 
therefore this establishes a new method of calculation of the
integrals $U(\alpha,\beta)$.
\end{enumerate}

Another important observation is that the integration-by-parts 
provides 
recurrence relations 
in \emph{several} variables for the functions $U$, that is, 
a system of partial difference equations.
In general, the numerical solution of such a system is,
from the point of view of the numerical calculation,
a task comparable to
the solution of a system of partial derivative equations
or
to the multidimensional integration, so that high-precision results may be
very difficult to obtain if the number of variables is high.
Therefore, if we desire to obtain high-precision results, 
we must consider only difference equations in only \emph{one} variable.

In section \ref{introx} we show where to introduce such single parameter;
in section \ref{constsysdif} we show how to obtain 
recurrence relations (that is, difference equations) in this parameter 
from the integration-by-parts identities \itref{intbyparts}.

\subsection{Difference equations in one exponent}\label{introx}

Let us consider, for instance, a master integral of the form 
\begin{equation}\label{inti}
B=
\int
\dfrac
{\dk1 \dots \dk{\NK}}
{\displaystyle D_1 D_2 \ldots  D_\ND} \ .
\end{equation}
We raise one denominator to power $x$.
If we choose $D_1$, we define a function $U_{D_1}$ 
\begin{equation}
U_{D_1}(x)=
\int \dfrac
{\dk1 \dots \dk{\NK}}
{\displaystyle D_1^x D_2 \ldots  D_\ND} \ ,
\end{equation}
where the subscript shows that the exponent has been introduced in the
denominator
$D_1$. The value of the integral \itref{inti} is recovered as $U_{D_1}(1)=B$.

Using integration-by-parts identities,
we look for an identity
which relates $U_{D_1}$ with integrals
with a smaller number of denominators.
As $B$ is a master integral,
an identity which directly transforms $U_{D_1}(x)$ into simpler integrals
cannot be found;
instead, 
as we will see in the next section, one finds identities of the form
\begin{equation}\label{equsimple}
\sum_{j=0}^\R p_j(x) U_{D_1}(x+j)=F(x)\ ,
\end{equation}
where $p_j(x)$ are polynomials in $x$, and $F(x)$ is a known function.
$U_{D_1}$ appears with arguments shifted by an integer.
This identity is a linear difference equation of order $R$
in the variable $x$ satisfied by the function $U_{D_1}(x)$.

The right-hand side function $F(x)$ is in general
a linear combination of functions analogous to $U_{D_1}$,
derived
from master integrals containing $D_1$,
but with  some of the denominators ${D_2, D_3, \ldots, D_\ND}$ missing.

We note that it is possible
to raise to $x$ another denominator, $D_2$, $D_3$, \ldots, etc.
(or equivalently to permute the numbering of the lines of the diagram);
for each denominator $D_j$ a function $U_{D_j}(x)$ will be defined.
In general the functions $U_{D_j}(x)$ are different each one
and satisfy different difference equations,
but for $x=1$ they all have the same value $U_{D_j}(1)=B$.
This fact is particularly useful for checking the consistency
of the calculations.

\subsection{Construction of systems of difference equations}
\label{constsysdif}

Let us suppose that,
by means of the procedure described in section~\ref{identificamaster}, 
we establish that there are $L$ master integrals,
with a number of denominators ranging from $\NK$ to $\ND$.

We note that some master integrals may have the first denominators missing.
Then we group the master integrals in $\ND-\NK+1$ sets $S_m$ such that
each set includes the integrals which contain
the denominators $D_i$ with $i \ge m$.
Each set will contain $L_m$ master integrals $B_{ml}$, 
\begin{equation}
B_{ml}=
\int {\dk1 \dots \dk{\NK}}
\dfrac
 {\prod_{j=1}^{\NPS-n} {\indpk}_j^{\beta_{jml}} }
{D_{m} D_{i_2} D_{i_3} \ldots D_{i_n} }\ , 
\quad
m<i_2<\ldots<i_n \ ,
\end{equation}
where
$ m=1,\ldots,\ND-\NK+1$,
$ l=1, \ldots,L_m $ and
$\sum_{m} L_m =L$;
the number of denominators $n$ ranges from $\NK$ to $\ND-m+1$.
Note that $n$, the indices $i$ and the exponents $\beta$
of the various integrals depend on $m$ and $l$.
The ordering $B_{m1}$, $B_{m2}$, \ldots, $B_{mL_m}$ follows the ordering
of master integrals with the same first denominator $D_m$
provided by the algorithm~\ref{algsys}.
Changing the exponent of $D_m$ from 1 to $x$ 
in each master integral, we define the ``master functions''
\begin{equation}\label{definitionu}
U_{ml}(x)=
\int {\dk1 \dots \dk{\NK}} \  
\dfrac
 {\prod_{j=1}^{\NPS-n} {\indpk}_j^{\beta_{jml}} }
{D_{m}^{x} D_{i_2} D_{i_3} \ldots D_{i_n} } \ ; 
\end{equation}
$U_{ml}(1)=B_{ml}$ recovers the original master integral.
Now we must find the difference equations satisfied
by the functions $U_{ml}(x)$.
We follow a method quite similar
to that used for the reduction of a generic integral to master integrals.
We build a system of identities obtained from integration-by-parts
using \eqref{intbyparts}, modifying the integrand 
$V'_{ni\alpha\beta}$ of \eqref{genericvind} 
by adding $x-1$ to the exponent of the first
denominator:
\begin{equation}\label{genericvindnn}
V'_{ni\alpha\beta}(x)=\dfrac
{\prod_{j=1}^{\NPS-n} {\indpk}_j^{\beta_{j}} }
{D_{i_1}^{x-1+\alpha_1} D_{i_2}^{\alpha_2} \ldots D_{i_n}^{\alpha_n} },
\quad n\le\ND, \quad
\alpha_j,\beta_j \ge 0\ .
\end{equation}
In this way all the integrals become functions of $x$
\begin{equation}\label{generictermvpn}
U_{ni\alpha\beta}(x)=
\int {\dk1 \dots \dk{\NK}} \  V'_{ni\alpha\beta}(x) \ ,
\end{equation}
and 
the identities \itref{intbyparts} become difference equations between
these functions.
Integrals which have a different first denominator $D_{m}$
never appear in the same identity, so that it is convenient
to build and solve separately the systems of the identities
with different values of $m$.
We do this with the algorithm
\begin{alg}\label{algsys4}
Consider the algorithm~\ref{algsys} with the following modifications:
\begin{enumerate}
\item \label{cond01} Set $i_1=m$.
\item
Add $x-1$ to the exponent of $D_{m}$.
\item\label{cond03}
Ignore the addition of $x-1$ to $\MDEN$ in 
the definition of set \itref{groupdef}. 
\item \label{cond04}
Add the following conditions to the order of priority of the extraction of
integrals:
\begin{enumerate}
\item
the generic functions $U_{ni\alpha\beta}$ 
must be extracted before of the functions $U_{ml}$;
\item
the functions $U_{ml}(x-1)$ (generated by the algebraic identities)
must be extracted before of the functions $U_{ml}(x+i)$, where 
$i\ge 0$.
\end{enumerate}
\end{enumerate}
\end{alg}
As result of the solution of the system, we find lots of
identities analogous to \eqref{sumc3}, not interesting,
which express
generic functions $U_{ni\alpha\beta}$
as combinations of
master functions $U_{m l}$ with argument possibly shifted,
and rational functions of $x$, $D$, masses and scalar products of external
momenta as coefficients;
we also find much more interesting identities 
containing only master functions $U_{m l}$.

Let us suppose, for simplicity, that each master integral $B_{ml}$
contains a different combination of denominators.
With a suitable choice of the constants $a_i$ and $b_i$,
the solution of the system provides difference equations satisfied by 
the functions $U_{ml}$,
with the structure 
\begin{equation}\label{diffu1}
\sum_{i=0}^{\R_l} p_{il}(x) U_{ml}(x+i) = 
\sum_{k=1}^{l-1} \sum_{j=0}^{Q_{lk}} q_{jkl}(x) U_{mk}(x+j) \ , \quad
 l=1,\ldots,L_m \ ,\quad 
\end{equation}
where $p_{il}$ and $q_{jkl}$ are polynomials in $x$, $D$, 
masses and scalar products of external momenta.
The right-hand side of the $l$th equation contains the functions
from $U_{m1}$ to $U_{m,l-1}$.
Therefore the set of $L_m$ equations \itref{diffu1} 
forms a triangular system of difference equations.
The triangular structure is particularly useful for simplifying 
the numerical solution:
the equations are solved in ascending order,
one at a time, for $l=1, 2, \ldots, L_m$;
when the equation for $U_{ml}$ has to be solved,
the functions $U_{m1}$,\ldots,$U_{m,l-1}$ in the right-hand side
are already known.
Note that some functions may be missing from the right-hand side;
for example, if the master function $U_{ml}$ has $\NK$ denominators, 
the right-hand side is zero.

Now we consider the more general case where $G$ different master integrals
have the same denominator $(D_{i_1}\cdots D_{i_n})$ and different numerators.
These master integrals correspond to master functions 
with contiguous indices   
$U_{m,l'+1}$, $U_{m,l'+2}$, \ldots, $U_{m,l'+G}$
as effect of the ordering of master integrals.
The solution of the system provides for  
these functions
a subsystem of simultaneous $G$ difference equations of the kind
\begin{equation}\label{diffu2}
\sum_{g=1}^{G} \sum_{i=0}^{\R_{hg}} p_{igl'h}'(x) U_{m,l'+g}(x+i) = 
\sum_{k=1}^{l'} \sum_{j=0}^{Q_{hl'k}} q'_{jkl'h}(x) U_{mk}(x+j) \ , \quad h=1,\ldots,G\ ,
\end{equation}
not in triangular form,
where the left-hand side of each equation contains all the master functions
$U_{m,l'+1}$, \ldots, $U_{m,l'+G}$.
We prefer not deal with the solution of 
the subsystem of simultaneous difference equations,
so that we transform it into triangular form; this is not obligatory, 
is only convenient to simplify the subsequent numerical solution.
We make the replacement $x\to x+c$
in the equations \itref{diffu2}, with $c=1,2,\ldots$,
generating new identities which are inserted in the subsystem,
and taking care of extracting the terms containing
the function $U_{m,l'+j}$ before of the terms containing $U_{m,l'+k}$
if $j>k$.
This procedure is repeated until one obtains a set of $G$ equations
in triangular form,
\begin{equation}\label{diffu1a}
\sum_{i=0}^{\R'_h} p_{il'h}''(x) U_{m,l'+h}(x+i) = 
\sum_{k=1}^{l'+h-1} \sum_{j=0}^{Q'_{hl'k}} q_{jkl'h}''(x) U_{mk}(x+j) \ , \quad h=1,\ldots,G
\ .
\end{equation}
Another advantage of the transformation into triangular form is that 
one obtains 
a difference equation 
containing the function $U_{m,l'+1}$, but not containing
the other functions $U_{m,l'+2}$, $U_{m,l'+3}$,\ldots,
so that $U_{m,l'+1}$,
which corresponds to the master integral with 1 as numerator,
can be found independently.
Unfortunately the transformation into triangular form has a price:
one obtains
for $U_{m,l'+1}$
an equation of order $\R'_1 >\max_h R_{h1}$ 
which has coefficients $q_{jkl'1}''(x)$ much more complicated and cumbersome 
than the coefficients $q_{jkl'h}'(x)$ of the equations of the
subsystem \itref{diffu2}.
If $\R'_1$ is large (say $\R'_1>10$) these coefficients become huge and 
difficult to obtain 
so that it may be more convenient in these cases 
to solve the system of simultaneous equations \itref{diffu2} directly; 
this will be described in some next paper.

Once all the subsystems of simultaneous equations corresponding
to the various groups of master integrals with equal denominators 
are transformed into triangular form, the whole system takes 
the form \itref{diffu1}.

Concerning the choice of $a_i$ and $b_i$, all the considerations
done in section~\ref{secab} remain valid;
for each combination of $n$ denominators $\{D_{i_1}\ldots D_{i_n}\}$
there is a minimal subsystem $\groupl{n}{\tilde a_i}{\tilde b_i}$
whose solution allows one to obtain the equations \itref{diffu2}.
For the diagrams so far examined $\tilde a_i$ turns out to be always equal to
the maximum number of scalar products
appearing in  the master integrals (typically $1\ldots 3$);
typical values of $\tilde b_i$ are $0,1,2$.

\section{Solutions of difference equations with 
        {\hskip 0.8pt}fac\-torial series}
\label{secsolfact}

By using the algorithms of section \ref{difequfey}, 
the triangular system of difference equations 
satisfied by the master functions is worked out. 
Now we consider the solution of each difference equation.

Let us suppose, for instance, that the master function $U(x)$ defined as  
\begin{equation}\label{init1}
U(x)=
\int {\dk1 \dots \dk{\NK}} \  
\dfrac
{\prod_{j=1}^{\NPS-\ND} \indpk_j^{\beta_{j}}}
{\displaystyle D_1^x D_2 \ldots  D_\ND} \ ;
\end{equation}
satisfies a difference equation of order $R$
\begin{equation}\label{equdifref}
\sum_{i=0}^\R p_i(x) U(x+i)=F(x)\ ,
\end{equation}
where $p_i(x)$ are polynomials and $F(x)$ is some known function.
The solution of this nonhomogeneous equation can be written as
\begin{equation}
U(x)=\UOMOG(x)+\UNOMOG(x)\ ,
\end{equation}
where $\UNOMOG$ is a particular solution of the nonhomogeneous equation
\itref{equdifref}
and $\UOMOG$ is the general solution of the homogeneous equation
\begin{equation}\label{equr1}
\sum_{i=0}^\R p_i(x) \UOMOG(x+i)=0\  .
\end{equation}
The general solution of \eqref{equr1} can be written as
\begin{equation}\label{equhom}
\UOMOG(x)=\sum_{j=1}^\R \tilde\omega_j(x) \UOMOG_j(x) \ ,
\end{equation}
where $\tilde\omega_j(x)$ are periodic functions of period 1,
and  $\{\UOMOG_1(x),\ldots,\UOMOG_R(x)\}$ is a
fundamental system of independent solutions of the
homogeneous equation.
In the following, recalling 
from \cite{Milne} 
for the ease of the unfamiliar reader 
the essential matter on 
the solution of linear difference equations with factorial series,
we describe how to obtain factorial series expansions
of $\UOMOG$ and $\UNOMOG$.

\subsection{Factorial series} 
The series
\begin{multline}\label{remfac}
\sum_{s=0}^\oo\frac{a_s \Gamma(x+1)}{\Gamma(x-\K+s+1)} 
    =\frac{\Gamma(x+1)}{\Gamma(x-\K+1)}\left(
     a_0+\frac{a_1}{x-\K+1} +\frac{a_2}{(x-\K+1)(x-\K+2)} +\ldots \right)\  
\end{multline}
is known as factorial series of the first kind~\cite{Milne12}, or
series of inverse factorials~\cite{Milne271}, 
or faculty series~\cite{Knopp446}. 
We refer to it in brief as factorial series.
The series converges for every point in a half-plane which is limited
on the left by $\Re x =\lambda$ (excluding $x=\K-1, \K-2,\ldots $).
The number $\lambda$ is the \emph{abscissa of convergence}.
If $\lambda=\oo$ the series is everywhere divergent.

As coefficients of the series encountered in this work behave for large $s$
as $|a_s|\sim s! s^\alpha$,
it is useful (especially for numerical applications) to define the reduced coefficients $a'_s=a_s/s!$.
For large $s$  the generic term of the sum tends to 
$ a'_s\Gamma(x+1) s^{\K-x}$,
so that the factorial series has the same convergence properties as the
Dirichlet series $\sum_{s=0}^\oo a'_s s^{\K-x}$.

\subsection{Operators $\PI$ and $\RHO$}

Given an arbitrary number $m$ one defines the operator $\RHO$ as\footnote{
In Ref.\cite{Milne} the operators $\PI$ and $\RHO$ 
are defined more generically as
$\RHO^m U(x)=(\Gamma(x-r+1)/\Gamma(x-r-m+1))U(x-m)$ and
$\PI U(x)=(x-r)(U(x)-U(x-1))$,
where $r$ is a fixed number.
Here for simplicity we have set $r=0$.
}
\begin{equation}
\RHO^m U(x)=\frac{\Gamma(x+1)}{\Gamma(x-m+1)}U(x-m) \ .
\end{equation}
This operator has the property 
\begin{equation}
\RHO^m\RHO^n U(x)= \RHO^{m+n} U(x)\ ;
\end{equation}
if the operand is the unity, we omit it and write
\begin{equation}
\RHO^m 1 = \RHO^m = \frac{\Gamma(x+1)}{\Gamma(x-m+1)} \ .
\end{equation}
One defines the operator $\PI$ as
\begin{equation}
\PI U(x)=x(U(x)-U(x-1)) \ .
\end{equation}
The following properties can be proven:
\begin{align}
\begin{split}\label{xpiro}
[\PI,\RHO]U(x)&=\RHO U(x)\ ,\\
p(\PI) \RHO^m U(x) &= \RHO^m p(\PI+m) U(x)\ ,
\end{split}\\
\notag\\
\begin{split}\label{xpirorel12}
xU(x)&=(\PI+\RHO)U(x), \\
x^n U(x)&=\sum_{k=0}^n \left(
\sum_{j=k}^n (-1)^{j-k}\binom{n}{j}S_{jk} \PI^{n-j} \right) \RHO^k U(x) \ , \\
p(x) U(x)&=\left[p(\PI)+p_1(\PI)\RHO +p_2(\PI)\RHO^2/2! +
\ldots \right]U(x) \ , 
\end{split}
\end{align}
where $p$ is a polynomial, $p_n$ is 
\begin{equation*}
p_n(\lambda)=\Delta_n p(\lambda)= 
\sum_{j=0}^n  (-1)^j\binom{n}{j}p(\lambda-j) \ ,
\end{equation*}
and $S_{jk}$ are the Stirling's numbers of second kind~\cite{Abramowitz}.
Using these operators an expansion in factorial series becomes
an expansion in powers of $\RHO^{-1}$
so that we will able to obtain solutions in factorial series 
of difference equations 
in the same manner as  power series solutions  of differential
equations are obtained.

\subsection{Solution of the homogeneous difference equation}
\label{solfactser}
Let us consider the homogeneous difference equation \itref{equr1} of order $\R$
\begin{equation}\label{diff01}
p_0(x) \UOMOG(x) +p_1(x) \UOMOG(x+1) + \ldots + p_\R(x) \UOMOG(x+\R) =0 \ ;
\end{equation}
making the replacement $x\to x-\R$
and defining $q_{\R-i}(x)=p_i(x-\R)$ the equation becomes
\begin{equation}\label{diff1}
q_0(x) \UOMOG(x) +q_1(x) \UOMOG(x-1) + \ldots + q_\R(x) \UOMOG(x-\R) =0 \ .
\end{equation}
Using the previous definitions of operators 
now we search for a formal solution in factorial series.
We make the change of variable $\UOMOG(x)=\mu^x \VOMOG(x)$ in \eqref{diff1},
where $\mu$ is an unspecified parameter, obtaining
\begin{equation}\label{diff2}
\mu^\R q_0(x) \VOMOG(x) +\mu^{\R-1} q_1(x) \VOMOG(x-1)
+ \ldots + q_\R(x) \VOMOG(x-\R) =0 \ .
\end{equation}
Now we multiply the equation by $x(x-1)(x-2)\ldots (x-\R+1)$
and we observe that $xV(x-1)=\RHO V(x)$, $x(x-1)V(x-2)=\RHO^2V(x)$, etc.;
the equation takes the form
\begin{equation}\label{diff3}
\left[ \phi_0(x,\mu)  +\phi_1(x,\mu) \RHO + \ldots +
\phi_\R(x,\mu) \RHO^\R \right] \VOMOG(x) =0 \ ,
\end{equation}
where $\phi_j$ are polynomials in $x$ and $\mu$.
The multiplication by $x$ is 
equivalent to the multiplication by $\PI+\RHO$,
therefore substituting the relations \itref{xpirorel12} in \eqref{diff3}
one obtains the \emph{first canonical form} of the difference equation:
\begin{equation}\label{cano1}
\left[f_0(\PI,\mu) +f_1(\PI,\mu)\RHO +f_2(\PI,\mu)\RHO^2
 + \ldots + f_{m+1}(\PI,\mu)\RHO^{m+1}\right]\VOMOG(x) =0 \ ,
\end{equation}
where $f_i$ are polynomials in $\PI$ and $\mu$.
In the case of the difference equations encountered in this work
we find that $f_{m+1}(\PI,\mu)$ is independent of $\PI$, 
but not independent of $\mu$; therefore $f_{m+1}(\PI,\mu)=f_{m+1}(\mu)$.
The algebraic equation in $\mu$
\begin{equation}\label{equchar}
f_{m+1}(\mu)=0
\end{equation}
is the \emph{characteristic equation}\footnote{
If the coefficients of the equation are
$ p_i(x)=\sum_{j=0}^g p_{ij} x^{g-j} $, the characteristic equation
has the explicit expression $\sum_{i=0}^\R p_{i0} \; \mu^i =0 $.}.
It turns out that the characteristic equation has always $R$ solutions different
from zero.
Let $\mu_1$, $\mu_2$, \ldots, $\mu_\lambda$ be the
$\lambda$ \emph{distinct} solutions of \eqref{equchar}.
For each of these values, $\mu=\mu_i$, $i=1,\ldots,\lambda$, 
the first canonical form \itref{cano1} takes the form
\begin{equation}\label{cano2}
\left[f_0(\PI) +f_1(\PI)\RHO +f_2(\PI)\RHO^2
 + \ldots + f_{m}(\PI)\RHO^{m}\right]\VOMOG(x) =0 \ .
\end{equation}
Now we try to satisfy the equation \itref{cano2} 
in $\VOMOG$ with the factorial series
\begin{equation}\label{serrho}
\VOMOG(x)=\sum_{s=0}^\oo\frac{a_s \Gamma(x+1)}{\Gamma(x-\K+s+1)} 
    =\sum_{s=0}^\oo a_s \RHO^{\K-s} = a_0\RHO^\K +a_1\RHO^{\K-1} +\ldots \ , 
\end{equation}
whose asymptotic behaviour for large $x$ is $\VOMOG(x)\approx a_0 x^\K$.
Substituting \eqref{serrho} in \eqref{cano2} one obtains the recurrence
relations
\begin{align}
\begin{split}\notag
&a_0 f_m(\K+m)=0 \ ,\\
&a_1 f_m(\K+m-1) +a_0 f_{m-1}(\K+m-1)=0 \ ,
\end{split} \\
\begin{split}\label{systa2}
&\cdots \\
&a_s f_m(\K+m-s) +a_{s-1} f_{m-1}(\K+m-s) +\ldots  
+a_{s-m} f_0(\K+m-s)=0  \quad  (s\ge m) \ . \qquad 
\end{split}
\end{align}
Supposing that $a_0\not=0$, the equation 
\begin{equation}\label{indic}
 f_m(\K+m)=0\ 
\end{equation}
is the \emph{indicial equation}.
Let $\K_1, \K_2, \ldots, \K_\nu$ the roots of this equation.
In the case of the difference equations encountered in this work
all the roots turn out to be distinct, and $\nu$ turns out to be 
the multiplicity of $\mu_i$.
For each of these values of $\K$ we solve the system of recurrence relations.
If there are no roots differing by a positive integer,
$f_m(\K+m-s)\not=0$ for $s\ge1$, and therefore the recurrence relation
\itref{systa2} provides  the coefficient $a_s$ 
of the factorial series for every $s$.
If there are roots differing by a positive integer (so-called
\emph{congruent} roots)
we find $f_m(\K+m-s_0)=0$ for some $s_0$, so that 
the term $a_{s_0} f_m(\K+m-s_0)$
vanishes from the relation \itref{systa2} used to obtain $a_{s_0}$.
The remaining terms of this relation always vanish\footnote{
Otherwise $\VOMOG(x)$ would be a derivative with respect to $\K$
of a factorial series, with asymptotic behaviour $\sim x^\K \log^n(x)$.
}
in the case of the difference equations encountered in this work,
therefore 
the value of $a_{s_0}$ remains undetermined and can be chosen at will
(usually one puts $a_{s_0}=0$).
For details on the convergence of the factorial series so obtained, 
see section~\ref{sumfact2}.

In order to obtain the general solution of the difference equation,
for each distinct solution of the characteristic equation
$\mu_i$, $i=1,\ldots,\lambda$, we find an indicial equation,
whose solutions are $\K_{ij}$, $j=1,\ldots,\nu_i$.
For each solution of the indicial equation we solve the system of recurrence
relations between the coefficients $a^{(i,j)}_s$ and we find $\nu_i$
solutions of \eqref{cano2},
\begin{equation}
\VOMOG_{ij}(x)=\sum_{s=0}^\oo\frac{a^{(i,j)}_s
\Gamma(x+1)}{\Gamma(x-\K_{ij}+s+1)} \ , \qquad j=1,\ldots,\nu_i \ .
\end{equation}
Multiplying by $\mu_i^x$ we find solutions of the difference equation
\itref{diff01}.
In total we find $\sum_{i=1}^\lambda \nu_i= \R$ different solutions.
The general solution of the homogeneous difference equation \itref{diff01}
will be a linear combination of the these solutions
with  periodic functions $\tilde\omega_{ij}(x)$ of period one as coefficients,
\begin{equation}\label{resho}
\UOMOG(x) =\sum_{i=1}^\lambda \sum_{j=1}^{\nu_i} 
\tilde\omega_{i j}(x) \mu_i^x \VOMOG_{i j}(x) \ . 
\end{equation}

\subsection{Solution of the nonhomogeneous difference equation}
Let us consider the nonhomogeneous difference equation \itref{equdifref}
of order $\R$
\begin{equation}\label{diff01n}
p_0(x) \UNOMOG(x) +p_1(x) \UNOMOG(x+1) + \ldots + p_\R(x) \UNOMOG(x+\R) 
=F(x) \ ;
\end{equation}
making the same replacements as \eqref{diff01}
the equation becomes
\begin{equation}\label{diff1n}
q_0(x) \UNOMOG(x) +q_1(x) \UNOMOG(x-1) + \ldots + q_R(x) \UNOMOG(x-\R) 
=F'(x) \ ,
\end{equation}
where $F'(x)=F(x-R)$.
In the difference equations encountered in this work, 
the right-hand side can be written as a sum of factorial series expansions.
We assume for simplicity that $F'(x)=\mu^x T(x)$  
and $T(x)$ has the factorial series expansion
\begin{equation}\label{expa1b}
T(x)=c_0\RHO^\K +c_1\RHO^{\K-1} +c_2\RHO^{\K-2} +\ldots \ ,
\end{equation}
where $\mu$, $K$ and $c_i$ are known.

Making the substitution $\UNOMOG(x)=\mu^x \VNOMOG(x)$ in \eqref{diff1n}
and using the relations (\ref{xpiro}-\ref{xpirorel12}) 
we obtain the canonical form
\begin{equation}\label{cano1b}
\left[f_0(\PI) +f_1(\PI)\RHO +f_2(\PI)\RHO^2
 + \ldots + f_m(\PI)\RHO^m\right]\VNOMOG(x) =T(x) \ .
\end{equation}
The expansion in factorial series of $\VNOMOG(x)$ has the form
\begin{equation}\label{serroh}
\VNOMOG(x)=a_0\RHO^{\K-m} +a_1\RHO^{\K-m-1} +a_2\RHO^{\K-m-2} +\ldots \ ;
\end{equation}
inserting this expansion in \eqref{cano1b} and equating the coefficients
of the powers of $\RHO$ one finds the recurrence relations 
\begin{align}
\begin{split}\notag
&a_0 f_m(\K)=c_0 \ ,\\
&a_1 f_m(\K-1) +a_0 f_{m-1}(\K-1)=c_1 \ ,
\end{split}\\
\begin{split}\label{systb2}
&\cdots \\
&a_s f_m(\K-s) +a_{s-1} f_{m-1}(\K-s) +\ldots 
        +a_{s-m} f_0(\K-s)=c_s  \quad (s\ge m) \ . \quad
\end{split}
\end{align}
Solving this system we can find the coefficients $a_s$.
If $f_m(\K-s_0)=0$ for some $s_0$, the term $a_{s_0} f_m(\K-s_0)$
vanishes from the relation \itref{systb2} and, 
in the difference equations encountered in this work, 
the remaining terms form the trivial identity $c_s=c_s$,
so that the value of $a_{s_0}$ remains undetermined and can be chosen at will.
If the factorial series \itref{serroh} converges,
$\UNOMOG(x)=\mu^x \VNOMOG(x)$ is a particular solution
of the nonhomogeneous difference equation.

The case $F'(x)=\mu^x p(x) T(x)$, where $p$ is a polynomial, can be
reduced to the previous case by transforming $x$
into $\PI+\RHO$ in the polynomial 
and letting the operators act on the expansion of $T(x)$
following \eqref{xpiro}.

\section{Determination of arbitrary constants}\label{detcon}
The general solution of the difference equation \itref{equdifref} can be written as
\begin{equation}
U(x)=\sum_{j=1}^\R \tilde\omega_j(x) \UOMOG_j(x) +\UNOMOG(x)\ ,
\end{equation}
where $\tilde\omega_j(x)$ are periodic functions of period 1.
If we consider only integer values of $x$,
the values of $\tilde\omega_j(x)$ are independent of $x$
so that we can replace them with arbitrary constants $\C_j$:
\begin{equation}\label{equhom2}
U(x)=\sum_{j=1}^\R  \C_j \UOMOG_j(x) +\UNOMOG(x)\ .
\end{equation}

The unambiguous solution of the difference equation requires
the determination of the $R$ constants $\C_j$.
The value of these constants can be determined 
\begin{enumerate}
\item\label{c1}
from the large-$x$ behaviour 
of the solution \itref{equhom2} and of the integral \itref{init1},
by equating the first coefficients of the expansions in factorial series;
\item\label{c2} 
by equating the values at $x=0$ of \eqref{equhom2} and \eqref{init1}.
\end{enumerate}
As we will see in the following sections,
the method~\ref{c1} may provide the values of all the constants,
but the determination of the coefficients of the factorial series 
turns out to be a simple task only for euclidean massive integrals
(see sections~\ref{euclsec0}, \ref{neuclsec0} and~\ref{ondeft}),
where it involves integrals with one-loop less,
and gets more complicated 
for non-euclidean integrals or integrals with zero masses
(see section~\ref{othercases});
on the contrary the method~\ref{c2}
has the limitation that it provides only one relation between the constants, 
but on the other hand 
it is not affected by the kind of 
the external momenta or masses (see section~\ref{valx0}).
\subsection{Large-$x$ behaviour of integrals: euclidean massive case}
\label{euclsec0}
In this section we work out a relation between the coefficients
of the expansion in factorial series of the master integral \itref{init1}
and simpler integrals with one-loop less.
We assume that all the external momenta are euclidean,
that is, 
the matrix of the scalar products $p_i \cdot p_j$ is semidefinite non-negative,
and no mass is zero.

Choosing the momentum routing of the integral \itref{init1}
such that $D_1=k_1^2+m_1^2$, we write it as 
\begin{equation}\label{init2}
U(x)=
\int   
\dfrac  {\dk1}{\displaystyle (k_1^2+m_1^2)^x } {g(k_1)} \ ,
\end{equation}
where  $g$ includes the contribution of the remaining $\NK-1$ loops 
\begin{equation}\label{intgi}
g(k_1)=
\int {{\dk2 \; \dots \dk{\NK}} \  
\dfrac
{\prod_{j=1}^{\NPS-\ND} \indpk_j^{\beta_{j}}}
{\displaystyle D_2 D_3 \ldots  D_\ND} \ } \ .
\end{equation}
Note that the function $g$ also depends on the external momenta $p_j$
and the masses $m_2$, $m_3$, $\ldots,$ $m_\ND$ of the other denominators.
From a graphical point of view, $g(k_1)$ corresponds to the original diagram
with the line $D_1$ cut. 
Introducing hyperspherical polar coordinates
for the integration over $k_1$
\begin{equation}
 d^D k_1= |k_1|^{D-1} d|k_1| \,d\Omega_D(\hat k_1) \ ,
\end{equation}
($\Omega_{D}=2\pi^{D/2}/\Gamma(D/2)$ is the $D$-dimensional solid angle)
and separating the angular and radial part,
\eqref{init2} becomes
\begin{equation}\label{intk2}\label{int00}
U(x) = \dfrac{1}{\Gamma(D/2)}
  \int_0^\oo\dfrac{d k_1^2 \;(k_1^2)^{D/2-1}} {(k_1^2+m_1^2)^x}
     f(k_1^2) \ ,
\end{equation}
where $f$ is the angular mean over $k_1$ of $g$
\begin{equation}\label{definf}
 f(k_1^2)=
     \dfrac{1}{\Omega_D}
 \int d\Omega_{D}(\hat k_1) \; g(k_1) \ .
\end{equation}

For large $x$ the factor $(k_1^2+m_1^2)^{-x}$ of \eqref{int00} peaks strongly
around $k_1^2=0$;
because of the assumption on external momenta and masses,
the function $f(k_1^2)$ has no singularities for $k_1^2\ge 0$,
and therefore we expect that the large-$x$ behaviour of the integral 
is controlled only by the behaviour of $f(k_1^2)$ near $k_1^2=0$.
Making the change of variable $k_1^2=m_1^2 \dfrac{u}{1-u}$ in \eqref{int00}
we obtain
\begin{equation}\label{intuf}
U(x)=\dfrac{(m_1^2)^{D/2-x}}{\Gamma(D/2)}\int^1_0 du \; u^{D/2-1}
(1-u)^{x-1-D/2}  \tilde f(u)\ ,
\end{equation}
where $\tilde f(u)=f(m_1^2 u/(1-u))$.
We expand $\tilde f(u)$ in $u$ 
\begin{equation}\label{expuf}
\tilde f(u)=u^\alpha(1-u)^\beta \sum_{s=0}^\oo b_s u^s \ ;
\end{equation}
$\alpha$ is an integer greater than or equal to zero ($f$ is regular
in the origin), while
the factor $(1-u)^\beta$ has been introduced for convenience.
Expressing the integrals over $u$ in terms of Beta function 
(which takes care of analytical continuation if the integral 
diverges at the endpoints), and choosing $\beta=D/2+1$, 
\eqref{intuf} takes the desired form of a factorial series
\begin{equation}\label{resu1}
U(x)=\mu_0^x\sum_{s=0}^\oo a_s \RHO^{\K_0-s} 
    =\mu_0^x\sum_{s=0}^\oo a_s
 \dfrac{\Gamma(x+1)}
       {\Gamma(x+1-\K_0+s)}\ ,
\end{equation}
where 
\begin{equation}\label{resu2}
\mu_0=1/m_1^2,\quad   \K_0=-D/2-\alpha, \quad \text{and  }\ 
a_s= b_s m_1^D \Gamma(s+D/2+\alpha)/\Gamma(D/2)\ .
\end{equation}
The coefficients $b_s$ of \eqref{expuf} can be easily expressed in terms of
the coefficients $f_s$ of the expansion in $k_1^2$ of $f(k_1^2)$
\begin{equation}\label{expuf2}
f(k_1^2)=(k_1^2)^\alpha \sum_{s=0}^\oo f_s k_1^{2s} 
\ ,
\end{equation}
($b_0=m_1^{2\alpha} f_0\ $, 
$\ldots$), 
so that the large-$x$ behaviour of $U(x)$ proves to be determined 
by the behaviour of $f(k_1^2)$ for small $k_1^2$.
For large $x$ the leading behaviour is given by the first term of \eqref{resu1}
\begin{equation}\label{expkf0}
U(x)\approx   (m_1^2)^{D/2-x+\alpha} x^{-D/2-\alpha} f_0
                   \dfrac{\Gamma(D/2+\alpha)}{\Gamma(D/2)}\ .
\end{equation}
In the frequent case of an integral with
numerator equal to one, $\alpha=0$ and  $f_0=f(0)$,
it becomes 
\begin{equation}\label{expkf00}
U(x)\approx   (m_1^2)^{D/2-x} x^{-D/2} f(0) \ .
\end{equation}

Now we equate \eqref{resu1} with the expansion 
in factorial series of the general solution \itref{equhom2}
(see \eqref{resho} and \eqref{serroh}),
obtaining
\begin{equation}\label{gen1}
\left(\frac{1}{m_1^2}\right)^x \sum_{s=0}^\oo a_s \RHO^{-D/2-\alpha-s} = 
 \sum_{j=1}^R \C_j \, \mu_j^x \sum_{s=0}^\oo  \hat a_{js} \RHO^{\K_j-s}
 +
 \sum_l   (\mu^{NH}_l)^x \sum_{s=0}^\oo a_{ls}^{NH} \RHO^{\K^{NH}_l-s}
\ .
\end{equation}
In this equation only the $R$ constants $\C_j$ are unknowns,
each one
corresponding to a solution $\UOMOG_j$ with a different pair of values
$\mu_j$ and $K_j$,
$\mu_j$ being one solution of the characteristic equation of the
homogeneous equation and $\K_j$ being the solution of the corresponding
indicial equation; we have set $\hat a_{j0}=1$.
We have also assumed that the expansion of the nonhomogeneous term contains
a sum of expansions with different pairs of values
$\mu^{NH}_l$  and $\K^{NH}_l$.

Fortunately, the number of constants $\C_j$ to find can be drastically reduced.
If the difference equation is homogeneous,
the constants $\C_{j}$ which may be different from zero are only
those such that the corresponding $\mu_j$ and $\K_j$
satisfy the
condition 
\begin{equation}\label{cond1}
 \mu_j=1/m_1^2\ , \quad  \K_j+D/2+\alpha=\text{integer}\le 0
 \qquad  \Rightarrow \qquad \C_j\not=0\  ;
\end{equation}
all the other constants must be zero.

In the case of nonhomogeneous equation, one must recall that 
it is part of a triangular system of difference equations;
the nonhomogeneous term receives contributions
from master integrals with a smaller number of denominators,
which in their turn satisfy other nonhomogeneous difference equations,
up to the simplest master integrals which satisfy homogeneous equations.
Clearly $\mu^{NH}_l=1/m_1^2$ and  
$\K^{NH}_l+D/2$ is always an integer,
therefore $\C_{j}$ may be different from zero if 
$\mu_j$ and $\K_j$ satisfy \eqref{cond1}, or the new condition 
\begin{equation}\label{cond1a}
 \mu_j=1/m_1^2\ , \quad  
 0<\K_j+D/2+\alpha=\text{integer}\le \max_l \K^{NH}_{l} +D/2+\alpha 
  \qquad \Rightarrow \qquad \C_j\not=0\  
\end{equation}
which implies cancellations between the first coefficients of the homogeneous 
and nonhomogeneous expansions.
We found that in the equations considered in this work, 
the condition \itref{cond1a} is never satisfied, 
and the condition \itref{cond1} is satisfied by 
a small number of pairs of $\mu_j$ and  $\K_j$, often only one;
if no pair satisfies the conditions,
$U(x)$ is completely determined by the nonhomogeneous term.

The values of the non-zero constants $\C_j$ are determined by comparing
the coefficients of the same powers of $\RHO$ of the two sides of \eqref{gen1}.
The required few coefficients $a_s$ 
are easily calculated from the first coefficients $f_s$
of the expansion in $k_1^2$
of $f(k_1^2)$.
\label{euclsec}

The required coefficients $f_s$ are calculated
by expanding numerators and denominators of \eqref{intgi} for small $k_1$,
and by performing the angular integration  over $k_1$
of \eqref{definf}. Angular integrals are straightforward as they contain
exclusively powers of scalar products containing $k_1$.
As a result, the coefficients $f_s$ will be expressed by integrals
over $k_2$,\ldots  $k_\NK$,
belonging to diagrams with \emph{one loop less}
(for example, if the numerator of \eqref{intgi} is unity 
then $f_0=f(0)=g(0)$).
These integrals can be expressed using algorithm~\ref{algsys}
as combinations of the master integrals of the new diagrams;
in their turn, 
these new master integrals can be calculated by inserting an exponent in a
denominator and building and solving new difference equations.
The arbitrary constants of their solutions can be found using
the large-exponent behaviours,
which can be expressed in terms of integrals with another loop less, and so on.
In this way we can explicitly calculate the values of all the arbitrary
constants $\C_j$. This fact is very important.
\subsection{Non-euclidean case: deformation of the radial path}
\label{neuclsec0}
Now we consider the non-euclidean case:
the matrix of the scalar products $p_i \cdot p_j$
is definite negative.
We write \eqref{intk2} as 
\begin{equation}\label{intkh}
U(x,P) =
     \dfrac{1}{\Gamma(D/2)}
\int_{l_0} \dfrac{d k_1^2 (k_1^2)^{D/2-1}} {(k_1^2+m_1^2)^x}
     f(k_1^2,P)\ ,
\end{equation}
where we explicitly show the dependence on the external momenta
$P=$ $\{p_1,$ $\ldots,p_{\NP}\}$,
and where $l_0$ indicates the path of the radial integration.
For the moment the path is assumed to be the positive axis in the complex plane.

The integral $U(x,P)$, considered as a function of the external momenta $P$,
is defined in the non-euclidean region by
an analytical continuation through a generic path in the complex $P$-space.
The path begins in some euclidean initial point $P_{in}$ and ends
in the desired non-euclidean final point $P_{end}$.
It is possible that in some intermediate point
of the continuation path in the $P$-space,
for a particular value of the external momenta $P_s$,
$f(k_1^2,P_s)$ is singular in a point $k_1^2=q^2>0$, just on the radial
integration path, breaking off the analytical continuation.
If this singularity in $k_1^2$ cannot be avoided by modifying
the continuation path
in the $P$-space,
then we are forced to deform\footnote{
The deformation of the radial integration path was 
first discussed in \cite{Levinehyp} for self-energy diagrams in $D=4$.
}
the integration path in the $k_1^2$-space
by turning around the singularity;
the singularity in $k_1^2$ itself moves in the complex $k_1^2$-plane
to a final point $k_{end}^2$ when we complete the analytical continuation
from $P_s$ to $P_{end}$. 
The final integration path $l_0$ starts in $k_1^2=0$, turns around
the singularity $k_{end}^2$ (usually on the negative axis) 
and comes back to $k_1^2\to +\oo$. 
In the general case of multiple singularities the path turns around 
all them.
The deformation of the radial path must be performed
when the values of the scalar products $p_i \cdot p_j$ 
get over some ``deformation thresholds''
in the non-euclidean region,
the exact point of these thresholds being depending on the values of masses and
the structure of the diagram.

One can show 
using Feynman parameters 
that the deformation thresholds are
the thresholds (anomalous thresholds included) in $P$ of $g(0,P)$,
the diagram obtained by eliminating the line $D_1$ and setting $k_1=0$ everywhere.
Possible denominators depending only on $k_1$ give ``additional''
deformation thresholds of the kind $p^2=-m^2$.

Once the deformation thresholds are determined,
we can split the non-euclidean region of the $P$-space into two regions:
\begin{enumerate}
\item one region below the deformation thresholds 
     which adjoins the euclidean region,
     where the integration path is the same as the euclidean region;
\item the remaining region above the deformation thresholds 
     where the integration path turns around some singularities.
\end{enumerate}

\subsubsection{Example: deformation of the path for the one-loop self-energy integral}
As an example, now we consider the simplest case,
the one-loop integral
\begin{equation}\label{hyp14r}
I=\int   
\dfrac  {\dk{}}{\displaystyle (k^2+m_1^2) ((p-k)^2+m_2^2)}
=
\dfrac{1}{\Gamma(D/2)}
\int_{l_0}    \dfrac{d k^2 (k^2)^{D/2-1}} {k^2+m_1^2}
    \dfrac{1}{\Omega_D}\int \dfrac{d\Omega_D(\hat k)}{(p-k)^2+m_2^2}\ .
\end{equation}
The angular integral is\footnote{
Here we generalize to arbitrary $D$ dimensions
the result obtained in \cite{Levinehyp} in the four-dimensional case.} 
\begin{equation}\label{hyp14}
\dfrac{1}{\Omega_D}\int \dfrac{d\Omega_{D}(\hat k)}{(p-k)^2+m_2^2}=
\dfrac{Z}{pk} \Phi(Z^2) \ ,
\end{equation}
where $\Phi(Z^2)$ is the hypergeometric function $F(1,2-D/2;D/2;Z^2)$,
\begin{equation}
Z^{\pm1}=\dfrac{p^2+k^2+m_2^2\mp R(p^2,k^2,-m_2^2)}{2p k}\ ,
\qquad p=\sqrt{p^2}\;, \quad k=\sqrt{k^2}\;,\quad
\end{equation}
and $R(x,y,z)$ is the usual two-body phase space square root
\begin{equation}\label{defrxyz}
R(x,y,z)=\sqrt{x^2+y^2+z^2-2xy-2xz-2yz}\ .
\end{equation}
The angular integral \itref{hyp14} has two branching points
in the complex $k^2$-plane, $k^2=(p \pm i m_2)^2$.
Considering \eqref{hyp14r}
in the euclidean case, $p^2>0$, the path of integration is always
the positive real axis.
If $p^2 < 0$, \eqref{hyp14r} must be analytically continued in the
complex $p-$plane from a generic initial euclidean point $p=p_{in}>0$
to the non-euclidean final point $p=i\sqrt{-p^2}$ which is
on the imaginary axis.
The initial point has $\Im p_{in} =0$. 
If $p^2 \ge -m_2^2$
no radial singularity appears so that 
\begin{equation}\label{hypbelow}
I= 
\dfrac{1}{\Gamma(D/2)}
\int_0^\oo \dfrac{d k^2 (k^2)^{D/2-1}} {k^2+m_1^2}
  \dfrac{Z}{pk}\Phi(Z^2)  \ .
\end{equation}
If $p^2<-m_2^2$,  the final point has $\Im p >m_2$ so that
any path connecting the initial and final point must cross the line
$\Im p =m_2$ in some point.
Let  $p_c=im_2+c$ be this point;
if we consider \eqref{hyp14} with momentum $p_c$, one of the branching points
falls in $k^2=c^2>0$ which is exactly on the path of integration.
Therefore the path of integration $l_0$ must be deformed avoiding the
branching point $(p-im_2)^2$ which is on the negative real axis.
One chooses as path of integration\cite{Levinehyp}
first the segment $((p-im_2)^2,0)$ above the
negative real axis, taking the square root $R$ with the plus sign
and using $Z^{-1}$ in the place of $Z$,
then the straight line $((p-im_2)^2,\oo)$ below the axis.
Therefore, if $p^2<-m_2^2$ \eqref{hyp14r} becomes 
\begin{equation}\label{hypabove}
I=
\dfrac{1}{\Gamma(D/2)}
 \int_0^{(p-im_2)^2}   \dfrac{d k^2 (k^2)^{D/2-1}} {k^2+m_1^2}
  \dfrac{Z^{-1}}{pk}\Phi(Z^{-2})  
+
\dfrac{1}{\Gamma(D/2)}
\int_{(p-im_2)^2}^\oo \dfrac{d k^2 (k^2)^{D/2-1}} {k^2+m_1^2}
  \dfrac{Z}{pk}\Phi(Z^2)  \ .
\end{equation}
This result will be used in section~\ref{1loopself}.

\subsection{Large-$x$ behaviour of integrals: non-euclidean massive case
 below and at the deformation threshold}
In the region below the deformation thresholds,
in the case of non-euclidean massive integrals,
the determination of the large-$x$ behaviour is quite similar
to that described in section~\ref{euclsec0}.

\label{ondeft}
At the deformation threshold
the determination of the large-$x$ behaviour
presents some peculiarities, since
some external momenta have on-mass-shell values.
In the integral \itref{intgi} denominators of the form $k_1^2-2 \bar p\cdot k_1$
appear, where $\bar p$ is some on-mass-shell momentum.
Considering the expansion in $k_1$ of $g(k_1)$,
these denominators vanish at $k_1=0$ and cannot be expanded 
in a way as straightforward as for off-mass-shell denominators; 
they must be included in the angular integrals,
complicating quite a lot the integration.

In order to explain the situation,
let us consider here the case of a one-loop diagram 
with $N+1$ external lines,
and the integral 
\begin{equation}
\int  \dfrac  {\dk{}}{\displaystyle
(k^2+m_0^2)((p_1-k)^2+m_1^2)\ldots ((p_N-k)^2+m_N^2)} \ .
\end{equation}
Setting $k=0$ we see that the deformation threshold is $p_i^2=-m_i^2$,
$i=1$, \ldots, $N$.
Now we set the external momenta to these on-mass-shell values,
and we consider the integral
\begin{equation}\label{intkn1}
W(x)=\int \dfrac{\dk{}}{(k^2+m_0^2)^x
(k^2-2 p_1\cdot k)\cdots (k^2-2 p_N\cdot k)} \ ,
\end{equation}
whose we want to calculate the large-$x$ leading behaviour. 
Following the notation of section~\ref{euclsec0} we write 
\begin{equation}\label{hyp000}
W(x) =\int_0^\oo\dfrac{d k^2 (k^2)^{D/2-1}} {(k^2+m_0^2)^x}
     \dfrac{f(k^2)}{\Gamma(D/2)}\ ,
\end{equation}
\begin{equation}\label{hypint1}
f(k^2)=\dfrac{1}{\Omega_D}\int \dfrac{d\Omega_D(\hat k)}
{(k^2-2 p_1\cdot k)\cdots (k^2-2 p_N\cdot k)}\ .
\end{equation}

Now we extract
the leading behaviour of $f(k^2)$ for $k^2 \to 0$.
Inserting Feynman parameters $x_i$, $i=1$, $\ldots$, $N-1$ one finds
\begin{equation}\label{hypfe}
f(k^2)=\frac{\Gamma(N)}{\Omega_D}
\int {dx_1 \ldots dx_{N-1}} \int \dfrac{d\Omega_D(\hat k)}
{(k^2-2P(x_i)\cdot k)^N}\ ,
\end{equation}
where $P(x_i)=\sum_{i=1}^N x_i p_i$ and $\sum_{i=1}^{N}x_i=1$.
The angular integral in $D$ dimensions 
can be expressed using a generalization of the formula \itref{hyp14}
\begin{equation}\label{hyp14f}
\dfrac{1}{\Omega_D}\int \dfrac{d\Omega_{D}(\hat k)}{((p-k)^2+m^2)^x}=
\left(\dfrac{Z}{pk}\right)^x F(x,x+1-D/2;D/2;Z^2) \ ;
\end{equation}
using properties of hypergeometric function one finds
for $k^2\to 0$
\begin{equation}\label{hypsvil0}
\dfrac{1}{\Omega_D}\int \dfrac{d\Omega_{D}(\hat k)}{(k^2-2P(x_i)\cdot k)^N} \approx 
(-P^2(x_i) k^2)^{-N/2} 
\dfrac{\Gamma(D/2)\Gamma(N/2)}{2\Gamma(N)\Gamma((D-N)/2)}\ ,\quad
\end{equation}
then
\begin{equation}\label{fk2a1}
f(k^2)\approx\dfrac{\Gamma(D/2)\Gamma(N/2)}{2\Gamma((D-N)/2)}
(k^2)^{-N/2} \int \dfrac{dx_1 \ldots dx_{N-1}}{\left(-P^2(x_i)\right)^{N/2}}
\ .
\end{equation}
We see that $f(k^2)$ is singular in $k=0$ and,
if $N>D$, the integral over $k^2$ of \eqref{hyp000}
diverges for small $k^2$;
\eqref{expkf0} with arbitrary $\alpha$ takes care of the necessary analytical
continuation.

Let us now consider a one-loop integral in $E$
dimensions with the same denominators as the angular integral \itref{hypint1}
\begin{equation}\label{fdefle}
L_E(p_i) \equiv \dfrac{1}{2}
\int\dfrac{[d^E q]}{
{(q^2-2 p_1\cdot q)\cdots (q^2-2 p_N\cdot q)}
}\ ;
\end{equation}
introducing Feynman parameters in the same way as for \eqref{hypfe} we obtain
\begin{equation}\label{fk2a2}
L_E(p_i)= \dfrac{\Gamma(N-E/2)}{2} \int \dfrac{dx_1 \ldots
dx_{N-1}}{\left(-P^2(x_i)\right)^{N-E/2}}\ .
\end{equation}
If we choose $E=N$ the integrals over $x_i$ of \eqref{fk2a1} and \eqref{fk2a2} 
become identical,
so that we can rewrite \eqref{fk2a1} as
\begin{equation}\label{connhyp2}
f(k^2)\approx{(k^2)^{-N/2}}\dfrac{\Gamma(D/2)}{\Gamma((D-N)/2)} 
L_N(p_i)\ .
\end{equation}
This result shows that the angular integral \itref{hypint1}
with $N$ on-mass-shell denominators
in the limit $k^2\to 0$
is proportional to a one-loop integral with the same denominators
(one less than \eqref{intkn1}),
calculated in a \emph{number of dimensions equal to the number of on-mass-shell
denominators}.
Moreover, using \eqref{connhyp2} and \eqref{expkf0}
(with $\alpha=-N/2$ according to \eqref{fk2a1})
one finds the large-$x$ leading behaviour of
\eqref{intkn1}:
\begin{equation}\label{expkf11}
W(x)\approx (m_0^2)^{D/2-N/2-x}  x^{-D/2+N/2}  L_N(p_i)\ .
\end{equation}
This result is similar to euclidean leading behaviour 
for a one-loop vacuum integral (\eqref{expkf00} with $f(0)=1$);
the differences are the change of the exponents and 
the multiplication by the factor $L_N(p_i)$
which is \emph{independent} of the number of dimensions $D$ of the integral
\itref{intkn1}.
As a consequence of the divergence of the radial integral
and of the necessary analytical continuation,
we observe an apparently paradoxical result:
if $N>D$ and $m_0=1$, from \eqref{expkf11} we see that
the integral $W(x)$ increases as the exponent $x$ increases,
while in the euclidean case the integral $U(x)$ decreases as $x$ increases.
The integral $L_N(p_i)$ may be calculated analytically,
for example, extracting the $k^2\to 0$ behaviour
from the known analytical expressions of the angular integral \itref{hypint1}
for $D=4$ and $N=1$~\cite{Levinehyp} and $N=2,3$~\cite{NChyp},
or numerically,
by inserting an exponent in one denominator of \eqref{fdefle}
and building and solving a difference equation,
or by using the identities of the section~\ref{valx0} 
(for example see \eqref{L1num}).
In section~\ref{resu} we will need 
the following values $\tilde L_N$ of $L_N(p_i)$ in the 
on-mass-shell and equal masses case $m_i=1$, $p_i^2=-1$ and  $(p_i-p_j)^2=-1$ 
for every $i$ and $j$:
\begin{equation}\label{valuesl}
\begin{split}
\tilde L_1=&{\sqrt{\pi}}/{2} \;,
\qquad 
\tilde L_2={\pi\sqrt{3}}/{9}\;,\\
\tilde L_3=&\left(\arctan{\sqrt{8}}-\arctan{\sqrt{3}}\right)\sqrt{{9\pi}/{8}} \;,
\quad 
\tilde L_4=0.172751462\ldots \;.\\ 
\end{split}
\end{equation}
The generalization of \eqref{expkf11} to multi-loop integrals is
straightforward; considering
\begin{equation}
W'(x)=\int \dfrac{\dk{}}{(k^2+m_0^2)^x
(k^2-2 p_1\cdot k) \cdots (k^2-2 p_N\cdot k) }
\,h(k)\ ,
\end{equation}
\begin{equation}
h(k)=\int 
\dfrac
{\dk2 \cdots \dk{\NK}}
{D_{N+2} \cdots D_\ND}\ ,
\end{equation}
the large-$x$ leading behaviour is 
\begin{equation}
W'(x) \approx (m_0^2)^{D/2-N/2-x}  x^{-D/2+N/2}
L_N(p_i) \, h(0)\ .
\end{equation}
The condition \itref{cond1} in this case must be
modified to
\begin{equation}\label{cond1mod}
 \mu_j=1/m_1^2\ , \quad  \K_j+D/2-N/2=\text{integer or half-integer}\le 0 
 \qquad \Rightarrow \qquad \C_j\not=0\  .
\end{equation}

\subsection{{\hskip 1pt} Non-euclidean case above the deformation threshold 
             and  massless case}
\label{othercases}
The cases described above, euclidean and non-euclidean
below the deformation thresholds,
are both united by the fact that the
large-$x$ behaviour of $U(x)$ depends only on the behaviour of $f(k_1^2)$
near the origin. In these cases it is easy to express this behaviour
in terms of simpler diagrams obtained by putting $k_1=0$
in the main diagram.
The situation gets more complicated
if $f(k_1^2)$ has other singularities on the integration
path: for the $i$th singularity placed in $q_i^2 \not= 0 $,
an additional contribution to the large-$x$ behaviour of $U(x)$ appears,
of the form $(q_i^2+m_1^2)^{-x} V(x)$, with $V(x)\sim x^\K$ for large $x$.
The precise form of $V(x)$ depends on
the behaviour of $f$ near the singularity;
unfortunately its determination requires the calculation of angular integrals
for $k_1^2 \not=0 $
(see \eqref{definf}) 
\begin{equation}
\int_{k_1^2\approx q_i^2} d\Omega_D(\hat k_1) \; g(k_1)\ ,
\end{equation}
which is difficult and  case-dependent.
In this work we have performed these calculations only in the one-loop example
of section~\ref{1loopself}.
In the case of more complicated diagrams
we have preferred to avoid the calculation of these angular integrals 
and to look at the problem from a different point of view,
using a method with broader applicability 
based on differential equations (see section~\ref{differentialequ}).

The cases where singularities of $f$ other than the origin appear are:
\begin{itemize}
\item
The non-euclidean massive case, above the deformation threshold:
the path of integration must be deformed, turning around a number of
singularities of $f(k_1^2)$. 
See section~\ref{m1m2nz} for an example.
\item
The zero mass case, where $m_1>0$ and some of the masses $m_2$, \ldots, $m_\ND$
are zero. If the external momenta are euclidean, 
singularities on the positive axis of $k_1^2$ may appear,
but this fact does \emph{not} require deformation of the path.
See section~\ref{m2zero} for an example.
\end{itemize}

The situation worsens if the mass $m_1$ of the
denominator raised to $x$ is zero.
In this case the behaviour of $U(x)$ for large $x$ depends
on the behaviour of $f(k_1^2)$ on the whole real axis, and not in some
isolated points.
This can be easily understood by noting that in this case
$U(x+D/2-1)$ is the Mellin transform of $f(k_1^2)$.
See section~\ref{m1zero} for an example.

\subsection{Zero exponent condition}
\label{valx0}
A very useful relation between the constants $\C_j$ arises from \eqref{equhom2}
calculated at $x=0$:
\begin{equation}\label{rel0}
U(0)-\UNOMOG(0)=\sum_{j=1}^\R \C_j\UOMOG_j(0)\ .
\end{equation}
$U(0)$ is an integral without the denominator $D_1$;
if the value of $U(0)$ can be found by some method,
as an example, by adding an exponent to $D_2$ and by solving a difference
equation,
and if at least one $\UOMOG_j(0)$ has a non-zero value,
\eqref{rel0} establishes one new relation between the constants $\C_j$.
The existence of this  relation must be verified
in each particular case.
The advantage of this relation 
over the relations found using the asymptotic behaviour of $U(x)$,
is that it is valid for every kind of integral, regardless of the 
value of masses or external momenta.

It is possible to construct identities analogous to \eqref{rel0},
by inserting in \eqref{equhom2} $x=-1,-2,\ldots $ instead of $x=0$. 
Unfortunately in all the analyzed cases  
the identities so found turned out to be equivalent to \eqref{rel0}.

\section{Evaluation of factorial series}\label{sumfact}
Once the constants $\C_j$ of \eqref{equhom2}
have been determined using the methods described in section \ref{detcon},
in order to obtain the value of the master integral $U(1)$
we must calculate the values of the homogeneous solutions
$\UOMOG_j(1)$ and of the nonhomogeneous solution $\UNOMOG(1)$.

\subsection{Convergence of factorial series and instabilities of recurrence
relations}\label{sumfact2}

Let us unify the notation by considering the solution $U^{(\alpha)}(x)$,
where $(\alpha)$ indicates one of the solutions of the 
homogeneous or nonhomogeneous equation,
and expand it in factorial series: 
\begin{equation}\label{vaser}
U^{(\alpha)}(x)=\left(\mu^{(\alpha)}\right)^x \sum_{s=0}^\oo a_s^{(\alpha)}
\dfrac{\Gamma(x+1)}{\Gamma(x+1-\K^{(\alpha)}+s)} \ .
\end{equation}

Convergence of the series depends on the value of the abscissa of
convergence $\lambda$.
Analyzing the large-$s$ behaviour of the coefficients
$a_s^{(\alpha)}$ one finds that
the series has $\lambda<\oo$
if none of the solutions $\mu_j$ of the characteristic equation \itref{equchar}
satisfies the condition
\begin{equation}\label{conv}
0<|\mu_j/\mu^{(\alpha)}-1|< 1 \ , \qquad j=1,\ldots,\R \ .
\end{equation}
If $\lambda=\oo$ the series is everywhere divergent 
and the expansion in factorial series is only formal;
in this case another method must be used to calculate $U^{(\alpha)}(1)$
(see section~\ref{Laplacesec} on the Laplace's transformation).

If $\lambda<\oo$ the convergence of the factorial series is
logarithmic, that is,
if $S_m(x)$ is the sum of the first $m$ terms of \eqref{vaser},
one finds that $|S_m(x)-U^{(\alpha)}(x)|\sim {m^{\lambda-x}}$ for
large $m$.
The series converges if $x>\lambda$,
more and more quickly
as $x$ increases.
As usually one finds $\lambda\sim 1$,
it is not convenient or possible to calculate $U^{(\alpha)}(1)$
directly by summing the factorial series.
Therefore, chosen a number $x_{max}$ conveniently large,
one calculates $U^{(\alpha)}(x)$ for $R$  contiguous values of $x$,
 $U^{(\alpha)}(x_{max})$,
 $U^{(\alpha)}(x_{max}+1)$, \ldots,
 $U^{(\alpha)}(x_{max}+R-1)$,
where the series converges faster,
and one uses repeatedly the corresponding recurrence relation, 
\eqref{equdifref} or \eqref{equr1},
in order to obtain the values of $U^{(\alpha)}(x)$
for
$x=x_{max}-1,x_{max}-2,\ldots$ up to $x=1$.
A drawback of this procedure is that the recurrence relation may be unstable,
so that each iteration causes a loss of precision.

Let $A^{-1}=\min_{j}|\mu_j/\mu^{(\alpha)}|$.
The recurrence relation is unstable if
\begin{enumerate}
\item  $A>1$: \label{instab1}
       in this case each iteration increases the error on $U^{(\alpha)}(x)$ of
       a factor $A$.
\item  $A=1$, and $\mu^{(\alpha)}$ is a root of \eqref{equchar} of multiplicity $m>1$:
       in this case $n$ iterations of the recurrence relation
       increase the error on $U^{(\alpha)}(x)$ of a factor $n^{m-1}$;
       this is a kind of instability weaker than the preceding one.
\end{enumerate}

If the recurrence relation is stable $x_{max}$ can be chosen large at will,
with no effect on the precision of $U^{(\alpha)}(1)$.
In the case of instability with $A>1$,
in order to obtain the result $U^{(\alpha)}(1)$ with
a number $E$ of exact digits,
the calculations of $U^{(\alpha)}(x_{max}+i)$
must be performed with a greater number of decimal digits
$C=E+x_{max}\log_{10}A $.
Supposing $a_s^{(\alpha)}\sim s!$, 
a rough estimate of the number $s_{max}$ of terms
of the series  needed to obtain
the sum with such a precision is
$s_{max}\sim A^{\frac{C}{C-E}}{x_{max}}$.
Performing the calculations with fixed precision arithmetic
with $C$ digits, it is convenient to choose a value of $x_{max}$ as low as
possible in order to obtain the greatest $E$,
compatibly with the rapid increases of $s_{max}$ and of the computing time
(see an example in section~\ref{numexa}).
Performing calculations with multiprecision arithmetic, $C$ can be chosen at will,
and $x_{max}$ can be increased;
a convenient choice which minimizes the estimate of $s_{max}$ is
$C/E\sim 1+\ln A$ (see an example in section~\ref{pair2}).
Fixed $C/E$, by varying $E$ one sees that $x_{max}\propto E$
and $s_{max} \propto E$.
Therefore the number of terms of the series  
(and even the computation time) is proportional to the
number of digits of precision of the results;
this is true even in the stable case provided that 
one chooses $x_{max}\propto E$.

\subsection{Truncated expansion in $\e$}\label{truncd}
We are interested in the results in the limit $D \to 4$;
therefore, defined $\e=(4-D)/2$, we expand all the quantities in $\e$,
truncating the expansions at the first $n_\e$ terms, and we perform all
the calculations using truncated series;
in this way the coefficients of all the powers of $\e$ are found numerically,
including the \emph{negative} powers.
This technique is perhaps not the most efficient, but is very versatile.
We have implemented in the program \SYS~ (see section \ref{calcprog}) 
the arithmetic of truncated series, so that the use of series 
becomes as simple as with ordinary numbers.

The time of computation of a factorial series 
grows approximately as $n_\e^2$,
and it is mainly due to multiplications.
The division between series is an operation 
less frequent than the multiplication as it
occurs only once in the calculation of each $a_s$ with
the recurrence relations \itref{systa2} or \itref{systb2}
or in the calculation of each $U(x)$ using the recurrence relations
\itref{equdifref} or \itref{equr1}.
As effect of cancellations of terms when series are summed,
and of divisions by series beginning with a non-zero power of $\e$,
first or last terms of the expansions in $\e$ may be lost;
in this case the calculations must be necessarily performed
with an initial number $n_\e$
of terms of the expansions greater than the
desired final number $n'_\e$ of terms
of $U(x)$. The number $n_\e$ is chosen empirically.
The loss of terms of the expansions in $\e$ is frequent when 
the recurrence relations
\itref{equdifref} or \itref{equr1} are used to obtain $U(x)$
for $x\le \lambda$; this is due to the fact that for such values of $x$
the expansion in $\e$ of $p_0(x)$ begins with a non-zero power of $\e$.
Furthermore, we found that it is critical to recognize 
the cancellation of the first coefficient of a series used as divisor 
when, because of the numerical errors, the coefficient is a very small number
instead of zero; in this case a numerical cutoff must be carefully used.

\section{Applications to simple one-loop integrals}\label{example1loop}
In this section we discuss in detail 
(as we assume the reader unfamiliar with these operator techniques)
the solution of the difference equations
for the one-loop vacuum and self-energy integrals.

\subsection{One-loop vacuum integral}\label{1loopv}
Defining 
\begin{equation}\label{inte0}
J(x)=\int \dfrac{\dk{}}{(k^2+m_1^2)^x}\ ,
\end{equation}
we want to calculate the master integral $J(1)$.
According the single identity of the set $\group{1}{0}{0}$
$J(x)$ satisfies the homogeneous difference equation
\begin{equation}
\label{equ1den}
m_1^2 (x-1) J(x)-(x-1-D/2)J(x-1)=0\ .
\end{equation}
We look for a solution\footnote{The exact solution
$J(x)=a_0\Gamma(x-D/2)/\Gamma(x)$ does not have the form of \eqref{remfac}.}
of this equation in the form of a factorial series
(see section~\ref{solfactser}),
\begin{equation}\label{factserese1}
J(x)=\mu^x V(x) =\mu^x \sum_{s=0}^\oo a_s \RHO^{K-s}\ .
\end{equation}
Introducing the operators $\PI$ and $\RHO$,
multiplying by $x$ and using \eqref{xpirorel12} 
we find the first canonical form of the
difference equation
\begin{equation}\label{fcf1}
 \left(
  (\mu m_1^2 -1 )\RHO^2
 +((2\mu m_1^2-1) (\PI - 1) + D/2 )\RHO
 +\mu m_1^2 \PI(\PI-1) 
 \right) V(x)=0 \ .
\end{equation}
The characteristic equation \eqref{equchar}, $f_2(\mu)=0$, 
gives the value $\mu=1/m_1^2$.
Substituting this value of $\mu$ in \eqref{fcf1} one obtains 
\begin{equation}
\left(
( \PI - 1 + D/2 ) \RHO 
+\PI(\PI-1)
\right) V(x) =0 \ . 
\end{equation}
The indicial equation \eqref{indic}, $f_1(\PI=1+\K)=0$, 
gives the value $\K=-D/2$.
Using the recurrence relation \itref{systa2} one finds
\begin{equation}\label{asex1}
a_s=
\dfrac{\Gamma(s+D/2)\Gamma(s+D/2+1)}{\Gamma(D/2)\Gamma(D/2+1)\Gamma(s+1)}
 a_0 \ .
\end{equation}
\label{vala0}
From \eqref{expkf00} one finds that $ a_0=(m_1^2)^{D/2}$.
The final result is 
\begin{equation}\label{resfin0}
J(x)=(m_1^2)^{D/2-x} \sum_{s=0}^\oo  
\dfrac{\Gamma(s+D/2)\Gamma(s+D/2+1)}{\Gamma(D/2)\Gamma(D/2+1)\Gamma(s+1)}
\dfrac{\Gamma(x+1)}{\Gamma(x+1+D/2+s)}\ .\quad
\end{equation}
The coefficients $a_s$ behave for large $s$ as $a_s/s! \propto s^{D-1}$,
therefore the term of the factorial series behaves as $s^{D/2-1-x}$,
so that the series has abscissa of convergence $\lambda=D/2$.
This value signals the divergence of the integral \itref{inte0}.
The recurrence relation  \itref{equ1den}, being of order one,
is numerically stable. Therefore it is possible to obtain $J(x)$
for a large integer $x \gg D/2$
by summing a few terms of the factorial series
and to use the recurrence relation to obtain $J(1)$.
For a numerical example in the limit $D\to 4$ see section~\ref{numexa}.

\subsection{One-loop self-energy integral}\label{1loopself}
Let us consider the one-loop self-energy integral 
\begin{equation}
I(1)=\int \dfrac {\dk{}} {D_1 D_2}\ ,
\end{equation}
where $D_1=k^2+m_1^2$ and $D_2=(p-k)^2+m_2^2$.
A simple inspection of the system of identities 
$\group{1 \ldots 2}{0 \ldots 1}{0}$
shows that there are three master integrals, 
$\int \dk{}/D_1D_2$, $\int \dk{}/D_1$ and $\int \dk{}/D_2$.
We must find three difference equations for 
\begin{equation}\label{inte1}
I(x)=\int \dfrac {\dk{}} {D_1^x D_2} \ ,
\qquad
J(x)=\int \dfrac {\dk{}} {D_1^x} \ ,
\qquad
K(x)=\int \dfrac {\dk{}} {D_2^x} \ .
\end{equation}
The solution of the system made up of the set of identities
$\group{1 \ldots 2}{0}{0 \ldots 1}$ gives the system of difference equations
\begin{multline}\label{diffequ1}
  (x-D)I(x-2)+
  (-p^2+m_2^2-m_1^2)(2 x-D-1)I(x-1)
 +R^2(p^2,-m_1^2,-m_2^2) (x-1) I(x)  \\
   +J(x-1) ((D/2-1) (p^2+m_2^2+m_1^2)-(x-2) (p^2+m_2^2))/m_1^2=0 \ , 
\end{multline}
\begin{equation}\label{diffequ1b}
m_1^2 (x-1) J(x)-(x-1-D/2)J(x-1)=0 \ ,
\end{equation}
\begin{equation}\label{diffequ1c}
m_2^2 (x-1) K(x)-(x-1-D/2)K(x-1)=0 \ ,
\end{equation}
where $R$ was defined in \eqref{defrxyz}.
The difference equation for $J(x)$ and $K(x)$ has been solved 
in the preceding section, so that we consider here only
the second-order difference equation for $I(x)$.
We discuss separately the massive case and 
the cases where one of the masses $m_1$ or $m_2$ is zero.

\subsubsection{Case 1: $m_1\not = 0$, $m_2\not = 0$}\label{m1m2nz}
Following section~\ref{solfactser}
we substitute $I(x)=\mu^x V(x)$ and $J(x)=\mu^x W(x)$ in \eqref{diffequ1},
multiply it by $x(x-1)$ and insert the operators $\PI$ and $\RHO$.
We obtain the canonical form of the equation 
\begin{equation}\label{chan1}
\left( f_3(\mu)\RHO^3 +f_2(\PI,\mu)\RHO^2 +f_1(\PI,\mu)\RHO +f_0(\PI,\mu)
\right)V(x)= 
\mu\left(g_3(\PI)\RHO^3+g_2(\PI)\RHO^2+g_1(\PI)\RHO \right)W(x) \ ;
\end{equation}
$f_i$ and $g_i$ are polynomials in $\PI$, $\mu$, $D$, $p^2$, $m_1^2$ and $m_2^2$,
whose explicit expressions are not shown for brevity.
The characteristic equation $f_3(\mu)=0$ has the two different roots
\begin{equation}
\mu=\mu_\pm=\dfrac{1}{(p\pm i m_2)^2 + m_1^2}\ 
\end{equation}
(we define $p=\sqrt{p^2}$, and if $p^2<0$  then $p=i\sqrt{-p^2}$).
The solution of the difference equation \itref{diffequ1} is 
\begin{equation}\label{solgeni}
I(x)=\C_{+}\IOMOG_{+}(x) + \C_{-}\IOMOG_{-}(x) +\INOMOG(x)\ ,
\end{equation}
where $\IOMOG_{\pm}$ are the solutions of the 
homogeneous equation corresponding
to $\mu_{\pm}$ and $\INOMOG$ is one solution of the nonhomogeneous equation.

Let us now consider the homogeneous solution $\IOMOG_{-}$.
We write
$\IOMOG_{-}(x) = \mum^x \VOMOG_{-}(x) $ and
we look for the coefficients of the expansion in factorial series of 
$\VOMOG_{-}(x)$
\begin{equation}\label{factserm}
\VOMOG_{-}(x)=\sum_{s=0}^\oo a_s^{-} \RHO^{\K_{-}-s}\ .
\end{equation}
Substituting $\mu=\mum$,
the homogeneous part of the canonical form \itref{chan1} becomes 
\begin{equation}\label{chan2}
\left( f_2^{-}(\PI)\RHO^2 +f_1^{-}(\PI)\RHO +f_0^{-}(\PI) \right)\VOMOG_{-}(x)
= 0\ ,
\end{equation}
\begin{equation}
\begin{split}
f_2^{-}(\PI)=& 2im_2p (2 \PI+D-5) \ ,\\
f_1^{-}(\PI)=& \left((p^2+m_1^2-m_2^2) (\PI+D-3)+2im_2p(3 \PI-4) \right) 
(\PI-1) \ ,\\
f_0^{-}(\PI)=&(p^2+m_1^2-m_2^2+2im_2p) \PI (\PI-1)^2 \ ;\\
\end{split}
\end{equation}
the indicial equation $f_2^{-}(\K_{-}+2)=0$ gives $\K_{-}=(1-D)/2$.
Fixed $a_0^{-}=1$, the other coefficients $a^{-}_s$ can be found 
using the recurrence relation \itref{systa2};
the behaviour of $a_s^{-}$ for large $s$ can be determined 
by considering the recurrence relation as a difference equation in $s$ for
$a^{-}_s$ and solving it; one finds  
\begin{equation}\label{asminf}
a_s^{-}/s!\approx C_1 s^{(3D-7)/2} + C_2 B^s/s\ ,
\end{equation}
where $B=\mum/(\mum-\mu_{+})$, while $C_1$ and $C_2$ are constants.
If $|B|>1$ the series \itref{factserm} never converges
(in fact the condition \itref{conv} is satisfied);
if $|B|\le 1$, the series converges with abscissa of convergence $\lambda=D-2$.
The solution $\IOMOG_{+}$ can be obtained in analogous way 
with the replacement $im_2p \to -im_2 p$.

Let us now consider the nonhomogeneous solution $\INOMOG(x)$.
We write $\INOMOG(x)=\mu^x \VNOMOG(x)$.
The value of $\mu$ must be taken from the expansion  in factorial
series of $J(x)$ obtained in section~\ref{1loopv}: $\mu=1/m_1^2$.
Substituting this value of $\mu$ in the canonical form,
\eqref{chan1} becomes
\begin{equation}\label{chan3}
\left( \hat f_3(\PI)\RHO^3 +\hat f_2(\PI)\RHO^2 
      +\hat f_1(\PI)\RHO   +\hat f_0(\PI)       \right)\VNOMOG(x) = 
 \left(\hat g_3(\PI)\RHO^3 +\hat g_2(\PI)\RHO^2 +\hat g_1(\PI)\RHO
\right)W(x)\ ,
\end{equation}
\begin{equation}
\begin{split}
\hat f_3(\PI)=&(p^2+m_2^2)^2\ ,\\
\hat f_2(\PI)=&     \left( 3 (p^2+m_2^2)^2     +2 m_1^2 (p^2-m_2^2)\right) \PI
                     -5 (p^2+m_2^2)^2  
		     +(D-5) m_1^2 (p^2-m_2^2)      \ ,\\
\hat f_1(\PI)=& \left[
     \left((p^2+m_2^2)^2+m_1^2 (p^2-m_2^2)\right)(3 \PI-4)  
      +m_1^2  (p^2+m_1^2-m_2^2) (\PI+D-3) 
    \right](\PI-1)\ ,\\
\hat f_0(\PI)=& R^2(p^2,-m_1^2,-m_2^2) \PI (\PI-1)^2   \ ,\\
\hat g_3(\PI)=& p^2+m_2^2 \ ,\\
\hat g_2(\PI)=& (p^2+m_2^2) (2 \PI-3) -(D/2)(p^2+m_1^2+m_2^2)+m_1^2 \ ,\\
\hat g_1(\PI)=& (p^2+m_2^2) (\PI-1)^2
 -\left((D/2)(p^2+m_1^2+m_2^2)-m_1^2\right)(\PI-1)  \ ,\\
\end{split}
\end{equation}
where $J(x)=(m_1^2)^{-x} W(x)$, $W(x)=\sum_{s=0}^\oo b_s \RHO^{\K_b-s}$,
$\K_b=-D/2$, and the coefficients $b_s$ are the coefficients $a_s$ 
defined in \eqref{asex1}. 

The right-hand side of \eqref{chan3} can be written as
$ \sum_{s=0}^\oo c_s \RHO^{\K_c-s} $,
where $\K_c=\K_b+3$, and the expression of $c_s$ is given by
\begin{equation}
c_s= b_s \hat g_3(\K_c-s)  +b_{s-1} \hat g_2(\K_c-s) +b_{s-2} \hat
          g_1(\K_c-s) \ .
\end{equation}
Then, the coefficients $\hat a_s$ of the expansion of
$\VNOMOG(x)=\sum_{s=0}^\oo \hat a_s \RHO^{\K_b-s}$ 
can be found by using the recurrence relation \itref{systb2};
the large-$s$ behaviour is 
\begin{equation}
\hat a_s/s!\approx C_1 s^{(3D-6)/2} +C_2 s^{D-2} +C_3 B_+^s/\sqrt{s} +C_4
   B_-^s/\sqrt{s}\ ,
\end{equation}
where $B_{\pm}=m_1^{-2}/(m_1^{-2}-\mu_{\pm})$ and $C_i$ are constants.
If $|B_+|>1$  or $|B_-|>1$ the expansion of $\VNOMOG$ 
never converges
(the condition \itref{conv} is satisfied);
if $|B_+|<1$ and $|B_-|<1$, the series converges,
and the abscissa of convergence is 
$\lambda=\max (D-2,D/2-1)$ if $C_1\not = 0$ and $C_2\not=0$.

Now we must determine the constants $\C_{+}$ and $\C_{-}$;
we compare the large-$x$ behaviours of the solutions $\IOMOG_{\pm}$ and
$\INOMOG$, 
\begin{align}
\label{lrn1}
 \IOMOG_{\pm}(x) &\approx \mu_{\pm}^x x^{(1-D)/2}\ , \\
\label{lrn1nomog}
      \INOMOG(x) &\approx 
\begin{cases}
(p^2+m_2^2)^{-1}(m_1^2)^{D/2-x} x^{-D/2}
&
\text{$p^2\not = -m_2^2$}\ ,\\
(2-D)(4m_2^2)^{-1} (m_1^2)^{D/2-x} x^{-D/2}
&
\text{$p^2 = -m_2^2$}\ ,
\end{cases}
\end{align}
obtained from the first term of factorial series,
with the large-$x$ behaviour of $I(x)$.
We point out that these behaviours and the values of $\C_{\pm}$
later inferred are 
rigorously valid only if all the factorial series have finite abscissa 
of convergence; however the deduction may be extended 
to the cases where the series have 
$\lambda=\oo$ with the help of 
the integral representations of section~\ref{2den}.

If $p^2\ge -m_2^2$ we are below the deformation threshold, 
and we can write $I(x)$ 
as radial integral (see \eqref{hypbelow})
\begin{equation}\label{Ia}
I(x)=\dfrac{1}{\Gamma(D/2)}\int_0^\oo 
\dfrac{dk^2 (k^2)^{D/2-1}}{(k^2+m_1^2)^x} \dfrac{Z}{pk}
     \Phi(Z^2) \ ;
\end{equation}
the large-$x$ behaviour is given by \eqref{expkf00}
\begin{equation}\label{ian}
I(x)\approx  (m_1^2)^{D/2-x} x^{-D/2} (p^2+m_2^2)^{-1}\ .
\end{equation}
The comparison of \eqref{ian} with \eqrefb{lrn1}{lrn1nomog} shows that,
as $\mu_{\pm}\not = 1/m_1^2$,
$\IOMOG_{\pm}$ do not contribute to $I$,
and that only $\INOMOG$ contributes, so that  
\begin{equation}\label{cond11eu}
\C_{+}=\C_{-}=0 \qquad (p^2 > -m_2^2,\  m_1 \not=0, \ m_2\not=0) \ .
\end{equation}
At the deformation threshold $p^2=-m_2^2$ one finds from \eqref{expkf11}
for large $x$
\begin{equation}\label{Ixth}
I(x) \approx (m_1^2)^{D/2-1/2-x} x^{(1-D)/2} \dfrac{\sqrt{\pi}}{2m_2}  \ ;
\end{equation}
as $\mu_{-}=1/m_1^2$ and $\K_{-}=-D/2+1/2$, \eqref{cond1mod} is satisfied,
so that $\IOMOG_{-}$ contributes to $I$, and $\C_{-}$ is different from zero.
Comparing \eqref{Ixth} and \eqrefb{lrn1}{lrn1nomog} one finds
\begin{equation}\label{cond11eu0}
\C_{-}=m_1^{D-1} m_2^{-1} \sqrt{\pi}/2\ ,\quad \C_{+}=0 
\qquad (p^2 = -m_2^2,\  m_1 \not=0, \ m_2\not=0) \ .
\end{equation}

If $p^2< -m_2^2$ we are above the deformation threshold,
and the radial path  of integration turns around the point $k^2=(p-im_2)^2$.
We split $I(x)=I_{a}(x)+I_b(x)$  (see \eqref{hypabove}),
where $I_{a}$ is given
by \eqref{Ia} and $I_b$ is 
\begin{equation}\label{Ib}
I_b(x)=\dfrac{1}{\Gamma(D/2)}
\int_{(p-im_2)^2}^0 \dfrac{dk^2 (k^2)^{D/2-1}}{(k^2+m_1^2)^x} 
   \left( 
   \dfrac{Z}{pk} \Phi(Z^{2})
    -\dfrac{Z^{-1}}{pk} \Phi(Z^{-2}) 
   \right)    \ . 
\end{equation}

Noting that for large $x$ the contribution to this integral comes
from the neighbourhood of the singularity, one finds 
\begin{equation}
I_b(x)\approx 
\mum^x x^{(1-D)/2}
\dfrac{\sqrt{\pi}}{m_2} \left(\dfrac{im_2/p}{\mum}\right)^{(D-1)/2}
\ ,
\end{equation}
so that  
\begin{equation}\label{cond11neu}
\C_{-}=\dfrac{\sqrt{\pi}}{m_2} \left(\dfrac{im_2/p}{\mum}\right)^{(D-1)/2} \ ,
\quad
\C_{+}=0 \qquad (p^2 < -m_2^2,\  m_1 \not=0, \ m_2\not=0) \ .
\end{equation}

\subsubsection{Case 2: $m_1\not= 0$, $m_2=0$}\label{m2zero}
In this case we have $\mup=\mum=1/(p^2+m_1^2)$.
The second-order difference equation \itref{diffequ1} simplifies to
the first-order equation 
\begin{equation}\label{diffequ10}
(D-x-1)I(x-1) +(x-1)(p^2+m_1^2)I(x) -(x-1)J(x) =0 \ , \ 
\end{equation}
whose solution is of the kind 
\begin{equation}
I(x)=\C\IOMOG(x)+\INOMOG(x)\ .
\end{equation}
In a manner analogous to the previous case, one obtains the 
expansion in factorial series
\begin{equation}
\IOMOG(x)=(p^2+m_1^2)^{-x} \sum_{s=0}^\oo a_s \RHO^{2-D-s}\ ;
\end{equation}
the large-$s$ behaviour of the coefficients is $a_s/s!\propto s^{2D-5}$,
so that the series has abscissa of convergence $\lambda=D-2$.
Considering the nonhomogeneous solution,
\begin{equation}
\INOMOG(x)=(m_1^2)^{-x} \sum_{s=0}^\oo \hat a_s
\RHO^{-D/2-s}\ ,
\end{equation}
the coefficients have the large-$s$ behaviour 
\begin{equation}
\hat a_s/s!\approx C_1 s^{(3D-6)/2} +C_2 s^{D-2} +B_0^s s^{-(D-2)/2}\ ,
\end{equation}
where $B_0=1+m_1^2/p^2$ and $C_i$ are constants.
If $p^2>-m_1^2/2$,  $|B_0|>1$  and the series does not converge;
if $p^2\le -m_1^2/2$  the series converges
with abscissa of convergence $\lambda=\max (D-2,D/2-1)$,
if $C_1\not= 0 $ and $C_2\not =0$.

Now we determine the constant $\C$.
The large-$x$ behaviour of the solutions is given by 
the first term of the factorial series
\begin{align}
\label{lrn2}
 \IOMOG(x) &\approx (p^2+m_1^2)^{-x} x^{2-D}\ , \\
\label{lrn2nomog}
\INOMOG(x) &\approx 
\begin{cases}
p^{-2}(m_1^2)^{D/2-x} x^{-D/2} + ? 
&
\text{$p^2  > 0$}\ ,\\
p^{-2}(m_1^2)^{D/2-x} x^{-D/2}  
&
\text{$p^2 < 0$}\ .
\end{cases}
\end{align}
In the whole region $p^2>0$ the factorial series expansion of $\INOMOG$ never
converges so that the large-$x$ behaviour shown above  is valid only in
asymptotic sense. Using the integral representations of section~\ref{2den}
one can show that
if $p^2>0$  the large-$x$ behaviour of $\INOMOG$ 
contains an additional contribution, 
exponentially small in comparison with the main contribution 
\begin{equation}\label{I0nbis}
\INOMOG(x)\approx p^{-2} (m_1^2)^{D/2-x} x^{-D/2}
    +\Gamma(D/2-1)(-p^2)^{1-D/2}(p^2+m_1^2)^{D-2-x} x^{2-D} .
\end{equation}

The large-$x$ behaviour of $I(x)$ can be deduced from \eqref{Ia}  and
\eqref{Ib}, as in the massive case.
But if $p^2>0$ there is a difference: 
the integrand of \eqref{Ia} is discontinuous in
the point $k^2=p^2$, in fact
\begin{equation}\label{Zdiscont}
Z = \dfrac{p^2+k^2-|p^2-k^2|}{2pk} =
\begin{cases}
k/p \quad\text{$k<p$}\ ,\\
p/k \quad\text{$k\ge p$}\ .
\end{cases}
\end{equation}
The presence of this discontinuity
gives rise 
to an exponentially small
additional contribution proportional to $(p^2+m_1^2)^{-x}$
in the large-$x$ behaviour of $I(x)$.
So we split the integral over $k^2$ into two parts,
\begin{multline}
I(x)=
\dfrac{1}{\Gamma(D/2)}
\int^\oo_0 
\dfrac{dk^2 (k^2)^{D/2-1}}{(k^2+m_1^2)^x}
     \dfrac{1}{p^2}\Phi\left(\dfrac{k^2}{p^2}\right) \\
    +
    \dfrac{1}{\Gamma(D/2)}
    \int^\oo_{p^2}  
    \dfrac{dk^2 (k^2)^{D/2-1}}{(k^2+m_1^2)^x}
\left(
 \dfrac{1}{k^2}\Phi\left(\dfrac{p^2}{k^2}\right)
-\dfrac{1}{p^2}\Phi\left(\dfrac{k^2}{p^2}\right)
\right)\ ,
\end{multline}
each one having a different large-$x$ behaviour.
One finds that the large-$x$ behaviour of $I(x)$ is identical to 
that of $\INOMOG(x)$ for $p^2>0$, \eqref{I0nbis}, so that 
\begin{equation}\label{cond11eu1}
\C=0 \qquad (p^2 > 0,\  m_1 \not=0, \ m_2=0) \ .
\end{equation}

If $p^2<0$ the large-$x$ behaviour of $I(x)$ can obtained from \eqref{Ib}
(with $m_2=0$).
One finds the same result found for $p^2>0$.
Therefore for $p^2<0$ the exponentially small contribution is present in the
large-$x$
behaviour of $I(x)$ but not in that of $\INOMOG(x)$:
it must come from $\C\IOMOG(x)$. Therefore the  constant $\C$ is 
\begin{equation}\label{cond11neu2}
\C=\Gamma(D/2-1)\left((p^2+m_1^2)/(-i\;p)\right)^{D-2} 
   \qquad (p^2 < 0,\  m_1 \not=0, \ m_2=0) \ .
\end{equation}

\subsubsection{Case 3: $m_1=0$, $m_2\not=0$}\label{m1zero}
In this case the denominator raised to $x$ has zero mass,
and the term containing $J$ in \eqref{diffequ1} disappears, so that
the difference equation becomes homogeneous;
the solution is of the kind 
\begin{equation}\label{solgeni3}
I(x)=\C_{+}\IOMOG_{+}(x) + \C_{-}\IOMOG_{-}(x) \ .
\end{equation}
$\IOMOG_{\pm}(x)$ are the same functions considered in section~\ref{m1m2nz}
with $m_1=0$, and their large-$x$ behaviour was given in \eqref{lrn1}.
But the large-$x$ behaviour of $I(x)$ cannot be found as in the 
massive case; in fact if $m_1=0$
the integrals \itref{Ia} and \itref{Ib} are
strongly divergent in $k^2=0$ for large $x$ and only
dimensional regularization can give a finite value to $I(x)$.
The large-$x$ behaviour of $I(x)$ must be determined by other methods,
for example by writing $I(x)$ as integral over one Feynman parameter and 
using the saddle-point method;
one finds  
\begin{equation}\label{cond11neu3}
\C_{\pm}=\dfrac{\sqrt{\pi}}{2m_2\sin(\pi D/2)} 
\left(\mp ip\mu_{\pm}/m_2\right)^{(1-D)/2}
\qquad ( m_1 =0, \ m_2\not=0) \ . \quad
\end{equation}

\subsubsection{Numerical example}\label{numexa}
Let us describe in some detail the calculation of $I(1)$, $J(1)$ and $K(1)$
in the case $-p^2=m_1^2=m_2^2=1$.
As $m_1=m_2$ then $K(x)=J(x)$, and we consider only $J(x)$ and $I(x)$.
The root of characteristic equation of the difference equation for $J(x)$ 
(\eqref{diffequ1b})
is $\mu=1$;
the roots of the characteristic equation of the homogeneous part
of the equation  for $I(x)$ (\eqref{diffequ1}) are $\mum=1$ and $\mup=-1/3$.
$I(x)$ is written as sum of the homogeneous and nonhomogeneous part, 
\eqref{solgeni};
from \eqref{cond11eu0} one finds $\C_{+}=0$,
so that $\IOMOG_{+}$ does not contribute.
Therefore we have to evaluate three functions:
$J(x)$, $\IOMOG_{-}(x)$ and $\INOMOG(x)$
for $x=1$ and $D\to 4$.
The factorial series expansions of all the three functions have
abscissa of convergence $\lambda \to 2$ for $D \to 4$ and do not converge for $x=1$.
Following the procedure
of section~\ref{sumfact} we evaluate the series for a large value
$x=x_{max}$ and, as the equation for $I(x)$ is of second order,
even for $x=x_{max}+1$.
The convergence becomes faster by increasing $x_{max}$, but,
as  $A=|\mum/\mup|=3>1$, the recurrence relation for $I(x)$ \eqref{diffequ1},
rewritten in order to obtain $I(x-2)$ from $I(x-1)$ and $I(x)$,
is unstable.
Each application of the recurrence relation increases the error of
the value of $I(x)$ of about a factor $3$.
Therefore we must choose a value of $x_{max}$ which is a compromise between
speed of computation and loss of precision in the result.
We choose $x_{max}=8$.
Calculations are performed by setting $D=4-2\e$ and expanding in $\e$,
truncating the expansions at the first three terms.
Doing the calculations with 19 digits,
the convergence is attained
for all the factorial series for $x=8$ (and consequently for $x=9$) in about 4000 terms.
The values for $x<8$ are calculated by using repeatedly
the recurrence relations \itref{diffequ1} and  \itref{diffequ1b}.
The application of the unstable recurrence relation
enlarges the error of about
a factor $3^7\approx 2000$, corresponding to a loss of about 3 significant
digits in $I(1)$.
Values of $J(x)$, $\IOMOG_{-}(x)$ and $\INOMOG(x)$  are shown in
Table~\ref{tableij},
with only the first two terms of the expansion in $\e$,
and the coefficients with only 6 digits to save space.
The value of $J(1)$ calculated agrees with the exact result
$-\e^{-1} +\gamma-1+O(\e)$
within the precision of the calculation ($\gamma$ is the Euler's constant).
Inserting $\IOMOG_{-}(1)$, $\INOMOG(1)$ 
and the value $\C_{-}=\sqrt{\pi}/2$ (from \eqref{cond11eu0}) in \eqref{solgeni}
one finds  
\begin{equation}\label{vali1}
I(1)= (1-2\times 10^{-17})\e^{-1} -0.391 015 029 135 750 3 +O(\e)\ ,
\end{equation}
with an error of $4 \times10^{-16}$ on the constant term
in comparison with the exact result $\e^{-1} +2-\pi/\sqrt{3}-\gamma+O(\e)$.

A numerical value of $\C_{-}$ can be obtained independently, 
using the identity \itref{rel0}
\begin{equation}
I(0)=\C_{-}\IOMOG_{-}(0)+\INOMOG(0)\ ;
\end{equation}
inserting the values of $I(0)=K(1)=J(1)$, $\IOMOG_{-}(0)$ and $\INOMOG(0)$
one finds  
\begin{equation}\label{L1num}
\C_{-}= 0.886 226 925 452 757 8 +5\times10^{-15} \e +O(\e^2)\ ,
\end{equation}
with an error of $2 \times10^{-16}$ on the constant term.
Using this numerical value in the place of the analytical value one obtains
a value of $I(1)$ with the same precision as \eqref{vali1}.
\begin{table}
\begin{center}
\begin{tabular}{rrrr}
\hline
$x$& $J(x)$ &$\IOMOG_{-}(x)$ & $\INOMOG(x)$   \\ \hline
9  & $0.017857 +0.033442 \e$ &$  0.047713        +0.089160 \e $&
     $ -0.008928   -0.007323 \e$ \\
8  & $0.023809 +0.040621 \e$ &$  0.059006        +0.100337 \e $&
     $ -0.011904   -0.007663 \e$ \\
7  & $0.033333 +0.050203 \e$ &$  0.075609        +0.113249 \e $&
     $ -0.016666   -0.007159 \e$ \\
6  & $0.05     +0.062805 \e$ &$  0.101857        +0.126598 \e $&
     $ -0.025      -0.003929 \e$ \\
5  & $0.083333 +0.076898 \e$ &$  0.148085        +0.133074 \e $&
     $ -0.041666   +0.009010 \e$ \\
4  & $0.166666 +0.070464 \e$ &$  0.245635        +0.090010 \e $&
     $ -0.083333   +0.067207 \e$ \\
3  & $0.5      -0.288607 \e$ &$  0.548843        -0.442122 \e $&
     $ -0.25       +0.534955 \e$ \\
2  & $\e^{-1}  -0.577215\fe$ &$  0.282094\e^{-1} +0.519388 \fe$&
     $ -0.25\e^{-1}+0.144303 \fe$ \\
1  & $-\e^{-1} -0.422784\fe$ &$  0.282094\e^{-1} -1.645293 \fe$&
     $  0.75\e^{-1}+1.067088 \fe$ \\
0  & $0\phantom{.123456\fe}$ &$ -0.564189\e^{-1} -0.238530 \fe$&
     $ -0.5\e^{-1} -0.211392 \fe$ \\
\hline
\end{tabular}
\end{center}
\caption{Values of $J(x)$, $\IOMOG_{-}(x)$ and $\INOMOG(x)$.}
\label{tableij}
\end{table}
\begin{table}
\begin{center}
\begin{tabular}{rrc}
\hline
$x_{max}$& terms & finite part of $I(1)$  \\ \hline
$30$  &$   125$ & $-0.3910008887063124$\\
$25$  &$   154$ & $-0.3910149952724784$\\
$20$  &$   217$ & $-0.3910150292106927$\\
$15$  &$   395$ & $-0.3910150291388126$\\
$10$  &$  1470$ & $-0.3910150291357554$\\
$ 9$  &$  2454$ & $-0.3910150291357472$\\
$ 8$  &$  4439$ & $-0.3910150291357503$\\
$ 7$  &$ 13086$ & $-0.3910150291357507$\\
$ 6$  &$ 36210$ & $-0.3910150291357507$\\
\hline
\end{tabular}
\end{center}
\caption{Dependence of the finite part of $I(1)$ on $x_{max}$.}
\label{tablei1}
\end{table}
In Table~\ref{tablei1} we show for different choices of $x_{max}$
the number of terms of series
necessary to evaluate $J(8)$, $\IOMOG_{-}(8)$ and $\INOMOG(8)$
with 19 digits of precision 
and the obtained values of the finite part of $I(1)$;
it is evident that by increasing $x_{max}$ the series converges faster,
but the precision degrades because of the increasing number of applications
of the unstable recurrence relation.

\section{Solutions of difference equations by means of La\-pla\-ce's
transformation}\label{Laplacesec}

The expansion in factorial series is certainly the most direct method of
solution of difference equations with polynomial coefficients;
however, for some values of masses and external momenta of the diagram, 
as we have seen in the above example,
the abscissa of convergence of factorial series may become infinite.
In this case the factorial series become divergent for every value of $x$
and therefore useless, 
so that another method of solution must be used:
the Laplace's transformation method\cite{Milne15}.

This method is described in section \ref{laplacedisc},
and applied to the systems of difference equations in section
\ref{trasys}, \ref{consyslaplace} and \ref{inicondf}.
Techniques used for integrating the differential equations 
obtained from the application of the method 
are described in section \ref{solvedifequ}.
The application to simple one-loop integrals is shown in 
section \ref{example1looplapla}. 

\subsection{Transformation of a difference equation}\label{laplacedisc}
Let us consider the difference equation 
\begin{equation}\label{diffequl}
p_0(x) U(x) + p_1(x) U(x+1) + \ldots + p_N(x) U(x+N) = 0\ ,
\end{equation}
where $p_i(x)$ are polynomials in $x$ of maximum degree $P$.
The Laplace's transformation method consists in the substitution 
\begin{equation}\label{subLaplace}
U(x)=\int_l dt\; t^{x-1} v(t) \ ,
\end{equation}
where $l$ is a line of integration suitably determined and where
$v(t)$ is found from a certain differential equation.
Writing the coefficients as
\begin{equation}
p_k(x)= A_{k0} +\sum_{i=1}^P A_{ki} \prod_{j=0}^{i-1} (x+k+j) \ ,
\end{equation}
substituting \eqref{subLaplace} in \eqref{diffequl} and integrating by parts
 one finds
\begin{equation}
\sum_{k=0}^N p_k(x) U(x+k) = 
 \int_l dt \; t^{x-1}  \sum_{i=0}^P \Phi_i(t) (-t)^{i} v^{(i)}(t) 
 +[I(x,t)]_l \ ,
\end{equation}
where
\begin{equation}
\Phi_i(t)=\sum_{k=0}^N A_{ki}t^k\ ,
\end{equation}
\begin{equation}
I(x,t)=\sum_{i=0}^{P-1} (-1)^i v^{(i)}(t) 
     \sum_{m=0}^{P-1-i}  
     \left(\dfrac{d}{dt}\right)^m \left( \Phi_{m+i+1}(t) t^{x+m+i} \right)\ .
\end{equation}

\eqref{subLaplace} provides a solution of the difference equation
\itref{diffequl} if
$v(t)$  is a solution of the differential equation 
\begin{equation}\label{diffequv}
\sum_{i=0}^P \Phi_i(t) (-t)^{i} v^{(i)}(t) =0 \ ,
\end{equation}
provided that the line of integration $l$ be chosen so that $I(x,t)$
has the same value at each endpoint of the line, if the line is open.
Note that
the difference equation \itref{diffequl}
has order $N$ with coefficients of degree $P$, while 
the differential equation \itref{diffequv}
has order $P$ with coefficients of degree $N+P$.

The singular points of this differential equation are 
$0$, $\oo$ 
and the zeros $t_i$ 
(of multiplicity $m_i$) 
of the \emph{characteristic equation}
\begin{equation}\label{char1v}
\Phi_P(t)=0\ ;
\end{equation}
these points turn to be always regular singular points
in the case of the differential equations encountered in this work.

We choose as lines of integration the lines
which begin in the origin and end in one of the singular points $t_i$.
This is a convenient choice.
One can show that $I(x,t)=0$ at each endpoint of such lines,
if the integral over $t$ of \eqref{subLaplace} is finite.
Under these conditions, $U(x)$ is a solution of 
the difference equation.

We can construct a set of $\sum_i m_i=N$ functions $U_{ij}(x)$ 
which form a fundamental system of solutions 
of the difference equation  \itref{diffequl} by defining
\begin{equation}
U_{ij}(x)=\int_{l_i} dt \; t^{x-1} v_{ij}(t) \qquad j=1,\ldots,m_i \ ,
\end{equation}
where $v_{ij}(t)$ is one of the $m_i$ solutions of the differential equation
singular in $t_i$, and $l_i$ is the line which begins in $t=0$ and ends
in $t=t_i$.

It is important to 
note that the characteristic equation \itref{char1v} of the differential 
equation and the characteristic equation of the difference equation
\itref{diffequl}
turn out to be identical, so that we can readily identify
the singular points $t_i$ of the differential equation
with the solutions $\mu_i$ of \eqref{equchar}.

Now we consider the nonhomogeneous equation 
\begin{equation}\label{diffequl2}
\sum_{k=0}^N p_k(x) U(x+k) = \sum_{k=0}^{N'} q_k(x) T(x+k) \ ,
\end{equation}
where $q_k(x)$ are polynomials
and $T(x)$ is a solution of some difference equation.
A particular solution $\UNOMOG(x)$ is found 
by substituting into the equation 
\begin{equation}\label{subLaplacew}
      T(x)=\int_{l_T} dt \; t^{x-1} w(t)      \ , \qquad 
\UNOMOG(x)=\int_{l_T} dt \; t^{x-1} v_{NH}(t) \ ,
\end{equation}
where $l_T$ is a known line of integration
and $w(t)$ is a known function, solution of a differential equation
analogous to \eqref{diffequv},
obtained from the difference equation satisfied by $T(x)$.
Provided that $I(x,t)=0$ at the endpoints of $l_T$, 
one finds the nonhomogeneous differential equation
\begin{equation}\label{diffequv2}
\sum_{i=0}^P    \Phi_i(t) (-t)^{i} v^{(i)}_{NH}(t) =
\sum_{i=0}^{P'} \Psi_i(t) (-t)^{i} w^{(i)}(t) \ ,
\end{equation}
whose solution gives $v_{NH}$.
Then $\UNOMOG$ is found using \eqref{subLaplacew}.

\subsection{Transformation of the system of difference equations}\label{trasys}
In section~\ref{constsysdif} we described the construction of the system
of difference equations.
The algorithm used yields a system in triangular form;
applying to this system the Laplace's transformation
one obtains a system of differential equations with
the same triangular structure and the same ease of solution.
There is, however, a complication:
the $l$th difference equation 
\begin{equation}\label{diffequl2a}
\sum_{k=0}^N p_k(x) U_l(x+k) = \sum_{j=0}^{l-1}\sum_{k=0}^{N'_j} q_{jk}(x)
U_j(x+k) \ 
\end{equation}
of order $N$ and with polynomial coefficients of degree $P$
is transformed, using the Laplace's transformation,
into a differential equation of order $P$
\begin{equation}\label{diffequv2a}
\sum_{i=0}^P \Phi_i(t) (-t)^{i} v_l^{(i)}(t) =
\sum_{j=0}^{l-1} \sum_{i=0}^{P'_j} \Psi_{ij}(t) (-t)^{i} v_j^{(i)}(t) \ 
\end{equation}
and coefficients of degree $N+P$.
As effect of the particular algorithm of construction and solution of the
system of identities, $N$ is usually small (typically $1\sim 4$) while
$P$ may be large ($3\sim 30$).
Therefore the differential equation may have a high order.
Calculations in some test cases have shown that
solution of high order equations slows down the calculations  and is source
of undesired numerical errors,
so that, if possible, it is better to avoid it.

We have discovered that this difficulty can be overcome by applying
the Laplace's transformation to the identities obtained
by integration-by-parts \emph{before} the insertion in the system of
identities,
building a system of ``transformed'' identities between
the ``transformed'' integrals $v(t)$  instead of the real integrals $U(x)$.
The solution of the system of transformed identities
will provide a system of differential equations of smaller order;
a small price to pay is the (possible) appearance of spurious singular points
in the equations so obtained (see section~\ref{mobile}).
More in detail, a generic integral is transformed into 
\begin{equation*}
U_{ni\alpha\beta}(x)=\int 
\dk1 \ldots \dk{\ND}
\dfrac
{\prod_{j=1}^{\NPS-n} {\indpk}_j^{\beta_{j}} }
{D_{i_1}^{x+\alpha_1} D_{i_2}^{\alpha_2} \cdots D_{i_n}^{\alpha_n}}
= \int_l dt \; t^{x-1} v_{ni\alpha\beta}(t)\ ,
\end{equation*}
where the line $l$ is unspecified;
the values of the functions $I(x,t)$ at the endpoints of the line are
always zero because of dimensional regularization.
Therefore, a generic integration-by-parts identity,
\begin{equation*}
\sum_{ni\alpha \beta} (x \; r_{ni\alpha \beta} +s_{ni\alpha \beta})
U_{ni\alpha\beta}(x)=0 \ ,
\end{equation*}
($r_{ni\alpha \beta}$ and $s_{ni\alpha \beta}$ are independent of $x$)
becomes the transformed identity 
\begin{equation}\label{idet}
\sum_{ni\alpha \beta} \left( (s_{ni\alpha\beta}
-r_{ni\alpha\beta}\alpha_1)v_{ni\alpha\beta}(t)
 -t\;r_{ni\alpha\beta} \dfrac{dv_{ni\alpha\beta}}{dt}(t)  \right)
       t^{\alpha_1} =0 \ 
\end{equation}
which is a differential equation between the functions $v_{ni\alpha\beta}(t)$.

\subsection{Construction of the system of differential
equations}\label{consyslaplace}
We want to build a triangular system of differential equations 
\begin{equation}\label{triform}
\sum_{i=0}^{P_l} p_{il}(t) 
 v^{(i)}_{ml}(t)
=
\sum_{k=1}^{l-1}\sum_{j=0}^{P_{lk}} q_{jkl}(t) 
v^{(j)}_{mk}(t)\ , \quad l=1,\ldots,L_m'\ ,
\end{equation}
between a set of ``master transformed'' functions $v_{ml}(t)$
($m=1,\ldots,\ND-\NK+1$)
analogous to the system of difference equations \itref{diffu1}
between the master functions $U_{ml}(x)$ discussed in section~\ref{constsysdif}.
For this reason we devise an algorithm of construction and solution 
of the system of identities similar to the algorithm~\ref{algsys4}:
\begin{alg}\label{algsys6}
Consider the algorithm~\ref{algsys4} with the following modifications:
\begin{enumerate}
\item After the step~\ref{quiins1} of algorithm~\ref{algsys}, transform
      each integration-by-parts identity using \eqref{idet};
      in the subsequent steps replace everywhere 
      the integrals 
      $\int \dk1\ldots\dk\NK W_{ni\alpha\beta}$
      with the functions $v_{ni\alpha\beta}(t)$.
\item
The transformed identity obtained does not depend on the
index $\alpha_1$ of $W$ 
in the step~\ref{quissset} of algorithm~\ref{algsys}
because it appears as an overall factor
$t^{\alpha_1}$;
in order to restore the total number of different identities
we have decided (somewhat arbitrarily) to differentiate $\alpha_1$ times 
with respect to $t$ each transformed identity.
\item Ignore step~\ref{cond04} of algorithm~\ref{algsys4}.
\item Derivatives of master transformed functions have priority
      of extraction lower than other generic transformed integrals.
\item
Add  the new entry \emph{the greatest derivative}
to the list of priorities
after the entry~\ref{cf6b} of algorithm~\ref{algsys}.
\end{enumerate}
\end{alg}

With a suitable choice of the parameters $a_i$ and $b_i$ 
(see the end of section~\ref{constsysdif}),
by means of this algorithm 
we can identify the master transformed functions $v_{ml}(t)$
as the functions which satisfy equations of non-zero order,
and we can work out a set of differential equations among them.
If each function $v_{ml}(t)$ corresponds to an integral
with a different combination of denominators,
the system is obtained directly in the triangular form \itref{triform};
if, on the contrary, there are
different master transformed functions $v_{m,l+1}(t)$,\ldots,$v_{m,l+G}(t)$
corresponding to integrals containing the same combination of denominators,
the algorithm provides a set of $G$ simultaneous differential equations
containing all these $G$ functions, 
which are conveniently transformed into 
triangular form using a procedure quite analogous 
to that described in section~\ref{constsysdif}.
As final result one obtains a system of differential equations 
with the triangular form \itref{triform}. 
It is important to note that the functions
\begin{equation}
F_{ml}(x)=\int_l dt \; t^{x-1} v_{ml}(t)
\end{equation}
are not necessarily identical to the master functions $U_{ml}(x)$ defined in
\eqref{definitionu}.
While $v_{ml}(t)$ is a master transformed function and
satisfies a differential equation of non-zero order,
$F_{ml}(x)$ satisfies a difference equation of different order,
which may be zero; if so, $F_{ml}(x)$ is not a \emph{master} function.
This fact frequently occurs when many master integrals
have the same combination of denominators.
For these reasons the number and the structure of 
the master transformed functions $v_{ml}(t)$ must be found independently
of the master functions $U_{ml}(x)$. See section~\ref{pair2} for an example.

\subsection{Correspondence  with factorial series and initial conditions}
\label{inicondf}
Let $U(x)=\sum_\alpha U_\alpha(x)$ be the solution of
a generic difference equation;
the initial conditions for the integration of the differential equation
can be determined by comparing the integral representation of $U_\alpha(x)$
with the factorial series expansion 
\begin{equation}\label{factserua}
U_\alpha (x) = \int_0^{\mu_\alpha} dt \; t^{x-1} v_\alpha(t) 
             = \mu_\alpha^x 
\sum_{s=0}^\oo a_{s\alpha} \RHO^{\K^F_\alpha-s} \ .
\end{equation}
The behaviour of $v_\alpha(t)$ near the singular point $t=\mu_\alpha$
\begin{equation}\label{va0k}
v_\alpha(t)\approx A_{0\alpha} (\mu_\alpha-t)^{\K^L_\alpha}, \qquad t\approx
\mu_\alpha\ ,
\end{equation}
can be deduced from the known behaviour of $U_\alpha(x)$ for large $x$.
Substituting \eqref{va0k} in \eqref{factserua}, integrating over $t$
and comparing the result with the large-$x$ leading behaviour of the first
term of the series 
$U_\alpha(x) \approx \mu_\alpha^x a_{0\alpha} x^{\K^F_\alpha}$
one finds $K^L_\alpha$ and  $A_{0\alpha}$: 
\begin{equation}\label{aA0}
\K^L_\alpha=-\K^F_\alpha-1 \ , \qquad
A_{0\alpha}=
a_{0\alpha}/(\mu_\alpha^{\K^L_\alpha} \Gamma(\K^L_\alpha+1))\ .
\end{equation}
If necessary,  
the subsequent coefficients of the series expansion of $v_\alpha(t)$ 
can be deduced in the same way from the coefficients $a_{s\alpha}$.

\subsection{Integrating differential equations}\label{solvedifequ}
The choice of a effective numerical method of integration of the
differential equations 
obtained by applying Laplace's transformation is not simple.
In fact we must consider that:
\begin{itemize}
\item
The initial and final point of the path of integration $l$ 
are singular points of the solutions; 
the numerical methods usually used to integrate differential equations 
(for example, the Runge-Kutta method) cannot be used in singular points.
\item
Master functions, coefficients of differential equations
and (sometimes) singular points 
depend on $D$ and 
are represented by truncated series in $\e$.
\item
The method must be able to provide very high-precision values,
and the time of computation must grow linearly with
the number of exact digits of the result;
by using fixed-order  methods (like the Runge-Kutta method) the time grows
exponentially.
\end{itemize}
Therefore we have decided to solve the differential equations
by using power series,
expanding the general solution around a number of selected points
near or on the path $l$ and equating the sums of the series 
in some intermediate points.
Each power series will be evaluated inside 
the respective circle of convergence,
so that the number of terms of the series necessary to attain 
a precision of $E$ digits in the results (and also the computation time)
will be proportional to $E$.

\subsubsection{Integration over the path}
The line of integration can have any shape,
but for convenience it is assumed here to be the segment $[0,\mu]$.
The segment is subdivided into $M$ intervals
$[t_{M+1},t_{M}]$,
$[t_{M},t_{M-1}]$, \ldots,
$[t_{2},t_{1}]$,
where $t_1=\mu$ and $t_{M+1}=0$;
the (possible) singular points placed on the segment are
$0= t^{\sng}_P<\ldots<t^{\sng}_2<t^{\sng}_1<\mu$.
The choice of a line of integration which passes through singular points
of the differential equation,
instead of avoiding them,
may be convenient: it allows one to check if $v(t)$ is regular
or singular in these points 
and it speeds up the calculations,  avoiding the use of
complex numbers.
Let us consider the first interval $[t_2,t_1]$.
The solution is expanded around the point $t_1=\mu$ (which may be a singular point).
The first coefficients of the expansions of the $P$ solutions corresponding
to the roots of the indicial equation are obtained using \eqref{aA0}.
All the subsequent coefficients are obtained using recurrence relations
obtained by substituting the expansions into the differential equation.
Then the power series are evaluated in the point $t_2=t_1-r_1/2$,
where $r_1$ is the radius of convergence of the series
and the factor $1/2$ has been chosen in order to minimize the total time
of calculations.
In the next interval the solution is expanded around the
regular point $t_2$,
the first $P$ coefficients of the expansion are obtained 
from the already known values of $v$ and
its derivatives in $t_2$, the subsequent coefficients are obtained using
the recurrence relations,
and the power series is evaluated in a point $t_3=t_2-r_2/2$.
This process involving expansions around regular points
continues for $m$ steps until a point $t_m$ is reached,
placed inside the circle of convergence of the series with center the
singular point $t^{\sng}_1$,
at a distance less than one half of the radius $r_1^{\sng}$ of the circle.
Now we expand the solution around the singular point $t^{\sng}_1$.
We write the general solution as $v(t)=\sum_{j=1}^P c_j v_j(t)$,
where each $v_j$ corresponds to one of the $P$ solutions
of the indicial equation.
The values of $c_j$ are found by solving the linear system\footnote{
It is important to note that if $m+1$ solutions of the indicial equation
coincide for $D\to 4$, the Wronskian determinant of the linear system
is proportional to $(D-4)^m$;
this causes a loss of $m$ terms in the expansions in $D-4$ of $c_j$
and, consequently, in the expansions of  $v$ in the rest of integration, and
therefore in $U(x)$.
This mishap sometimes occurred in the final singular point $t=0$
in the calculations of section~\ref{resu}.
}
$\sum_{j=1}^P c_j v^{(i)}_j(t_m)=v^{(i)}(t_m)$, $i=1,\ldots,P$.
The right-hand sides contain the already known values of $v$ and its
derivatives in $t_m$;
the necessary values of partial solutions $v_j(t_m)$ are worked out by summing
the corresponding expansions about $t^{\sng}_1$, whose
coefficients are found using recurrence relations.
Then the singular point is got over by evaluating $v$ (and its derivatives) 
in the regular point $t_{m+1}=t^{\sng}_1 -r_1^{\sng}/2$.
The process is repeated until the next singular point $t^{\sng}_2$ is reached,
then $t^{\sng}_3$, $t^{\sng}_4$, etc., up to the final point $t^{\sng}_P=t_{M+1}=0$.

The integration of $v(t)$ over $t$ needed to obtain $U(x)$ 
can be easily carried out 
by integrating the expansions in series in the corresponding intervals:
\begin{equation}\label{usum}
U(x)=\int_0^\mu dt \; t^{x-1} v(t) = \sum_{i=1}^{M} \int_{t_i}^{t_{i+1}}
dt \; t^{x-1} 
\sum_{j=1}^P \sum_{s=0}^\oo a_s^{(i,j)} (t-\bar t_i)^{\K_{ij}+s} \ .
\end{equation}
The integrals 
\begin{equation}
I(j,s)= 
\int_{\bar t+a}^{\bar t+b} dt \; t^j (t-\bar t)^{k+s} 
= 
\int_a^b dy \; (y+\bar t)^j y^{k+s}
\end{equation}
which appear in \eqref{usum} 
can be expressed in terms of incomplete Beta function
(note that $k$ is not an integer).
If $j=0$ the integral is immediate; 
if $j$ is positive integer
the value can be efficiently computed using
the recurrence relation
\begin{equation}
I(j,s)=I(j-1,s+1)+\bar t I(j-1,s) \ .
\end{equation}

\subsubsection{Singular points depending on $D$}\label{mobile}
The coefficients $a^{(i,j)}_s$ and the exponents $\K_{ij}$
in \eqref{usum} depend on $D$;
therefore all quantities are expanded around $D=4$, and truncated series
are used in the calculation, as described in section~\ref{truncd}.
A new feature, characteristic of the differential equations
obtained by solving the system of transformed identities,
is the appearance of spurious apparent singular points $\bar t$,
not corresponding to any solution $\mu$ of the characteristic equations of 
the difference equations obtained by solving the system of
`original' identities.
These apparent singular points are depending on $D$,
and correspond to regular points of the solution of differential equation,
in contrast with regular singular points,
which are independent of $D$.
In general the line of integration can be deformed 
in order to avoid these spurious singular points; 
but if for $D\to 4$ one or more of
these \emph{mobile} points tends to one of the endpoints of the
line, $t=\mu$ or $t=0$, we cannot avoid it,
and we encounter difficulty in working out the solution 
near these coalescing points.
Let us explain this fact, by considering 
a homogeneous differential equation with polynomial coefficients
\begin{equation}\label{equhm0}
\sum_{i=0}^P p_i(t) v^{(i)}(t)=0\ ,
\qquad 
p_i(t)= t^i \sum_{j=0}^{g_i} p_{ij} t^{j}\ ,
\end{equation}
and supposing for simplicity 
that the equation has only one apparent singular point $t_0(D)$ such that
\begin{equation}\label{critp}
  t_0(D)=O(D-4) \quad \text{for}\quad D\to 4 \ .
\end{equation}
The coefficients of the expansion $v(t)=\sum_{s=0}^\oo a_s t^{s+\K}$
can be found using the recurrence relation  
\begin{equation}\label{recas1}
a_s=-\dfrac{\sum_{j=1}^m a_{s-j} f_j(\K+s-j) }{f_0(\K+s)}\ ,
\end{equation}
where $a_i \equiv 0$ if $i<0$,
$m=\max_i g_i$, 
$
f_j(k) =\sum_{i=0}^P k(k-1)\cdots (k-i+1) p_{ij} \ ,
$
and 
$\K$ is one of the roots of indicial equation $f_0(\K)=0$
(note the analogy with the solution of a difference equation with 
expansions in factorial series).

As the point $t_0$ is a regular point,
the coefficients $a_s$, which are functions of $D$,
have values in $D=4$ finite and, in general, different from zero;
unfortunately, there are problems for calculating them.
As a consequence of \eqref{critp}, the function 
$f_0$ vanishes if $D=4$, and the other $f_j$ do not vanish 
(but the sum in the numerator of \eqref{recas1} always vanishes for $D=4$), 
so that the recurrence relation \itref{recas1} turns out to be very unstable,
with a degree of instability proportional to $1/(D-4)$.
Performing the calculations of $a_s$ using truncated expansions in $D-4$,
each iteration of \eqref{recas1} (with increasing $s$)
yields one new coefficient $a_s$, whose expansion in $D-4$ has a number of terms
reduced by one in comparison 
with 
$a_{s-1}$;
after a few iterations, 
the given number of terms of the expansion in $D-4$ is exhausted.
We found that a solution of the problem is to modify the recurrence relation to
\begin{equation}\label{recas2}
a_{s-1}^{(n+1)}=-\dfrac{\sum_{j=2}^m a_{s-j}^{(n)} f_j(\K+s-j)  + a_s^{(n)}
f_0(\K+s)}{f_1(\K+s-1)}\ ,
\end{equation}
so that the denominator is $f_1$, which does not vanish for $D=4$.
The new recurrence relation requires 
as input the value of the $s$th coefficient before that 
its value is obtained;
therefore \eqref{recas2} must be seen as part of an iterative process:
\begin{enumerate}
\item set $a_s^{(0)}=0$ for $s=1,2,\ldots$, $s_{{max}}$;
\item \label{rpoint}
      apply \eqref{recas2} for $s=1,2,\ldots$, $s_{{max}}$
      obtaining the coefficients $a_s^{(1)}$;
\item repeat 
      $n$ times
      the step~\ref{rpoint}
      until $|a_s^{(n+1)}-a_s^{(n)}|=O((D-4)^m)$ for every $s$, where $m$
      is the desired number of term of the expansion in $D-4$;
      the convergence is guaranteed in about $m$ steps by the
      fact that $f_0= O(D-4)$. 
\end{enumerate}      

Analogous modifications must be made to the recurrence relation 
in the case of nonhomogeneous equations, 
or in case of two or more spurious singular points
coalescing to the same endpoint.
An example of equation with one mobile singular point is
the third-order equation\footnote{
The equation \itref{equmob3} was found by solving the system of transformed
identities. Applying the Laplace's transformation directly to the difference
equation \itref{equmob2} one gets a higher order differential equation which 
does not have
the mobile singular point. This equation can be also derived from \eqref{equmob3}
by writing $(3t d\Phi(t)/dt +4(D-4)\Phi(t))/(4(D-1)t-(D-4))=0$ where
$\Phi(t)$ is the left-hand side of \eqref{equmob3}.}
\begin{multline}\label{equmob3}
-2(t-1)(8t+1)(4(D-1)t-D+4))t^3v_{\ref{figself}\text h}''' 
+\bigl( -32(D-1)(5D-7)t^3 \\
        +8(D-1)(25D-63)t^2 
         +(-4D^2 +100D -192)t -(D-4)(9D-22)  \bigr)t^2v_{\ref{figself}\text h}'' \\
+\bigl( -32(D-1)(D-2)(3D-5)t^3 
         +32(D-1)(D-3)(5D-12)t^2 \\
          +(18D^3 -70D^2 -44D +288)t
	  +(D-4)(-13D^2 +72D-100) \bigr)tv_{\ref{figself}\text h}' \\
+2(D-4)(D-3)^2\bigl(16(D-1)t^2 +8Dt -3D+10\bigr) v_{\ref{figself}\text h}=0\ , 
\end{multline}
which is the homogeneous part of the equation satisfied
by the function $v_{\ref{figself}\text h}(t)$, corresponding to the integral
of Fig.~\ref{figself}h of section~\ref{resu}, 
used in the calculation of \eqref{treh}
(\eqref{equmob3} remains the same for both choices of the line $D_1$). 
The singular points are $t=1$, $-1/8$, $0$  and $(D-4)/(4D-4) $.
The last singular point, which satisfies the condition \itref{critp},
is a regular point with exponents $0$, $1$ and $3$.
The characteristic equation of the  homogeneous part of
the corresponding difference equation,
\begin{multline}\label{equmob2}
(x-D+1) (x-2 D+4) (2 x-3 D+6) (3 x-4 D+10)  U_{\ref{figself}\text h}(x-1) \\
+ 2(x-D+2)\bigl( 21x^3  +( 136 - 67 D )x^2 
	  + ( 243 + 64 D^2 - 253 D )x \\  -8(D-1)(D-2)(2D-5) \bigr)
	                                      U_{\ref{figself}\text h}(x) \\
- 8x (x-D+3) (2 x-3 D+7) (3 x-4 D+7) U_{\ref{figself}\text h}(x+1)  =0 \ ,
\end{multline}
has only the roots $1$ and $-1/8$.

\subsection{Applications to simple one-loop integrals}
\label{example1looplapla}
Now we consider the solution with Laplace's transformation
of the difference equations analyzed in the examples
of sections~\ref{1loopv} and~\ref{1loopself}.
\subsubsection{One-loop vacuum integral}\label{1loopv_dif}
Considering the integral $J(x)$ of \eqref{inte0},
the endpoints  of the line of integration are the origin  
and the root of the characteristic equation $\mu=1/m_1^2$, so that  we write 
\begin{equation}\label{inint}
J(x)=\int_0^{1/m_1^2} dt \; t^{x-1} v_J(t) \ ,
\end{equation}
where $v_J$ satisfies the differential equation
\begin{equation}\label{equde1}
-t (m_1^2 t-1) {v'_J}(t) + (D/2-m_1^2 t) v_J(t) =0 \ .
\end{equation}
The solution is 
\begin{equation}\label{v000}
v_J(t)= C (1/m_1^2-t)^{D/2-1} t^{-D/2}  \ ;
\end{equation}
the constant $C$ can be deduced from 
the value of $a_0$ (see section~\ref{vala0}),
the value of $\K=-D/2$ and the relation \itref{aA0}.
One finds
\begin{equation}
C= (m_1^2)^{D/2-1}/\Gamma(D/2) \ .
\end{equation}

\subsubsection{One-loop self-energy integral}\label{2den}
Here we consider $I(x)$ of \eqref{inte1}, and in particular
the case of non-zero masses.
The homogeneous solution can be written using the Laplace's transformation as 
\begin{equation}
\IOMOG_{\pm}(x)=\int_0^{\mu_{\pm}} dt \; t^{x-1} \vOMOG(t)  \ .
\end{equation}
The function $\vOMOG(t)$ satisfies the differential equation
\begin{equation}\label{equde2}
-t \Phi_1(t) \vOMOG'(t) + \Phi_0(t) \vOMOG(t) =0\ ,
\end{equation}
where 
\begin{equation}\label{Phi10}
\begin{split}
\Phi_1(t)
         &= R^2(p^2,-m_1^2,-m_2^2) (t-\mup)(t-\mum) \ ,\\
\Phi_0(t)&=-t^2 R^2(p^2,-m_1^2,-m_2^2) +(D-1)(p^2+m_1^2-m_2^2)t +2-D \ .\qquad
\end{split}
\end{equation}
The solution of this equation is
\begin{equation}
\vOMOG(t)= C \left((\mup-t)(\mum-t)\right)^{(D-3)/2} t^{2-D} \ .
\end{equation}
Values of $C$ such that $\IOMOG_{\pm}$ are the same functions 
considered in section~\ref{m1m2nz} are 
\begin{equation}
C_{\pm}= \mu_{\pm}^{(D-1)/2} 
   \left(\mu_{\mp}-\mu_{\pm}\right)^{(3-D)/2}/\Gamma((D-1)/2) \ .
\end{equation}
Considering now the nonhomogeneous solution
\begin{equation}
\INOMOG(x)=\int_0^{1/m_1^2} dt \; t^{x-1} \vNOMOG(t) \ ,
\end{equation}
the function $\vNOMOG$ satisfies the differential equation
\begin{equation}\label{equde3}
-t \Phi_1(t) \vNOMOG'(t) + \Phi_0(t) \vNOMOG(t) = t\phi_1(t) v_J'(t) - \phi_0(t)
v_J(t)\ ,
\end{equation}
where $\Phi_0$ and $\Phi_1$ are given in \eqref{Phi10},
$\phi_0$ and $\phi_1$ are
\begin{equation}
\begin{split}
\phi_1(t)&= -t(p^2+m_2^2)/m_1^2 \ , \\
\phi_0(t)&= t(D(p^2+m_1^2+m_2^2)/(2m_1^2)-1) \ , \\
\end{split}
\end{equation}
and $v_J$ is given in \eqref{v000}.

\subsubsection{Numerical example}
We consider the calculation using Laplace's transformation
of $J(1)$, $\IOMOG_{-}(1)$ and $\INOMOG(1)$, 
numerically calculated in section~\ref{numexa}.
The corresponding differential equations are 
\eqref{equde1}, \eqref{equde2} and \eqref{equde3}.
The singular points of the system are $t=1$, $-1/3$ and $0$.
The equations are integrated using the method described
in section~\ref{solvedifequ}.
The integral over $t$ is divided into 4 intervals,
with endpoints $0$, $1/8$, $1/4$, $1/2$, $1$;
a cutoff $\lambda \ll 1$ is conveniently introduced in the first interval
$[\lambda,1/8]$ because the solutions are not regular in $t=0$.
Doing the calculations with 19 digits,
the convergence of the expansions in $t$ is attained in about 80 terms;
the finite values of $J(x)$, $\IOMOG_{-}(x)$ and $\INOMOG(x)$ for $x=3$ and $x=4$
are obtained by calculating the integrals with \eqref{usum};
the unstable recurrence relations are used only
to calculate the values for $x\le 2$, where the integrals are divergent,
so that the error on $I(1)$ is reduced by two orders of magnitude with
respect to \eqref{vali1}.

\section{Application to multi-loop diagrams}\label{resu0}
After the self-energy diagram discussed in section~\ref{1loopself},
now we consider more complicated diagrams:
the vacuum and the self-energy diagrams up to three
loop, shown in Figs.~\ref{figvac} and~\ref{figself},
and the vertex and box diagrams up to two loops, shown in 
Figs.~\ref{figvert} and~\ref{figbox}
(we have considered all diagrams such that the scalar integral \itref{inti}
is always a master integral which does not factorize in a product of simpler master integrals).
A complete discussion of all these diagrams would be rather long 
and we postpone it to future papers. 
However, to give an idea of the kind and complexities of the equations
involved in the calculations, we will show some results.

First of all, we distinct in each diagram $g$ the 
topologically different lines 
which are indicated in the figures with a number;
of course in diagrams without numbers all lines are topologically equivalent.
For each topologically different line $l$ we set $D_1$ equal to the denominator
of such line and we consider the difference equation satisfied by the
scalar master function $U_{gl}(x)$ and 
the differential equation satisfied by the Laplace-transformed
function $v_{gl}(t)$, both  corresponding to the
scalar master integral, 
\begin{equation}
U_{gl}(x)= \int_l dt \; t^{x-1} v_{gl}(t) =
\int \dfrac {\dk1 \dots \dk{\NK}}
{\displaystyle D_1^x D_2 \ldots  D_\ND} \ . 
\end{equation}
\begin{figure}
\begin{center}
   \begin{picture}(380,120)(0,0)
   \thicklines
   {
   \put(257,079){\text{1}}   
   \put(262,062){\text{2}}   
\LARGE
   \put(031,032){\text{(a)}}   
   \put(106,032){\text{(b)}}   
   \put(181,032){\text{(c)}}   
   \put(255,032){\text{(d)}}   
   \put(332,032){\text{(e)}}   
   }
   \put(060,080){\circle*{7}}
   \put(040,080){\circle{40}}
   \put(115,080){\circle{40}}
   \put(095,080){\line(1,0){40}}
   \put(190,080){\circle{40}}
   \qbezier(170,080)(190,100)(209.5,080)
   \qbezier(170,080)(190,060)(209.5,080)
   \put(265,080){\circle{40}}
   \put(265,099.2){\line(+3,-5){17}}
   \put(265,099.2){\line(-3,-5){17}}
   \put(340,080){\circle{40}}
   \put(340,080){\line(0,1){20}}
   \put(340,080){\line(+5,-3){17}}
   \put(340,080){\line(-5,-3){17}}
   \end{picture}
\end{center}
 \caption{Vacuum diagrams up to three loops.}
 \label{figvac}
 \end{figure}
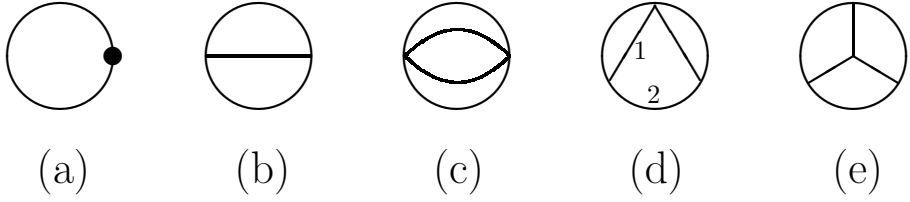
\begin{figure}
\begin{center}
   \begin{picture}(370,480)(0,0)
   \thicklines
   {
   \put(082,458){\text{$p$}}
   \put(211,476){\text{1}}   
   \put(247,472){\text{2}}   
   \put(227,442){\text{3}}   
   \put(301,476){\text{1}}   
   \put(321,457){\text{2}}   
   \put(121,396){\text{1}}   
   \put(157,392){\text{2}}   
   \put(137,362){\text{3}}   
   \put(211,396){\text{1}}   
   \put(247,392){\text{2}}   
   \put(227,362){\text{3}}   
   \put(301,396){\text{1}}   
   \put(317,362){\text{2}}   
   \put(031,316){\text{1}}   
   \put(052,306){\text{2}}   
   \put(047,282){\text{3}}   
   \put(121,316){\text{1}}   
   \put(144,297){\text{2}}   
   \put(211,316){\text{1}}   
   \put(234,321){\text{2}}   
   \put(248,311){\text{3}}   
   \put(230,299){\text{4}}   
   \put(227,282){\text{5}}   
   \put(301,316){\text{1}}   
   \put(322,298){\text{3}}   
   \put(337,312){\text{2}}   
   \put(337,282){\text{5}}   
   \put(297,282){\text{4}}   
   \put(031,236){\text{1}}   
   \put(070,227){\text{2}}   
   \put(047,202){\text{3}}   
   \put(121,236){\text{1}}   
   \put(157,232){\text{2}}   
   \put(211,236){\text{1}}   
   \put(222,226){\text{2}}   
   \put(232,208){\text{3}}   
   \put(301,236){\text{1}}   
   \put(322,215){\text{4}}   
   \put(337,202){\text{5}}   
   \put(340,227){\text{3}}   
   \put(328,240){\text{2}}   
   \put(297,202){\text{6}}   
   \put(072,148){\text{1}}   
   \put(092,122){\text{3}}   
    {\footnotesize \put(090,149){\text{2}}}   
   \put(113,151){\text{4}}   
   \put(165,156){\text{1}}   
   \put(187,144){\text{3}}   
   \put(202,152){\text{2}}   
   \put(173,131){\text{4}}   
   \put(256,156){\text{1}}   
   \put(249,131){\text{2}}   
   \put(268,138){\text{3}}   
   \put(272,122){\text{4}}   
   \put(072,068){\text{1}}   
   \put(098,060){\text{2}}   
   \put(092,042){\text{3}}   
   \put(160,064){\text{1}}   
   \put(182,082){\text{2}}   
   \put(193,057){\text{3}}   
   \put(179,048){\text{4}}   
   \put(201,041){\text{5}}   
   \put(250,064){\text{1}}   
   \put(272,082){\text{2}}   
\LARGE
   \put(040,425){\text{(a)}}   
   \put(130,425){\text{(b)}}   
   \put(220,425){\text{(c)}}   
   \put(310,425){\text{(d)}}   
   
   \put(040,345){\text{(e)}}   
   \put(130,345){\text{(f)}}   
   \put(220,345){\text{(g)}}   
   \put(310,345){\text{(h)}}   
   
   \put(040,265){\text{(i)}}   
   \put(130,265){\text{(j)}}   
   \put(220,265){\text{(k)}}   
   \put(310,265){\text{(l)}}   
   
   \put(040,185){\text{(m)}}   
   \put(130,185){\text{(n)}}   
   \put(220,185){\text{(o)}}   
   \put(310,185){\text{(p)}}   
   
   \put(085,105){\text{(q)}}   
   \put(175,105){\text{(r)}}   
   \put(265,105){\text{(s)}}   
   
   \put(085,025){\text{(t)}}   
   \put(175,025){\text{(u)}}   
   \put(265,025){\text{(v)}}   
   }

   \put(050,460){\circle{40}}
   \put(030,460){\line(-1,0){10}}
   \put(070,460){\line(+1,0){10}}
   \put(140,460){\circle{40}}
   \put(120,460){\line(1,0){40}}
   \put(120,460){\line(-1,0){10}}
   \put(160,460){\line(+1,0){10}}
   \put(230,460){\circle{40}}
   \put(210,460){\line(-1,0){10}}
   \put(250,460){\line(+1,0){10}}
   \qbezier(210,460)(228,462)(230,480)
   \put(320,460){\circle{40}}
   \put(320,440){\line(0,+1){40}}
   \put(300,460){\line(-1,0){10}}
   \put(340,460){\line(+1,0){10}}
   \put(050,380){\circle{40}}
   \qbezier(030,380)(050,400)(070,380)
   \qbezier(030,380)(050,360)(070,380)
   \put(030,380){\line(-1,0){10}}
   \put(070,380){\line(+1,0){10}}
   \put(140,380){\circle{40}}
   \put(120,380){\line(-1,0){10}}
   \put(160,380){\line(+1,0){10}}
   \qbezier(120,380)(130,390)(140,400)
   \qbezier(120,380)(138,382)(140,400)
   \put(230,380){\circle{40}}
   \put(210,380){\line(-1,0){10}}
   \put(250,380){\line(+1,0){10}}
   \qbezier(210,380)(227,383)(230,400)
   \put(210,380){\line(1,0){40}}
   \put(320,380){\circle{40}}
   \qbezier(300,380)(318,382)(320,400)
   \qbezier(340,380)(322,382)(320,400)
   \put(300,380){\line(-1,0){10}}
   \put(340,380){\line(+1,0){10}}
   \put(050,300){\circle{40}}
   \put(030,300){\line(-1,0){10}}
   \put(070,300){\line(+1,0){10}}
   \put(030,300){\line(+1,0){40}}
   \put(050,300){\line(0,+1){20}}
   \put(140,300){\circle{40}}
   \put(120,300){\line(-1,0){10}}
   \put(160,300){\line(+1,0){10}}
   \qbezier(140,280)(160,300)(140,320)
   \qbezier(140,280)(120,300)(140,320)
   \put(230,300){\circle{40}}
   \put(210,300){\line(-1,0){10}}
   \put(250,300){\line(+1,0){10}}
   \qbezier(210,300)(227,303)(230,320)
   \qbezier(210,300)(233,304)(240,317)
   \put(320,300){\circle{40}}
   \put(300,300){\line(-1,0){10}}
   \put(340,300){\line(+1,0){10}}
   \qbezier(300,300)(317,303)(320,320)
   \put(320,280){\line(0,+1){40}}
   \put(050,220){\circle{40}}
   \put(030,220){\line(-1,0){10}}
   \put(070,220){\line(+1,0){10}}
   \qbezier(030,220)(047,223)(050,240)
   \qbezier(050,240)(052,224)(064,233)
   \put(140,220){\circle{40}}
   \put(120,220){\line(-1,0){10}}
   \put(160,220){\line(+1,0){10}}
   \qbezier(120,220)(137,223)(140,240)
   \qbezier(120,220)(137,217)(140,200)
   \put(230,220){\circle{40}}
   \put(210,220){\line(-1,0){10}}
   \put(250,220){\line(+1,0){10}}
   \put(230,200){\line(0,+1){20}}
   \put(230,230){\circle{20}}
   \put(320,220){\circle{40}}
   \put(300,220){\line(-1,0){10}}
   \put(340,220){\line(+1,0){10}}
   \put(320,200){\line(0,+1){40}}   
   \qbezier(320,240)(322,224)(334,233)
   \put(095,140){\circle{40}}
   \put(075,140){\line(-1,0){10}}
   \put(115,140){\line(+1,0){10}}
   \qbezier(085,157)(088,150)(092,143)
   \qbezier(075,140)(100,140)(105,157)
   \put(185,140){\circle{40}}
   \put(165,140){\line(-1,0){10}}
   \put(205,140){\line(+1,0){10}}
   \put(185,120){\line(0,+1){40}}
   \put(165,140){\line(+1,0){20}}
   \put(275,140){\circle{40}}
   \put(255,140){\line(-1,0){10}}
   \put(295,140){\line(+1,0){10}}
   \put(275,159.2){\line(+3,-5){17}}
   \put(275,159.2){\line(-3,-5){17}}
   \put(095,060){\circle{40}}
   \put(075,060){\line(-1,0){10}}
   \put(115,060){\line(+1,0){10}}
   \put(085,043){\line(0,+1){34}}
   \put(105,043){\line(0,+1){34}}
   \put(185,060){\circle{40}}
   \put(165,060){\line(-1,0){10}}
   \put(205,060){\line(+1,0){10}}
   \put(185,060){\line(0,-1){20}}
   \put(185,060){\line(+5,+3){17}}
   \put(185,060){\line(-5,+3){17}}
   \put(275,060){\circle{40}}
   \put(255,060){\line(-1,0){10}}
   \put(295,060){\line(+1,0){10}}
   \put(261,074){\line(+1,-1){28}}
   \qbezier(261,046)(267,052)(273,058)
   \qbezier(289,074)(283,068)(277,062)
   \qbezier(273,058)(271,064)(277,062)
   \end{picture}
\end{center}
 \caption{Self-energy diagrams up to three loops.}
 \label{figself}
 \end{figure}
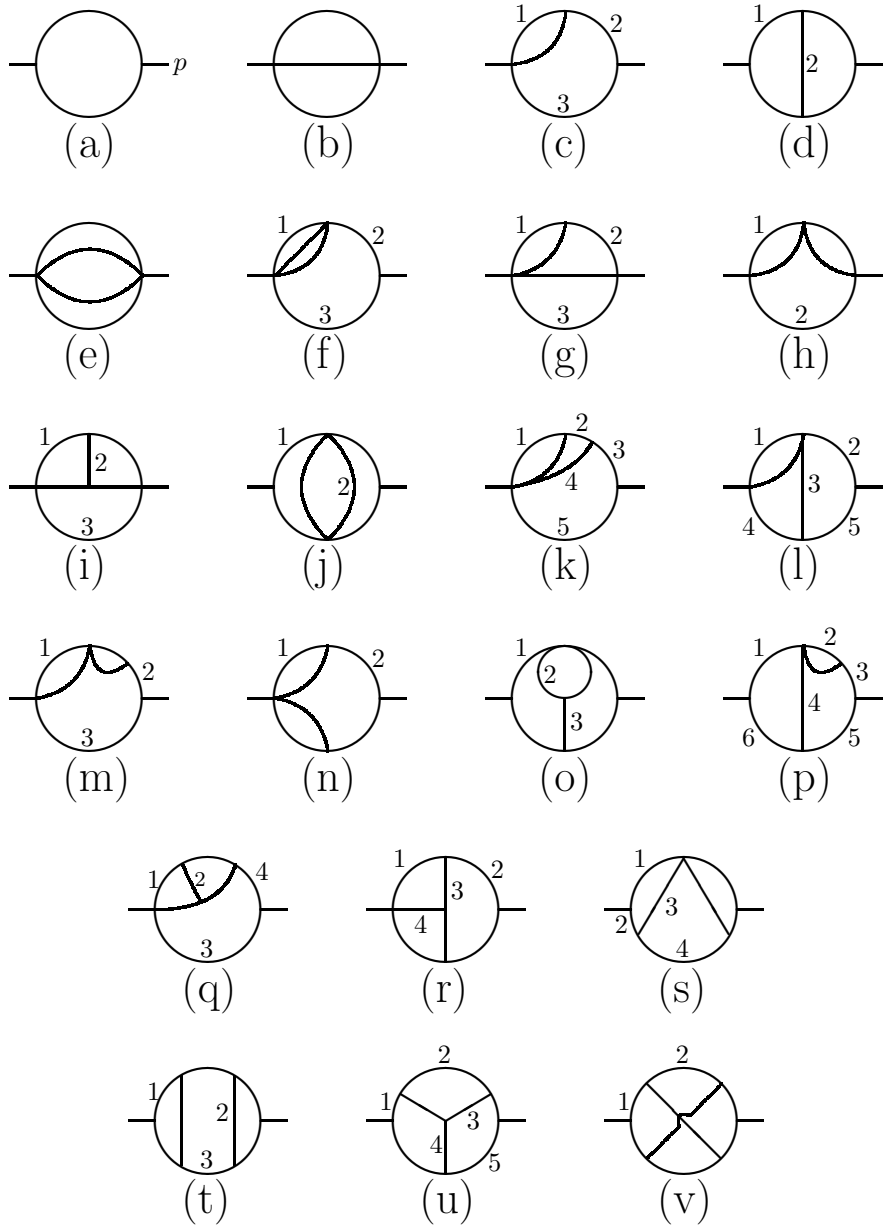
\begin{figure}
\begin{center}
   \begin{picture}(300,200)(0,0)
   \thicklines
   {
   \put(009,130){\text{$p_1$}}
   \put(082,130){\text{$p_2$}}
   \put(011,180){\text{$p_1-p_2$}}
   \put(135,160){\text{1}}   
   \put(147,149){\text{2}}   
   \put(235,160){\text{1}}   
   \put(242,148){\text{2}}   
   \put(255,143){\text{3}}   
   \put(259,160){\text{4}}   
   \put(035,060){\text{1}}   
   \put(043,048){\text{2}}   
   \put(056,042){\text{3}}   
   \put(138,063){\text{1}}   
   \put(127,046){\text{2}}   
   \put(145,049){\text{3}}   
   \put(154,042){\text{4}}   
\LARGE
   \put(040,120){\text{(a)}}   
   \put(140,120){\text{(b)}}   
   \put(240,120){\text{(c)}}   
   \put(040,020){\text{(d)}}   
   \put(140,020){\text{(e)}}   
   \put(240,020){\text{(f)}}   
   }
   \put(030,140){\line(+1,0){40}}
   \put(030,140){\line(+3,+5){20}}
   \put(070,140){\line(-3,+5){20}}
   \put(030,140){\line(-5,-3){10}}
   \put(070,140){\line(+5,-3){10}}
   \put(050,173){\line(0,+1){10}}
   \put(130,140){\line(+1,0){40}}
   \put(130,140){\line(+3,+5){20}}
   \put(170,140){\line(-3,+5){20}}
   \put(130,140){\line(-5,-3){10}}
   \put(170,140){\line(+5,-3){10}}
   \put(150,173){\line(0,+1){10}}
   \qbezier(130,140)(150,155)(169,140)
   \put(230,140){\line(+1,0){40}}
   \put(230,140){\line(+3,+5){20}}
   \put(270,140){\line(-3,+5){20}}
   \put(230,140){\line(-5,-3){10}}
   \put(270,140){\line(+5,-3){10}}
   \put(250,173){\line(0,+1){10}}
   \qbezier(230,140)(240,152)(250,140)
   \put(030,040){\line(+1,0){40}}
   \put(030,040){\line(+3,+5){20}}
   \put(070,040){\line(-3,+5){20}}
   \put(030,040){\line(-5,-3){10}}
   \put(070,040){\line(+5,-3){10}}
   \put(050,073){\line(0,+1){10}}
   \put(050,073){\line(0,-1){33}}
   \put(130,040){\line(+1,0){40}}
   \put(130,040){\line(+3,+5){20}}
   \put(170,040){\line(-3,+5){20}}
   \put(130,040){\line(-5,-3){10}}
   \put(170,040){\line(+5,-3){10}}
   \put(150,073){\line(0,+1){10}}
   \put(140,057){\line(+1,0){20}}
   \put(230,040){\line(+3,+5){20}}
   \put(270,040){\line(-3,+5){20}}
   \put(230,040){\line(-1,-1){10}}
   \put(270,040){\line(+1,-1){10}}
   \put(250,073){\line(0,+1){10}}
   \put(230,040){\line(+5,+3){29}}
   \put(270,040){\line(-5,+3){17}}
   \qbezier(246,054)(243,056)(240.5,057.5)
   \qbezier(246,054)(254,058)(252.5,050.5)
   \end{picture}
\end{center}
 \caption{Vertex diagrams up to two loops.}
 \label{figvert}
 \end{figure}
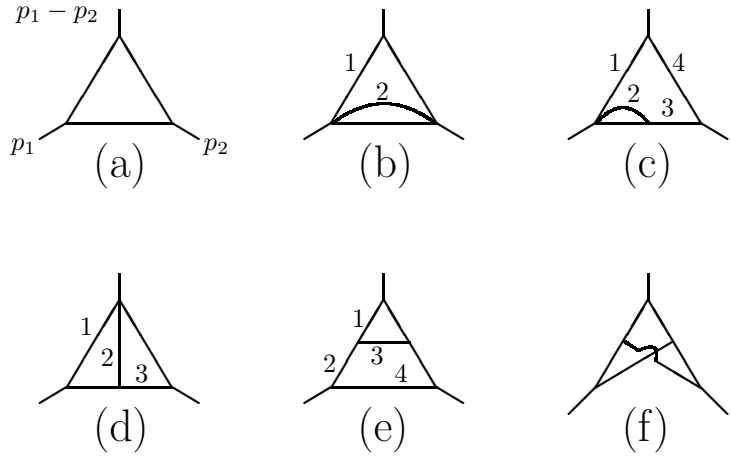
\begin{figure}
\begin{center}
   \begin{picture}(370,200)(0,0)
   \thicklines
   {
   \put(009,130){\text{$p_1$}}
   \put(082,130){\text{$p_3$}}
   \put(004,192){\text{$p_1-p_2$}}
   \put(063,192){\text{$p_2-p_3$}}
   \put(138,182){\text{1}}   
   \put(113,158){\text{2}}   
   \put(138,150){\text{3}}   
   \put(204,158){\text{1}}   
   \put(234,158){\text{2}}   
   \put(318,182){\text{1}}   
   \put(293,158){\text{2}}   
   \put(318,142){\text{3}}   
   \put(318,160){\text{4}}   
   \put(342,167){\text{5}}
   \put(342,147){\text{6}}
   \put(043,082){\text{1}}   
   \put(018,058){\text{2}}   
   \put(067,058){\text{3}}
   \put(033,050){\text{4}}   
   \put(052,042){\text{5}}   
   \put(103,082){\text{1}}
   \put(087,058){\text{2}}   
   \put(194,082){\text{1}}
   \put(189,062){\text{2}}
   \put(183,042){\text{3}}   
   \put(207,047){\text{4}}
   \put(264,082){\text{1}}
   \put(277,058){\text{2}}
   \put(248,058){\text{3}}
   \put(313,082){\text{1}}
   \put(333,082){\text{2}}
   \put(298,058){\text{3}}   
\LARGE
   \put(040,120){\text{(a)}}   
   \put(130,120){\text{(b)}}   
   \put(220,120){\text{(c)}}   
   \put(310,120){\text{(d)}}   
   \put(035,020){\text{(e)}}   
   \put(105,020){\text{(f)}}   
   \put(175,020){\text{(g)}}   
   \put(245,020){\text{(h)}}   
   \put(315,020){\text{(i)}}   
   }
   \put(030,140){\line(+1,0){40}}
   \put(030,140){\line(0,+1){40}}
   \put(070,140){\line(0,+1){40}}
   \put(030,180){\line(+1,0){40}}
   \put(030,140){\line(-1,-1){10}}
   \put(030,180){\line(-1,+1){10}}
   \put(070,140){\line(+1,-1){10}}
   \put(070,180){\line(+1,+1){10}}
   \put(120,140){\line(+1,0){40}}
   \put(120,140){\line(0,+1){40}}
   \put(160,140){\line(0,+1){40}}
   \put(120,180){\line(+1,0){40}}
   \put(120,140){\line(-1,-1){10}}
   \put(120,180){\line(-1,+1){10}}
   \put(160,140){\line(+1,-1){10}}
   \put(160,180){\line(+1,+1){10}}
   \qbezier(120,140)(140,155)(160,140)
   \put(210,140){\line(+1,0){40}}
   \put(210,140){\line(0,+1){40}}
   \put(250,140){\line(0,+1){40}}
   \put(210,180){\line(+1,0){40}}
   \put(210,140){\line(-1,-1){10}}
   \put(210,180){\line(-1,+1){10}}
   \put(250,140){\line(+1,-1){10}}
   \put(250,180){\line(+1,+1){10}}
   \put(210,180){\line(+1,-1){40}}
   \put(300,140){\line(+1,0){40}}
   \put(300,140){\line(0,+1){40}}
   \put(340,140){\line(0,+1){40}}
   \put(300,180){\line(+1,0){40}}
   \put(300,140){\line(-1,-1){10}}
   \put(300,180){\line(-1,+1){10}}
   \put(340,140){\line(+1,-1){10}}
   \put(340,180){\line(+1,+1){10}}
   \put(300,180){\line(+2,-1){40}}
   \put(025,040){\line(+1,0){40}}
   \put(025,040){\line(0,+1){40}}
   \put(065,040){\line(0,+1){40}}
   \put(025,080){\line(+1,0){40}}
   \put(025,040){\line(-1,-1){10}}
   \put(025,080){\line(-1,+1){10}}
   \put(065,040){\line(+1,-1){10}}
   \put(065,080){\line(+1,+1){10}}
   \qbezier(025,040)(035,055)(045,040)
   \put(095,040){\line(+1,0){40}}
   \put(095,040){\line(0,+1){40}}
   \put(135,040){\line(-1,+2){20}}
   \put(135,080){\line(-1,-1){10}}
   \put(095,040){\line(+1,+1){23}}
   \qbezier(117.5,063)(115,073)(124,070)
   \put(095,080){\line(+1,0){40}}
   \put(095,040){\line(-1,-1){10}}
   \put(095,080){\line(-1,+1){10}}
   \put(135,040){\line(+1,-1){10}}
   \put(135,080){\line(+1,+1){10}}
   \put(165,040){\line(+1,0){40}}
   \put(165,040){\line(0,+1){40}}
   \put(205,040){\line(0,+1){40}}
   \put(165,080){\line(+1,0){40}}
   \put(165,040){\line(-1,-1){10}}
   \put(165,080){\line(-1,+1){10}}
   \put(205,040){\line(+1,-1){10}}
   \put(205,080){\line(+1,+1){10}}
   \put(185,080){\line(+1,-1){20}}
   \put(235,040){\line(+1,0){40}}
   \put(235,040){\line(0,+1){40}}
   \put(275,040){\line(0,+1){40}}
   \put(235,080){\line(+1,0){40}}
   \put(235,040){\line(-1,-1){10}}
   \put(235,080){\line(-1,+1){10}}
   \put(275,040){\line(+1,-1){10}}
   \put(275,080){\line(+1,+1){10}}
   \put(255,080){\line(0,-1){40}}
   \put(305,040){\line(+1,0){40}}
   \put(305,040){\line(0,+1){40}}
   \put(345,040){\line(-1,+2){20}}
   \put(345,080){\line(-1,-2){8}}
   \put(325,040){\line(+1,+2){8}}
   \qbezier(332.5,056)(327,064)(336,064)
   \put(305,080){\line(+1,0){40}}
   \put(305,040){\line(-1,-1){10}}
   \put(305,080){\line(-1,+1){10}}
   \put(345,040){\line(+1,-1){10}}
   \put(345,080){\line(+1,+1){10}}
   \end{picture}
\end{center}
 \caption{Box diagrams up to two loops.}
 \label{figbox}
 \end{figure}
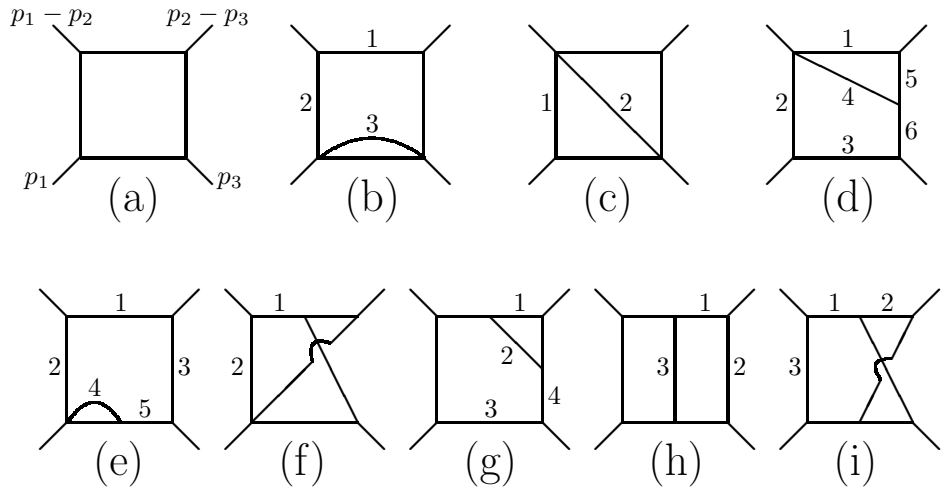

\def\baselinestretch{}
{
\begin{table}
\begin{center}
\vskip -17pt
\begin{tabular}{lcrr|crr}
\hline
\scriptsize{diagram}             & $n_b$ &   $R_C$     & $S$          
& $n_b'$ &  $R_{C'}'$     & $S'$  \\ \hline
\ref{figvac}a     &   1   & $1_1$       &  1           
&   1   & $1_1$        & 1      \\
\ref{figvac}b     &   1   & 2           &  1           
&   1   & 2            & 1      \\
\ref{figvac}c     &  1,2,1& $5_1$       &  4           
&   1   & $2_1$        & 2      \\
\ref{figvac}d1-2  &   1   & 2,5         &  1,1         
&   1   & 2,2          & 1,1    \\
\ref{figvac}e     &   1   & 2           &  1           
&   1   & 2            & 1      \\
\ref{figself}a   &   1   & 2           &  1           
&   1   & $2_1$         & 1      \\
\ref{figself}b   &  1,2,1& $5_1$       &  4           
&   1   & $2_1$         & 2      \\
\ref{figself}c1-3&   1   & 2,5,2       &  1,1,1       
&   1   & 2,2,$2_1$     & 1,1,1  \\
\ref{figself}d1-2&   1   & 2,2         &  1,1         
&   1   & 2,2           & 1,1    \\
\ref{figself}e   & 1,5,5 & $12_4$      &  11          
& 1,5,5 & $4_1$         & 3      \\
\ref{figself}f1-3& 1,2,1 & $5_1$,15,2  &  4,4,1       
&   1   & $2_1$,$5_1$,$2_1$& 2,2,1  \\
\ref{figself}g1-3& 1,2,1 & 2,15,$5_1$  &  1,4,4       
&   1   & 2,5,$2_1$     & 1,2,2  \\
\ref{figself}h1-2&1,2,2,1& $10_1$,$10_1$&  6,6        
&   1   & $2_1$,$2_1$ & 3,3    \\
\ref{figself}i1-3& 1,2,1 & 8,10,$5_1$  &  4,4,4       
&  1,1  & 3,3,$3_1$     & 3,2,3  \\
\ref{figself}j1-2& 1,2,1 & 8,$5_1$     &  4,4         
&  1,1  & $3_1$,$3_1$   & 3,3    \\
\ref{figself}k1-5&   1   & 2,5,5,2,2   &  1,1,1,1,1   
&   1   & 2,2,2,2,$2_1$& 1,1,1,1,1\\
\ref{figself}l1-5& 1,2,1 &$5_1$,8,8,8,10& 4,4,4,4,4   
&  1,1  & $3_1$,$3_1$,3,3,3& 3,3,3,3,2\\
\ref{figself}m1-3&   1   & 2,7,2       &  1,1,1       
&   1   & 2,2,$2_1$     & 1,1,1\\
\ref{figself}n1-2&   1   & 2,5         &  1,1         
&   1   & 2,2           & 1,1\\
\ref{figself}o1-3&   1   & 2,2,5       &  1,1,1       
&   1   & 2,2,3         & 1,1,1 \\
\ref{figself}p1-6&   1   & 2,2,5,2,2,2 &  1,1,1,1,1,1 
&   1   & 2,2,3,2,2,2   & 1,1,1,1,1,1 \\
\ref{figself}q1-4&   1   & 2,2,2,5     &  1,1,1,1     
&   1   & 2,2,$2_1$,3   & 1,1,1,1 \\
\ref{figself}r1-4& 1,2,2 & 6,9,9,6     &  5,5,5,5     
&  1,1  & 3,$4_2$,$4_1$,3& 3,3,3,3 \\
\ref{figself}s1-4& 1,2,2 & 6,9,6,9     &  5,5,5,5     
&  1,1  &$3_1$,$4_2$,3,$4_1$&3,3,3,3 \\
\ref{figself}t1-3&   1   & 2,2,6       &  1,1,1       
&   1   & 2,2,3         &1,1,1 \\
\ref{figself}u1-5&   1   & 6,2,2,6,2   &  1,1,1,1,1   
&   1   & 3,2,2,3,2     &1,1,1,1,1 \\
\ref{figself}v1-2& 1,1,1 & 8,8         &  3,3         
&   1   & 3,3           &2,2 \\
\ref{figvert}a   &   1   & 2           &  1           
&   1   & $2_1$         & 1      \\
\ref{figvert}b1-2& 1,2,1 & 8,$5_1$     &  4,4         
&  1,1  & $3_1$,$3_1$   & 3,3     \\
\ref{figvert}c1-4&   1   & 2,2,5,2     &  1,1,1,1     
&   1   & $2_1$,2,3,$2_1$& 1,1,1,1 \\
\ref{figvert}d1-3& 1,2,2 & 6,6,9       &  5,5,5       
&  1,1  & $3_1$,3,$4_2$  & 3,3,3 \\
\ref{figvert}e1-4&   1   & 2,6,2,2     &  1,1,1,1     
&   1   & 2,3,2,$2_1$   & 1,1,1,1 \\
\ref{figvert}f   & 1,1,1 & 8           &  3           
&   1   & 3             & 2 \\
\ref{figbox}a    &   1   & 2           &  1           
&   1   & $2_1$         & 1  \\
\ref{figbox}b1-3 & 1,2,1 & 8,8,$5_1$   &  4,4,4       
&  1,1  & $4_1$,$4_1$,$3_1$& 3,3,3\\
\ref{figbox}c1-2 &1,4,3  & 12,$9_1$    &  8,8         
& 1,2,1 & $6_1$,$4_1$   & 5,4  \\
\ref{figbox}d1-6 & 1,3,1 &6,10,10,6,9,13& 5,5,5,5,5,5 
&  1,2  & $4_1,$$5_1$,$5_1$,4,$5_2$,$7_4$ & 4,4,4,4,3,5   \\
\ref{figbox}e1-5 &   1   & 2,2,2,2,5   &  1,1,1,1,1   
&   1   & $2_1$,$2_1$,$2_1$,2,3 & 1,1,1,1,1   \\
\ref{figbox}f1-2 &1,3,6,3& (23),(19)   &  (13),(13)   
&1,3,6,1& $16_4$,$13_1$ & 13,13 \\
\ref{figbox}g1-4 &   1   & 2,2,2,6     &  1,1,1,1     
&   1   & 2,2,$2_1$,4   & 1,1,1,1  \\
\ref{figbox}h1-3 & 1,2,2 & 10,6,6      &  5,5,5       
&  1,1  & $5_2$,$3_1$,3 & 3,3,3  \\
\ref{figbox}i1-3 &1,2,3,1& (20),12,10  &  7,7,7       
&  1,1  & $13_4$,5,$3_1$& 7,5,3  \\
\hline
\end{tabular}
\vskip -14pt
\end{center}
\caption{Number of master integrals and orders of the equations for the diagrams.}
\label{tablevacself}
\end{table}
}
\def\baselinestretch{\factor}

\subsection{Arbitrary case}\label{resuarb}
In the left half of Table~\ref{tablevacself},
considering arbitrary (non exceptional) values of masses and momenta,
we list\footnote{
Due to the present limitations of the program used for the
calculations, the values listed in this part of the table 
were calculated by giving arbitrary
rational values to square masses and scalar products of external momenta.
It is possible, but very unlikely, that the chosen values correspond
to some particular case so that the results obtained do not correspond
to the real arbitrary case.}
\begin{itemize}
\item The diagram considered and, if present, the indication
      of the possible values of the index of the topologically different lines.
\item The number $n_b$ of master integrals containing all the $\ND$ denominators,
      determined with the procedure  of section~\ref{identificamaster}, 
 subdivided according to the number of scalar products in the
numerator;
for example, 1,4,3 means that we have found  8 master integrals altogether,
of which 1 with no scalar product, 
4 with 1 scalar product and 3 with a product of 2
scalar products.
\item In the $R_C$ column we list the values of the order $R$
     of the difference equations in $x$ satisfied by $U_{gl}(x)$,
     for each possible choice of $D_1$ as one of the
     topologically distinct lines; 
     the index $C$, where present, indicates 
     (assuming values of external momenta below the deformation threshold)
     the number of constants $\C_j$ which are different from zero because
     the corresponding partial solutions of the homogeneous equation
     satisfy the condition \itref{cond1}.

\item The order $S$ of the differential equations in $t$ 
      satisfied by the function $v_{gl}(t)$, for each possible choice of $D_1$.
\end{itemize}

The orders shown within parenthesis
are estimated from the subsystems of simultaneous equations \itref{diffu2},
avoiding the transformation into triangular form; 
the index $C$ is not shown in these cases.
Analyzing the data of the table we observe that
\begin{enumerate}
\item The differential equation has order less than the 
      order of the difference equation;
      this is expected as $v_{gl}(t)$ is an object simpler than $U_{gl}(x)$.
\item The order of the differential equation in $t$
      is equal to the number of master integrals, 
      with the curious exception
      of the diagrams~\ref{figself}f3 and~\ref{figself}g1 
      where $S<n_b$.
\item The choice of the line heavily affects the order and the
      complexity of the equations,
      as in the case of the diagram~\ref{figself}f, where the difference
      equation may have order 5,15 or 2. 
\item \label{simil} Some similarities appear
  between vertex diagrams, self-energy diagrams
      and vacuum diagrams with different number of loops:\footnote{
The equivalence of recurrence relations very recently described in \cite{prep2}
may probably throw light on this.}
      \begin{enumerate}
      \item 
      between a vertex diagram and 
      the self-energy diagrams
      obtained by connecting two external vertices of the vertex diagram
      with a line and inserting an external momentum in this new line;
      \item
      between a self-energy diagram and 
      the vacuum diagram obtained by connecting its two external lines.
      \end{enumerate}
      The total number of master integrals 
      and the number of master integrals 
      subdivided according to the number of scalar products
      turn out to be the same,
      as well as the orders of the equations for the lines present in both
      diagrams.
      For example, compare the diagrams 
      (\ref{figvert}a, \ref{figself}d, \ref{figvac}e),
      (\ref{figself}a, \ref{figvac}b), 
      (\ref{figself}b, \ref{figvac}c),
      (\ref{figself}c, \ref{figvac}d),
      (\ref{figvert}b, \ref{figself}j, \ref{figself}l),
      (\ref{figvert}c, \ref{figself}o, \ref{figself}p),
      (\ref{figvert}d, \ref{figself}r, \ref{figself}s),
      (\ref{figvert}e, \ref{figself}t, \ref{figself}u) and 
      (\ref{figvert}f, \ref{figself}v).
       The similarities presumably exist even 
       between three-loop self-energy diagrams 
       and four-loop vacuum diagrams 
       (some preliminary results seem to confirm it).
       Perhaps there is a relation with the heuristic ``rule of the mapping''
       described in \cite{Tkachov} for massless self-energy diagrams.
\item The number of master integrals grows probably exponentially\footnote
{We consider here the master integrals containing \emph{one particular}
combination of denominators.
The total number of combinations \itref{defins} itself grows  
exponentially with $\ND$, clearly along with  
the total number of master integrals with every possible
combination of denominators. 
} 
      with the number of loops and external vertices, and may be large.
      Consider for example the class of $L$-loop ``sunset'' self-energy diagrams
      with $L+1$ denominators,
      shown for one, two and three loops respectively
      in Fig.~\ref{figself}a, \ref{figself}b and~\ref{figself}e.
      The number $n_b(L)$ of master integrals for $L=1$ to $L=5$ is
      $1,4,11,26,57$, respectively (the last two values come from preliminary
      analysis). These values seem to follow the law
      $n_b(L)=2^{L+1}-L-2=\sum_{i=2}^{L+1}\binom{L+1}{i}$,
      corresponding to the (alternative) choice of all the master integrals
      with numerator equal to one
      and with one or two as exponents of the denominators, 
      with at least two exponents one.
\end{enumerate}

\subsection{The test case}\label{resu}
As first test of our approach, 
we have considered in particular detail the case where all masses are equal,
$m_1=\ldots=m_\ND=1$, and all the external lines are on-mass-shell.
Denoting the incoming external momenta $-p_1$, $p_1-p_2$,\ldots,
$p_{\NP-1}-p_\NP$, $p_\NP$, 
we choose $p_i^2=-1$ and  $(p_i-p_j)^2=-1$ for every $i$ and $j$.
In the case of box diagrams this corresponds to setting the Mandelstam variables
$s=t=1$ and  $u=2$.
These values of masses and momenta have been chosen because 
they introduce symmetries which allow several consistency checks on equations
and results. 
Similarly to the arbitrary case, 
in the right half of Table~\ref{tablevacself},
for each diagram we list the number $n_b'$ 
of master integrals with all the denominators,
the order $R'$ of the difference equation satisfied by the function
$U_{gl}(x)$,
the order $S'$ of the differential equation satisfied by the corresponding
function $v_{gl}(t)$ and, if present, the number $C'$ of non-zero constants 
needed to determine the solution.

For these particular values of masses and momenta
the equations turn out to be, as expected, simpler than in the arbitrary case.
The homogeneous parts of the equations
which presented some similarities 
in the arbitrary case
(see the observation~\ref{simil} of section~\ref{resuarb})
here turn out to be \emph{identical};
of course the nonhomogeneous parts of these equations are quite different.

The number of constants $\C_j$ to find turns out to be
greater than or equal to that of the arbitrary case 
(except for diagram~\ref{figself}e);
in this connection the test case is more complicated than the arbitrary case.
With the exception of the diagram~\ref{figbox}d6, the calculation of
the scalar master integrals requires no more than two constants,
easily determined using the identities of section~\ref{valx0}
(which turn out always to provide one useful relation involving the constants)
and the large-$x$ leading behaviours.
One can prove that the values of external momenta
are always below or at the deformation threshold,
with the exception of some subdiagrams of
diagrams~\ref{figbox}f and~\ref{figbox}{i};
in the cases at deformation threshold the values
of the constants $\tilde L_1, \tilde L_2$ and $\tilde L_3$ 
given in \eqref{valuesl} are needed
(for example, for the diagrams~\ref{figself}a, \ref{figvert}a and~\ref{figbox}a).

The characteristic equations of all difference equations
have always the solution $\mu=1$.
We list the values of the other roots for some diagrams:
(Fig.~\ref{figvac}b: $\mu=-1/3$),
(\ref{figvac}c: $-1/8$),
(\ref{figvac}e: $-1/2$),
(\ref{figself}e: $-1/3$, $-1/15$),
(\ref{figself}i1: $-1/3$, $-1/2$),
(\ref{figself}i2: $-1/3$),
(\ref{figself}i3, \ref{figself}j2, \ref{figself}l1 and~\ref{figvert}b2 :
 $1\pm\sqrt{4/3}$),
(\ref{figself}v1-2 and \ref{figvert}f: $-1/3$, $1/9$),
(\ref{figvert}a: $-1/2$),
(\ref{figbox}a: $-3/5$),
(\ref{figbox}b1-2: $-1/2$, $-1/4\pm i\sqrt{1/8}$),
(\ref{figbox}b3: $3/4\pm \sqrt{27/32}$),
(\ref{figbox}f1: $-1/2$, $-1/3$, $-1$,
              $(1\pm\sqrt{3})/2$, $-0.098\pm0.050i$, $-0.934$, $3.632$),
(\ref{figbox}f2: $-1/2$, $-1/3$, $-1$,
              $(1\pm\sqrt{3})/2$, $0.049\pm0.099i$, $-2.607$,
	      $-0.819$, $1.756$, $12.07$),
(\ref{figbox}i1: $1/9$, $3/2$, $-1/2$, $-1/3$, $-1$) and
(\ref{figbox}i2: $1/9$, $3/2$, $-1/3$, $-3$).
Most of these values, written in the form $1/(q^2+1)$,
are due to presence of singularities at $k^2=q^2$ 
in the function $f(k^2)$ (see section \ref{othercases}).
All characteristic equations turn out to 
have one negative rational solution $-1<\mu<0$,
so that, according to section~\ref{sumfact2},
all recurrence relations are unstable;
the diagram~\ref{figself}e shows the greatest instability ($A=15$).
One can ask whether the difference equations of
all the  analyzed
diagrams
admit as solutions \emph{convergent} factorial series expansions.
The answer is negative.
In fact
the root $\mu=1/9$ of the diagrams~\ref{figself}v1-2 and~\ref{figvert}f,
the root $(1+\sqrt{3})/2$ and the complex roots
of diagrams~\ref{figbox}f1-2,
the roots $1/9$ and $3/2$ of diagrams~\ref{figbox}i1-2
(note, all the diagrams with lines crossed)
and
the root $3/4+\sqrt{27/32}$ of the diagram~\ref{figbox}b3
satisfy
the condition \itref{conv} (with $\mu^{(\alpha)}=1$) so that the factorial
series expansions of the solutions never converge.
Therefore we are forced to use Laplace's transformation for these diagrams
and these choices of
the line $D_1$,
and for all the diagrams which become these diagrams by deleting lines.
In all the other cases 
factorial series expansions can be used.

\subsection{Numerical results}
\label{resuvalues}
For each diagram shown in 
Figs.~\ref{figvac}, \ref{figself}, \ref{figvert} and~\ref{figbox},
except for the diagrams~\ref{figbox}f and~\ref{figbox}i 
(see section~\ref{boxdiagrams}),
we have calculated the values of the master integrals for $D=4-2\e$,
using the values of masses and momenta shown in the previous section. 

Calculations were carried out by using the program 
\SYS~ described in section~\ref{program}.
The master integrals of the diagrams 
\ref{figvac}b, \ref{figvac}e,
\ref{figself}a, \ref{figself}d,
\ref{figself}i, \ref{figself}j, \ref{figself}l,
\ref{figvert}a, \ref{figvert}e,
\ref{figbox}a and~\ref{figbox}c
were first calculated using expansions in factorial series;
these integrals were also recalculated using Laplace's transformation
in order to provide important checks of the calculations.
The master integrals of all the remaining diagrams 
were calculated using Laplace's transformation.
Excluding the simplest diagrams, calculations with Laplace's transformation 
turned out faster than calculations with factorial series;
this is due to the instabilities of the recurrence relations, which become
deeper increasing the number of loops, 
and which force calculations with factorial series 
to be performed 
with a larger number of digits.
To give an idea of the size of calculations,
the systems of difference equations between the master integrals of the diagrams 
\ref{figvac}e,
\ref{figself}d,
\ref{figself}t,
\ref{figself}u,
\ref{figself}v,
\ref{figvert}f,
\ref{figbox}g,
and~\ref{figbox}h are formed with
 44, 28, 245, 304, 362, 81, 139 and  158 equations, respectively;
note that in each system, in the right-hand side of 
the equation corresponding to the more complicated master integral
almost all the other master functions appear.
We made no use of the symmetries due to the particular values of masses and
momenta
in order
to simplify or reduce the number of the equations,
as the aim of program \SYS~ is to deal with calculations of multi-scale
integrals, lacking in such symmetries;
we used them only to check the final results. 
In order to guarantee results with at least 20 digits of precision,
calculations with factorial series were performed with precision
up to 77 digits (depending on the degree of instability),
while calculations with Laplace's transformation
were all performed with 38 digits of precision.

For example, the calculation from scratch 
of the integral $I(\ref{figself}\text t)$, \eqref{res3t},
with Laplace's transformation,
requested about 128 hours of CPU time
on a 133 MHz Pentium PC;
16 hours were used for the determination of the systems of difference
and differential equations, obtained by solving systems up to 43000 identities.
The solution of the systems (245 equations) yields, as a byproduct, 
also the values of all simpler master integrals, 
including~\eqrefb{res3e}{res3h}, 
\eqrefb{res3j}{res3n}, \eqref{res3p} and~\eqref{res3s}.
We stress that at this preliminary stage of development
we directed our efforts to devise tests and cross checks rather
than to speed up the program.

For brevity, for each diagram we list here only 
the values of the master scalar integrals
\begin{equation}\label{ival}
I(diagram)= 
\int \dfrac {\dk1 \dots \dk{\NK}}
{\displaystyle D_1 D_2 \ldots  D_\ND} 
\end{equation}
without scalar products and containing all the $\ND$ denominators.
As usual, the results have been normalized
with the division by $\Gammae\equiv\Gamma(1+\e)$ raised to
the number of loops of the diagram.
Coefficients are shown with only 13 digits to save space.
Values for $\e=0$ of all finite integrals have been checked by comparing them 
with numerical values obtained by performing Monte-Carlo integrations over
Feynman parameters,
or by performing low dimensional Gaussian integrations on integrands
obtained using dispersion relations and hyperspherical variables.
As additional consistency check 
we have repeated the calculation 
of some diagrams with different choices of the line $D_1$,
and we have checked that the results obtained are the same.
\subsubsection{Vacuum diagrams}
\begin{multline}
I(\ref{figvac}\text a)\Gammae^{-1}= -\e^{-1} -1 -\e -\e^2 -\e^3 -\e^4 +O(\e^5)\;,
\qquad \qquad \qquad \qquad \qquad \qquad \qquad \qquad 
\end{multline}
\begin{multline}\label{2b}
I(\ref{figvac}\text b)\Gammae^{-2}=
 -1.5 \e^{-2} -4.5 \e^{-1} -6.984139141966 -18.00878162355 \e \\
  -27.99422356368 \e^2 -72.00378659799 \e^3 -111.9974983355 \e^4
+O(\e^5)\;,
\end{multline}

\begin{multline}\label{2c}
I(\ref{figvac}\text c)\Gammae^{-3}=
2 \e^{-3} +7.666666666667 \e^{-2} +17.5 \e^{-1} +22.91666666667 \\
    +21.25179105129 \e -184.2300051053 \e^2 \\
    -661.1105861534 \e^3 -3685.054779382 \e^4
  +O(\e^5)\;,
\end{multline}

\begin{multline}
I(\ref{figvac}\text d)\Gammae^{-3}=
-\e^{-3} -5.666666666667 \e^{-2} -15.30161161726 \e^{-1} \\
 -46.07511172933 -148.3508545129 \e -394.1378145809 \e^2 \\
 -1375.669435211 \e^3 -3466.9749998996 \e^4
   +O(\e^5)\;,
\end{multline}

\begin{multline}\label{2e}
I(\ref{figvac}\text e)\Gammae^{-3}=
 2.404113806319 \e^{-1} -10.03527847977 +35.94478903214 \e \\
 -119.1503507802 \e^2 +379.7433345095 \e^3 -1183.320931551 \e^4
   +O(\e^5)\;.
\end{multline}
\eqref{2b},
first six terms of \eqref{2c} and 
first two terms of \eqref{2e} agree  
with the analytical expressions given in
\cite{Davy2b}, \cite{3-loop} and \cite{2e},
respectively; 
remaining terms and other results are new.

\subsubsection{Self-energy diagrams}
\begin{multline}\label{res3a}
I(\ref{figself}\text a)\Gammae^{-1}= \e^{-1} +0.186200635766 +0.021156303568\e 
 +0.001726745353 \e^2 \\
 +0.000109897792 \e^3 +0.000005730593\e^4 +O(\e^5)\;, 
\end{multline}
\begin{multline}\label{res3b}
I(\ref{figself}\text b)\Gammae^{-2}=
-1.5 \e^{-2} -4.25 \e^{-1} -7.375 -17.22197253479 \e \\
-29.55920705372 \e^2 -68.87789517038 \e^3 -118.2464846454 \e^4
   +O(\e^5)\;,
\end{multline}
\begin{multline}\label{res3c}
I(\ref{figself}\text c)\Gammae^{-2}=
 0.5 \e^{-2} +0.6862006357658 \e^{-1} -0.6868398873414 \\ +1.486398391913 \e 
 -2.938796587745 \e^2 +5.871086365958 \e^3 \\ -11.73616571449 \e^4
   +O(\e^5)\;,
\end{multline}
\begin{multline}\label{res3d}
I(\ref{figself}\text d)\Gammae^{-2}=
0.9236318265199 -1.284921671848 \e +2.689507626490 \e^2 \\
-5.338399227511 \e^3 +10.67136736912 \e^4 
   +O(\e^5)\;,
\end{multline}
\begin{multline}\label{res3e}
I(\ref{figself}\text e)\Gammae^{-3}=
 2 \e^{-3} +7.333333333333 \e^{-2} +16.02777777778 \e^{-1}\\
 +21.92956264368 +3.605127475161 \e -184.1413665431 \e^2 \\
 -838.2364324178 \e^3 -3647.102197031 \e^4  
   +O(\e^5)\;,
\end{multline}
\begin{multline}
I(\ref{figself}\text f)\Gammae^{-3}=
 -\e^{-3} -2.612634286982 \e^{-2} -3.906420490690 \e^{-1} \\
 +0.5840769314959 +1.76460041453 \e +107.0072031435 \e^2 \\
 +163.2855293783 \e^3 +1372.241466189 \e^4  
   +O(\e^5)\;,
\end{multline}
\begin{multline}
I(\ref{figself}\text g)\Gammae^{-3}=
 -\e^{-3} -5.5 \e^{-2} -15.48413914197 \e^{-1} -45.68793012675 \\
-149.1607636537 \e -392.298867227 \e^2 \\
-1380.125833167 \e^3 -3455.548194007 \e^4 
   +O(\e^5)\;,
\end{multline}
\begin{multline}\label{treh}\label{res3h}
I(\ref{figself}\text h)\Gammae^{-3}=
 -\e^{-3} -5.333333333333 \e^{-2} -16 \e^{-1} -43.91483126325 \\
-154.918028663 \e -374.0941853334 \e^2 \\
-1436.672712535 \e^3 -3281.940436319 \e^4 
   +O(\e^5)\;,
\end{multline}
\begin{multline}
I(\ref{figself}\text i)\Gammae^{-3}=
 2.404113806319 \e^{-1} -9.763424447585 +34.99888165588 \e \\
-116.0420477564 \e^2 +370.0407274069 \e^3 -1153.646312515 \e^4 
   +O(\e^5)\;,
\end{multline}
\begin{multline}\label{res3j}
I(\ref{figself}\text j)\Gammae^{-3}=
 0.3333333333333 \e^{-3} +0.5195339690991 \e^{-2}\\ +0.5753609494269 \e^{-1} 
 -2.981838135558 +12.56596204108 \e \\-44.85302351538 \e^2 
 +149.1742811721 \e^3 -477.1440886129 \e^4 
   +O(\e^5)\;,
\end{multline}
\begin{multline}
I(\ref{figself}\text k)\Gammae^{-3}=
 0.1666666666667 \e^{-3} +0.5931003178829 \e^{-2}\\ +0.06234894542402 \e^{-1} 
 -1.364667486582 +7.062482427894 \e \\-26.87419915573 \e^2 
 +91.91641284417 \e^3 -298.196943613  \e^4 
   +O(\e^5)\;,
\end{multline}
\begin{multline}
I(\ref{figself}\text l)\Gammae^{-3}=
 0.1666666666667 \e^{-3} +0.5931003178829 \e^{-2} \\+0.06234894542402 \e^{-1} 
 -1.158877567105 +6.268660583427 \e \\ -24.1193749759  \e^2 
 +82.9872059343  \e^3 -270.1688760103 \e^4 
   +O(\e^5)\;,
\end{multline}
\begin{multline}
I(\ref{figself}\text m)\Gammae^{-3}=
 0.3333333333333 \e^{-3} +0.8528673024324 \e^{-2}\\ -1.728169411584 \e^{-1}  
 +6.070141409747 -19.48651365516 \e \\ +61.38828756627 \e^2 
 -190.3302695306 \e^3 +583.8045381529 \e^4 
   +O(\e^5)\;,
\end{multline}
\begin{multline}\label{res3n}
I(\ref{figself}\text n)\Gammae^{-3}=
 0.3333333333333 \e^{-3} +0.8528673024324 \e^{-2} \\ -1.728169411584 \e^{-1}  
 +6.120359708375 -19.67063042467 \e \\ +62.00178253235 \e^2 
 -192.2586253184 \e^3 +589.7212544716 \e^4 
   +O(\e^5)\;,
\end{multline}
\begin{multline}
I(\ref{figself}\text o)\Gammae^{-3}=
 0.9236318265199 \e^{-1} -2.423491634417 +8.381349710069 \e \\
 -26.99362121677 \e^2 +85.10096322999 \e^3 -263.903318629 \e^4 
   +O(\e^5)\;,
\end{multline}
\begin{multline}\label{res3p}
I(\ref{figself}\text p)\Gammae^{-3}=
 0.9236318265199 \e^{-1} -2.116169718457 +6.929544685259 \e \\
 -21.50327837738 \e^2 +66.32213380401 \e^3 -202.887025717 \e^4 
   +O(\e^5)\;,
\end{multline}
\begin{multline}
I(\ref{figself}\text q)\Gammae^{-3}=
1.326448208272  -5.196648136965 \e +18.37758387804 \e^2 \\
-60.41191503661 \e^3 +191.5963941 \e^4 
   +O(\e^5)\;,
\end{multline}
\begin{multline}
I(\ref{figself}\text r)\Gammae^{-3}=
 1.341399241447 -5.197752955896 \e +18.38704656407 \e^2 \\
 -60.4233521301  \e^3 +191.614009625 \e^4 
   +O(\e^5)\;,
\end{multline}
\begin{multline}\label{res3s}
I(\ref{figself}\text s)\Gammae^{-3}=
 2.002500041105 -8.162562835907 \e +29.46716463085 \e^2 \\
 -98.13080591819 \e^3 +313.871881187 \e^4 
   +O(\e^5)\;,
\end{multline}
\begin{multline}\label{res3t}
I(\ref{figself}\text t)\Gammae^{-3}=
 0.2796089232826 -0.1380294113932   \e  +0.3194688268113 \e^2 \\
 -0.4399664109267 \e^3 +0.6650515012166 \e^4 
   +O(\e^5)\;,
\end{multline}
\begin{multline}
I(\ref{figself}\text u)\Gammae^{-3}=
 0.1826272375392 -0.06746690965803  \e  +0.1865462420623 \e^2 \\
 -0.2498713405447 \e^3 +0.3796187113121 \e^4 
   +O(\e^5)\;,
\end{multline}
\begin{multline}
I(\ref{figself}\text v)\Gammae^{-3}=
 0.1480133039584 -0.009263002043238 \e  +0.1053308537397 \e^2 \\
 -0.1224292041846 \e^3 +0.1898480457555 \e^4 
   +O(\e^5)\;.
\end{multline}
\eqref{res3b}, first four terms of \eqref{res3c}, 
first term of \eqref{res3d} and first six terms of \eqref{treh}
agree with the analytical expressions given in
 \cite{3abcd1}, \cite{3abcd3}, \cite{3abcd2} and \cite{3-loop,pol}, 
 respectively; 
 remaining terms and other results are new.
\subsubsection{Vertex diagrams}
\begin{multline}
I(\ref{figvert}\text a)\Gammae^{-1}=
0.671253105748 +0.1998957762816 \e +0.03189366853371 \e^2 \\
+0.003532937320333 \e^3 +0.0003018185047825 \e^4 
   +O(\e^5)\;,
\end{multline}
\begin{multline}
I(\ref{figvert}\text b)\Gammae^{-2}=
0.5 \e^{-2} +0.6862006357658 \e^{-1} -0.5916667014024 \\ +1.356196533114 \e 
-2.669112118814 \e^2 +5.336651358516 \e^3 \\ -10.66866283741 \e^4
   +O(\e^5)\;,
\end{multline}
\begin{multline}
I(\ref{figvert}\text c)\Gammae^{-2}=
0.671253105748 \e^{-1} -0.08774519609257 +0.7262375626947 \e \\
-1.32112948587 \e^2 +2.667431469376 \e^3 -5.332675337091 \e^4
   +O(\e^5)\;,
\end{multline}
\begin{multline}
I(\ref{figvert}\text d)\Gammae^{-2}=
0.937139527315 -1.27184968708 \e +2.69185047506 \e^2 \\
 -5.336932134961 \e^3 +10.67100342934 \e^4
   +O(\e^5)\;,
\end{multline}
\begin{multline}
I(\ref{figvert}\text e)\Gammae^{-2}=
0.2711563494022 +0.1833941077514 \e +0.05375101058769 \e^2 \\
+0.01446103368419 \e^3 +0.000746187372276 \e^4 
   +O(\e^5)\;,
\end{multline}
\begin{multline}
I(\ref{figvert}\text f)\Gammae^{-2}=
 0.173896742268 +0.1816664876962 \e +0.04440899181832 \e^2 \\
 +0.02231547385785 \e^3 -0.003079810479797 \e^4 
   +O(\e^5)\;.
\end{multline}

\subsubsection{Box diagrams}\label{boxdiagrams}
\begin{multline}
I(\ref{figbox}\text a)\Gammae^{-1}=
0.3455029252972 +0.4731008318818 \e +0.1519459537543 \e^2 \\
+0.0275179284554 \e^3 +0.00348492177519 \e^4
   +O(\e^5)\;,
\end{multline}
\begin{multline}
I(\ref{figbox}\text b)\Gammae^{-2}=
0.671253105748  \e^{-1} -0.06425178040942 +0.7393966927045 \e \\
-1.317187699112 \e^2 +2.668107343399 \e^3 -5.332500882459 \e^4 
   +O(\e^5)\;,
\end{multline}
\begin{multline}
I(\ref{figbox}\text c)\Gammae^{-2}=
0.9509235623171 -1.258189955235 \e +2.694588643167 \e^2 \\
-5.335347124508 \e^3 +10.67067120761 \e^4 
   +O(\e^5)\;,
\end{multline}
\begin{multline}
I(\ref{figbox}\text d)\Gammae^{-2}=
0.276209225359  +0.1937422320842 \e +0.06034849310181 \e^2 \\
+0.01640828588681 \e^3 +0.001301642516765 \e^4 
   +O(\e^5)\;,
\end{multline}
\begin{multline}
I(\ref{figbox}\text e)\Gammae^{-2}=
0.3455029252972 \e^{-1} +0.4347670080988    +0.17885718095363   \e \\
+0.05005382113385 \e^2 +0.006936292250698 \e^3 +0.002511559375421 \e^4
   +O(\e^5)\;,
\end{multline}
\begin{multline}
I(\ref{figbox}\text g)\Gammae^{-2}=
0.1723367907503 +0.2679578491711 \e +0.13552112755141 \e^2 \\
+0.04468531532833 \e^3 +0.008430602827459 \e^4 
   +O(\e^5)\;,
\end{multline}
\begin{multline}
I(\ref{figbox}\text h)\Gammae^{-2}=
0.1036407209893 +0.2142416932987 \e +0.14046068671363 \e^2 \\
+0.04437197236388 \e^3 
   +O(\e^4)\;.
\end{multline}
In the case of the diagrams~\ref{figbox}f and~\ref{figbox}i
(from which the diagram~\ref{figbox}f is derived with the deletion of a line)
the equations \itref{diffu1a}, obtained 
after the transformation of the subsystems of equations 
into triangular form, 
have orders relatively high ($\ge 10$)
and expressions of size greater than the limit of the computer used,
so that our program failed to work out them
(the coefficients of the equations would be 
polynomials in two variables of degree $\sim 100$);
we were able to find only the characteristic and indicial equations.
The calculation of these integrals without the transformation 
into triangular form will be considered in the next paper.
\subsection{Some examples of equations}
\label{pair1}
Here we discuss the equations satisfied by the functions
$I_{g}(x)= U_{gl}(x)$ 
for some simple diagrams with all lines topologically equivalent,
using the values of masses and momenta of section \ref{resu}. 
For the diagrams~\ref{figself}a and~\ref{figvac}b
the difference equations are
\begin{align}
\label{diffeqxa}
\bW1 I_{\ref{figself}\text a}(x)&=-\frac{1}{2}(D -2)I_{\ref{figvac}\text a}(x)  \ , \\
\label{diffeqxb}
\bW1 I_{\ref{figvac}\text b}(x)&=
-(D -2)I_{\ref{figvac}\text a}(1)I_{\ref{figvac}\text a}(x)  \ , 
\end{align}
where $\bW1$ is the operator 
\begin{equation}\label{bw1}
\bW1 I(x)= -3x I(x+1) +(2x-D+1)I(x) +(x-D+1)I(x-1) \ ,
\end{equation}
and $I_{\ref{figvac}\text a}(x)$ is the integral \itref{inte0} with $m=1$.
The identity of the homogeneous parts of the equations is an example of the
similarities between equations described in section~\ref{resuarb}.
The solutions of \eqrefb{diffeqxa}{diffeqxb} can be written as 
\begin{align}
I_{\ref{figself}\text a}(x)&=\C_{\ref{figself}\text a} \IOMOG_{\ref{figself}\text a}(x)
+\INOMOG_{\ref{figself}\text a}(x)\ , \\
I_{\ref{figvac}\text b}(x) &=\C_{\ref{figvac}\text b} \IOMOG_{\ref{figself}\text a}(x) 
+2I_{\ref{figvac}\text a}(1) \INOMOG_{\ref{figself}\text a}(x)\ . 
\end{align}
Numerical calculation of $\IOMOG_{\ref{figself}\text a}(1)$ and
$\INOMOG_{\ref{figself}\text a}(1)$ was described in section~\ref{numexa}.
The constant $\C_{\ref{figvac}\text b}$ may be found by comparing 
the large-$x$ behaviours
of the solutions  $\IOMOG_{\ref{figself}\text a}(x)\approx x^{-D/2+1/2}$,
$\INOMOG_{\ref{figself}\text a}(x)\propto x^{-D/2}$ 
(see \eqrefb{lrn1}{lrn1nomog}) with
\begin{equation}
I_{\ref{figvac}\text b}(x) \approx x^{-D/2} \int \dfrac{\dk2 }{(k_2^2+1)^2}
      =       x^{-D/2} I_{\ref{figvac}\text a}(2)   
\end{equation}
(see \eqref{expkf00}). One finds $\C_{\ref{figvac}\text b}=0$.
Therefore $I_{\ref{figvac}\text b}(1) =2I_{\ref{figvac}\text a}(1)
\INOMOG_{\ref{figself}\text a}(1)$;
we see that  the two-loop integral $I_{\ref{figvac}\text b}(1)$ 
factorizes into a product 
of a vacuum one-loop integral and \emph{a part} of a one-loop self-energy
integral.

\label{pair2}
Now let us consider the diagrams of Figs.~\ref{figself}b and~\ref{figvac}c.
Both diagrams have only one master integral with all the denominators,
the scalar integral
(compared with the four master integrals of the arbitrary case);
the corresponding master functions 
$I_{\ref{figself}\text b}(x)$ and  $I_{\ref{figvac}\text c}(x)$
satisfy the difference equations
\begin{align}
\label{twoll}
\bW2 I_{\ref{figself}\text b}(x)&=(D-2)^2 I_{\ref{figvac}\text a}(1) I_{\ref{figvac}\text a}(x)
\ , \\
\label{twoll2}
\bW2 I_{\ref{figvac}\text c}(x)&={\tfrac{3}{2}}(D-2)^2 I_{\ref{figvac}\text a}^2(1) I_{\ref{figvac}\text a}(x)
\ ,
\end{align}
where $\bW2$ is the operator 
\begin{multline}\label{bw2}\label{twol}
\bW2 I(x)= -8x(x-D+2)I(x+1) 
      +\bigl( 7x^2 +( 13 - 10D )x   \\
       + (3D-4)(D-1)\bigr) I(x) +(x-D+1)(x-3D/2+2) I(x-1)
        \ .
\end{multline}
Note again the identity of the homogeneous parts of the equations.
The characteristic equation has the solutions $1$ and $-1/8$.
The index associated to $\mu=1$ is $\K=-D/2$; therefore the condition 
\itref{cond1}
is satisfied and the corresponding solution of the homogeneous equation
$\IOMOG_{\ref{figself}\text b}(x)$ contributes to  $I_{\ref{figself}\text b}(x)$
and $I_{\ref{figvac}\text c}(x)$.
The solutions of \eqrefb{twoll}{twoll2} can be written as 
\begin{align}
I_{\ref{figself}\text b}(x)&= \C_{\ref{figself}\text b} \IOMOG_{\ref{figself}\text b}(x)
   +\INOMOG_{\ref{figself}\text b}(x)\ , \\
I_{\ref{figvac}\text c}(x)&= \C_{\ref{figvac}\text c} \IOMOG_{\ref{figself}\text b}(x)
   +\tfrac{3}{2} I_{\ref{figvac}\text a}(1) \INOMOG_{\ref{figself}\text b}(x)\ .
\end{align}
The constants may be obtained by comparing the large-$x$ behaviours
$\IOMOG_{\ref{figself}\text b}(x)$$\approx$ $x^{-D/2}$ and
$\INOMOG_{\ref{figself}\text b}(x)\propto x^{-D/2-1}$ with
$I_{\ref{figself}\text b}(x)\approx  x^{-D/2} I_{\ref{figself}\text a}(1)$ and
$I_{\ref{figvac}\text c}(x)\approx  x^{-D/2}$ $ I_{\ref{figvac}\text b}(1)$.
One finds $\C_{\ref{figself}\text b}=I_{\ref{figself}\text a}(1)$ and
          $\C_{\ref{figvac}\text c}=I_{\ref{figvac}\text b}(1)$.
We consider the calculation of
$\IOMOG_{\ref{figself}\text b}(1)$ and $\INOMOG_{\ref{figself}\text b}(1)$
with factorial series.
The recurrence relation \itref{twol} is unstable with $A=8$.
We fix a precision of the coefficients of the powers of $\e$ of the results
of $E=13$ digits, with $n'_\e=7$
terms of the expansions in $\e$; following section~\ref{sumfact}
we choose $x_{max}=25$, and we perform the calculations with $C=38$ digits.
Expansions in factorial series of 
$\IOMOG_{\ref{figself}\text b}(x)$ and $\INOMOG_{\ref{figself}\text b}(x)$ 
converge for $x=25$  in about 1800 terms.
Values for $x=1$ of the solutions are calculated using repeatedly
the recurrence relations \itref{twoll}. 
Two terms of the expansions in $\e$ are lost
going beyond the abscissa of convergence $\lambda=3$,
so that we must retain the first $n_e=9$ terms of the expansions.
One obtains
\begin{multline}
\C_{\ref{figself}\text b}\IOMOG_{\ref{figself}\text b}(1)\Gammae^{-2}=
0.09188814923697 \e^{-3} +0.194632539439 \e^{-2} \\
-0.045490472375 \e^{-1}  +6.50912255436 +10.43240978278 \e  \\
 +76.7023111407 \e^2  +118.6149739413 \e^3 +732.0021187015 \e^4 +O(\e^5)\ ,
\end{multline}
\begin{multline}
\INOMOG_{\ref{figself}\text b}(1)\Gammae^{-2}=
 -0.09188814923697 \e^{-3} -1.694632539439 \e^{-2} \\ -4.204509527625 \e^{-1} 
 -13.88412255436 -27.65438231756 \e \\ -106.2615181944 \e^2 
 -187.4928691117 \e^3 -850.2486033469 \e^4 +O(\e^5).
\end{multline}
Summing the results one finds \eqref{res3b}; note the cancellation of
the $\e^{-3}$ term, and the partial cancellations of digits in the
$\e^{-2}$, $\e^{-1}$ and constant terms. 
Considering now the solution of \eqref{twoll} with Laplace's transformation,
there are \emph{three} master transformed functions:
$v_{\ref{figself}\text b}(t)$, $w_{1}(t)$ and $w_{2}(t)$ defined by 
\begin{equation}
I_{\ref{figself}\text b}(x)= 
\int \dfrac{\dk1 \;\dk2}{(k_1^2+1)^x (k_2^2+1) ((p-k_1-k_2)^2+1)}= 
\int_0^1 dt \; t^{x-1} v_{\ref{figself}\text b}(t)  \ , \qquad\qquad\qquad
\end{equation}
\begin{equation}
K_\alpha(x) =\int \dfrac{\dk1 \;\dk2 \ (p\cdot k_\alpha)}
{(k_1^2+1)^x (k_2^2+1) ((p-k_1-k_2)^2+1)} 
=\int_0^1 dt \; t^{x-1} w_{\alpha}(t) \ ,  \qquad \alpha=1,2 \ .
\end{equation}
Clearly $w_1(t)\not = w_2(t)$ and $K_1(x)\not =K_2(x)$
as the insertion of $x$ breaks the symmetry $k_1 \leftrightarrow k_2$.
The function $v_{\ref{figself}\text b}(t)$ satisfies the differential equation
\begin{multline}
 t^2(1-t)(1+8t)v_{\ref{figself}\text b}''
 +t(8(1-D)t^2 +10(D -2)t +(5D/2 -6))v_{\ref{figself}\text b}' \\
+(3D-8)(D-3)(t+1/2) v_{\ref{figself}\text b} =(D-2)^2t 
I_{\ref{figself}\text a}(1) v_{\ref{figself}\text a}\ ,
\end{multline}
where $v_{\ref{figself}\text a}$ is $v_J$ of \eqref{v000} with $m=1$.
$K_1(1)$ and $K_2(1)$ are not master integrals, as  
$K_{1}(1)=K_{2}(1)=-\tfrac{1}{3}I_{\ref{figself}\text b}(1)$;
therefore $K_1(x)$ and $K_2(x)$ are not master functions
and satisfy difference equations of order \emph{zero},
for example 
\begin{equation}
(x+2-D)(-3K_{1}(x+1)
-I_{\ref{figself}\text b}(x+1)+I_{\ref{figself}\text b}(x))= 
 (2-D)I_{\ref{figself}\text a}(x+1)I_{\ref{figself}\text a}(1) \ .
\end{equation}
On the contrary, $w_1(t)$ and $w_2(t)$ satisfy 
\emph{first} order differential equations,
for example
\begin{equation}
 -6t(tw_{1}'+(D-2)w_{1}) +2t(1-t)v_{\ref{figself}\text b}' 
 +2((2-D)t +D-3)v_{\ref{figself}\text b}= 
  2(D-2)I_{\ref{figself}\text a}(1) tv_{\ref{figself}\text a} \ ,
\end{equation}
and therefore they are master functions.

\section{Differential equations in masses and momenta} \label{differentialequ}
As shown in the one-loop example, the calculation
of the constants $\C_j$, the factors multiplying the solutions
of the homogeneous equation,
presents different degrees of difficulty according to the values of masses
and momenta.
For values below the deformation threshold,
these factors (when non-zero)
are easily expressed in form of integrals with one loop missing,
but, above the deformation threshold or when some masses are zero,
as discussed in section~\ref{othercases}, the calculation of the factors
becomes more complicated.
An alternative to this calculation
is to calculate the master integrals
for
values of masses and momenta such that the calculation of the factors
is simple, and subsequently to build and integrate a system of differential
equations in the masses and momenta in order to reconstruct the integrals
for the requested values of masses and momenta.
The approaches to calculation of diagrams
based on construction of differential equations in masses or
momenta were introduced in \cite{difmas,difmom};
up to now they have been used in the analysis of particular diagrams
(e.g. massless box diagrams in \cite{prep1}) with a small number of master integrals
by building and solving small systems of differential equations.
Here we describe how to build and solve
systems of differential equations in masses and momenta
in way very similar to that of the
differential equations obtained by Laplace's transformation.
In this way even large systems containing hundreds of equations can
be built and solved, obtaining results expanded at will in $\e$.

\subsection{Differentiation of master integrals}\label{diffint}
A generic master integral is a function $B(m_i^2, p_i\cdot p_j)$
of
scalar products and masses.
We parametrize the trajectory from the initial point to the final point
in the space of scalar products and masses with a parameter $\Y$;
the scalar products and masses of the points of the trajectory
will be in general functions of $\Y$.
In analogy with the differential equations obtained
by using the Laplace's transformation, we introduce the parametrization
in such a way that
the initial point corresponds to $\Y=1$ and
the final point corresponds to $\Y=0$.
Derivatives with respect to $\Y$ of master integrals will be written as derivatives with respect
to scalar products and square masses:  
\begin{equation}\label{deriv3}
\dfrac{dB}{d\Y}= 
\sum_{i=1}^\NP \dfrac{dp_i^2}{d\Y} \dfrac{\partial B}{\partial p_i^2} 
+\sum_{i<j}     \dfrac{d(p_i\cdot p_j)}{d\Y} 
                \dfrac{\partial B}{\partial (p_i \cdot p_j)} 
+\sum_{i=1}^\ND \dfrac{dm_i^2}{d\Y} \dfrac{\partial B}{\partial m_i^2} \ .
\end{equation}

Differentiation with respect to masses of the integral
is trivial; differentiation with respect to scalar products
is less immediate.
Let us consider a diagram with $\NP$ independent external momenta $p_i$;
the scalar products $p_\alpha \cdot p_\beta$  are $\NPP=\NP (\NP+1)/2$.
The derivative with respect to $p_\alpha \cdot p_\beta$ has the general form
\begin{equation}\label{deriv1}
\dfrac{\partial}{\partial(p_\alpha\cdot p_\beta)}= \sum^\NP_{i=1}\sum^\NP_{j=1}
         a_{ij\alpha\beta} \; p_i \cdot
                \dfrac{\partial}{\partial p_j}\ ,
\end{equation}
for $(\alpha,\beta)$ equal to each one of 
the $\NPP$ possible different pairs of  integer.
The coefficients $a_{ij\alpha\beta}$ can be determined by imposing the
$\NPP^2$ conditions
\begin{equation}\label{condi1}
\dfrac{\partial(p_\gamma \cdot p_\delta) }{\partial(p_\alpha\cdot p_\beta)}=
\begin{cases}
1 \text{ if } p_\gamma\cdot p_\delta = p_\alpha \cdot p_\beta \ ,\\
0 \text{ otherwise } \ .
\end{cases}
\end{equation}
But the number of the coefficients $a_{ij\alpha\beta}$ is $\NP^2\NPP$ which 
is greater
than $\NPP^2$; therefore some of these coefficients are 
not constrained by these conditions and may be chosen arbitrarily.
We conveniently rewrite \eqref{deriv1} as
\begin{equation}\label{deriv2}
\dfrac{\partial}{\partial(p_\alpha\cdot p_\beta)}= 
\sum_{i=1}^\NP  a_{ii\alpha\beta} \; p_i\cdot \dfrac{\partial}{\partial p_i}
+\sum_{i<j} a_{ij\alpha\beta} \left(
 p_i\cdot \dfrac{\partial}{\partial p_j} +M_{ij} \; 
 p_j\cdot \dfrac{\partial}{\partial p_i}
\right)\ ,
\end{equation}
where the coefficients $a_{ij\alpha\beta}$
to be found are $\NPP^2$ and there are $\NP(\NP-1)/2$
arbitrary constants $M_{ij}$.
In order to calculate the coefficients it is convenient, 
fixed $\alpha$ and $\beta$, to apply \eqref{deriv2} to a generic scalar
product $p_k \cdot p_l$,
and to construct the tensor $A_{ijkl}$ formed by the factors multiplying
$a_{ij\alpha\beta}$ in the obtained expression.
Now, considering the various pairs of indices $\{i,j\}$ 
as one single index $ij$, $A$ becomes a square matrix $A_{ij,kl}$ 
of dimension $\NPP$. Inverting it we find the desired coefficients
\begin{equation}
  a_{kl,ij} = \left(A^{-1}\right)_{ij,kl} \ .
\end{equation}
As an example, for a vertex diagram $\NP=2$, and  the matrix $A$ is 
\begin{equation}
A=
\begin{pmatrix}
2p_1^2          & 2M_{12} p_1\cdot p_2     & 0 \\
p_1\cdot p_2    & p_1^2+M_{12} p_2^2       & p_1\cdot p_2 \\
0               & 2p_1\cdot p_2            & 2p_2^2 \\
\end{pmatrix}
\ ,
\end{equation}
where the indices 1,2,3 correspond to $ij=11,12,22$.

\subsection{Construction of the system of differential
 equations}\label{consys}
The algorithm used for the generation and solution of the system of identities
is the following:
\begin{alg}\label{algsys5}
Follow the algorithm~\ref{algsys} with the following modifications:
\begin{enumerate}
\item New identities which transform the derivatives with respect to $\Y$ 
      of master integrals into combinations of integrals obtained using 
      \eqref{deriv3} are added to the system.
\item Master integrals and their derivatives have a priority of extraction
      lower than other integrals.
\item Add  the new entry {``the greatest derivative''}
      to the list of priorities after the entry \ref{cf6b}.
\end{enumerate}
\end{alg}

Analogously to sections~\ref{identificamaster} and~\ref{constsysdif},
the previous algorithm allows one to determine
the list of the master integrals $B_n(y)$, 
and to obtain a system of differential equations between them.
Applying a procedure of transformation into triangular form 
of the subsystems of equations
between master integrals with the same denominators, 
the whole system of differential
equations takes the triangular form.
The arbitrary constants $M_{ij}$ turn out to be 
particularly useful as a check,
because the expression of derivatives \itref{deriv3} contains $M_{ij}$
(through \eqref{deriv2}),
while the differential equations obtained from the system of identities 
are obviously independent of them.

The integration of the system of differential equations
requires the knowledge of the values of the master integrals and
of some derivatives in the initial point $\Y=1$.
The values of the master integrals are obtained 
by solving a system of difference equations.
The derivatives of the master integrals in the initial point 
are expressed in terms of the 
master integrals of the system of difference equations
using the identities provided by the algorithm~\ref{algsys}.

Note that the master integrals of the system of difference equations
may be different in number and structure from 
the master integrals of the system of differential equations.
The system of differential equation is solved with the same procedure 
used for the Laplace's transformation method.
We stress that mobile apparent singular points, 
as described in section~\ref{mobile}, may appear
in the system of differential equations.

\subsection{Expansion in $\e$ and infrared divergences}
\label{infrared}
All quantities, coefficients of differential equations,
solutions, $\Gamma$ functions, etc. are expanded in $\e=(4-D)/2$.
Solution of differential equations using power series in $\Y$
with coefficients expanded in $\e$
does not cause difficulties, unless we deal with IR divergences.
If a master integral $B(\Y,\e)$ is IR finite in the initial point $\Y=1$,
but is IR divergent in the final point $\Y=0$,
it is not possible to obtain the correct dimensionally
regularized value, which contains an additional IR pole in $\e$,
by integrating the differential equation from the initial point,
where the integral is IR finite:
the coefficients of the expansion in $\e$ of $B(\Y,\e)$ tend to infinity as
$\Y \to 0$, so that $\lim_{\Y \to 0}B(\Y,\e) \not = B(0,\e)$.
In this unfortunate case  the only remedy is
to substitute the master integral IR divergent with another integral
(or combination of integrals) IR finite.
Our preference for master
integrals with scalar products in the numerator 
and all exponents 1 in the denominator
rather than
integrals with exponents greater than 1 in the denominator
was devised to limit the appearance of master integrals IR divergent.

In the case where $B(0,\e)$ is IR finite,
but some derivatives with respect to $\Y$ of $B(\Y,\e)$  
are IR divergent in $\Y=0$,
the problem is overcome by integrating the differential equation 
up to a small value $\Y\approx 0$,
suitable chosen so that the value of the master integral 
will be calculated with the requested precision.

\subsection{Examples of calculation of integrals with zero masses}\label{resumass0}
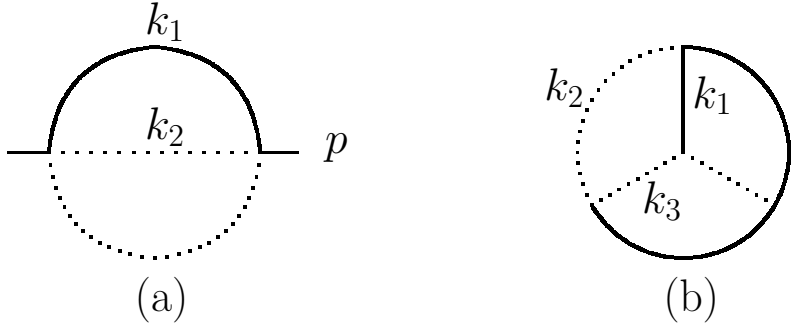
\begin{figure}
\begin{center}
   \begin{picture}(370,160)(0,0)
   \linethickness{1pt}
   {
   %\Large
   \LARGE
   \put(145,080){\text{$p$}}   
   \put(077,125){\text{$k_1$}}   
   \put(077,085){\text{$k_2$}}   
   \put(073,020){\text{(a)}}   

   \put(228,100){\text{$k_2$}}   
   \put(265,057){\text{$k_3$}}   
   \put(284,097){\text{$k_1$}}   
   \put(273,020){\text{(b)}}   
   }
   \put(040,080){\line(-1,0){15}}
   \put(120,080){\line(+1,0){15}}
   \qbezier[15](040,080)(080,080)(120,080)
   \qbezier(+120.0,+080.0)(+116.6,+116.6)(+080.0,+120.0)
   \qbezier(+080.0,+120.0)(+043.4,+116.6)(+040.0,+080.0)
   \qbezier[15](+040.0,+080.0)(+043.4,+043.4)(+080.0,+040.0)
   \qbezier[15](+080.0,+040.0)(+116.6,+043.4)(+120.0,+080.0)
   \qbezier[10](+280.0,+120.0)(+257.3,+119.3)(+245.4,+100.0)
   \qbezier[10](+245.4,+100.0)(+234.6,+080.0)(+245.4,+060.0)
   \qbezier(+245.4,+060.0)(+257.3,+040.7)(+280.0,+040.0)
   \qbezier(+280.0,+040.0)(+302.7,+040.7)(+314.6,+060.0)
   \qbezier(+314.6,+060.0)(+325.4,+080.0)(+314.6,+100.0)
   \qbezier(+314.6,+100.0)(+302.7,+119.3)(+280.0,+120.0)
   \qbezier[10](280,080)(297.3,070)(314.6,060)
   \qbezier[10](280,080)(262.7,070)(245.4,060)
   \qbezier(280,080)(280,100)(280,120)
\end{picture}
\end{center}
 \caption{Diagrams with massless lines (dashed).}
 \label{figder}
 \end{figure}
For illustrative purpose, we consider here only two simple diagrams;
applications to more complicated diagrams will be shown in future papers.
Let us consider the diagram of Fig.~\ref{figder}a, and the integral
\begin{equation}
\bar I(x)=\int \dfrac{[dk]}{(k_1^2+1)^x \; k_2^2\;(p-k_1-k_2)^2}\ ,
\quad p^2=-1\ ,
\end{equation}
where $[dk]=\dk1 \;\dk2$. It satisfies the homogeneous difference equation
\begin{equation}
x(x-2D+5)\bar I(x+1)-(x-3D/2+3)(x-D+2)\bar I(x)=0 \ ;
\end{equation}
the solution of this equation is 
\begin{equation}
\bar I(x)=\C 
 \dfrac{\Gamma(x-3D/2+3)\Gamma(x-D+2)}{\Gamma(x)\Gamma(x-2D+5)}   \ .
\end{equation}
The calculation of the constant $\C$ is not simple; in fact,
the value of $-p^2=1$ is above the deformation threshold $p^2=0$,
so that the large-$x$ behaviour of $\bar I(x)$ 
contains an
additional contribution due to the turning point of the integration path;
moreover, the identity \itref{rel0} is useless because $\bar I(0)=0$.

We replace the zero mass with  a square mass $\Y$ in the denominators
\begin{equation}
D_1=k_1^2+1 \ , \ D_2=k_2^2+\Y\ ,\  D_3=(p-k_1-k_2)^2+\Y \ ,
\end{equation}
then we build the system of identities following the algorithm~\ref{algsys5}.
The master integrals are
$I_1(\Y)=\int [dk]/{D_1 D_3}$,
$I_2(\Y)=\int [dk]/{D_1 D_2}$,
$I_3(\Y)=\int [dk]/{D_2 D_3}$,
$I_4(\Y)=\int [dk]/{D_1 D_2 D_3}$ and 
$I_5(\Y)=\int [dk](p \cdot k_2)/{D_1 D_2 D_3}$,
and satisfy the system of differential equations 
\begin{equation}
\begin{split}
  \Y I_1'&=(D-2)I_1 \ ,\qquad
2\Y I_2'=(D-2)I_2 \ ,\qquad
2\Y I_3'=(D-2)I_3 \ ,\\
8\Y (\Y -1)I_4''&=4\left((4D-13)\Y +7-2D\right)I_4' -2 (3D-8)(D-3)I_4 
               +(D-2)^2(I_1-I_2-I_3) \ , \\
6(D-2)I_5&=-4\Y (\Y -1)I_4'+2\left((D-3)\Y +5-2D\right)I_4 
               +(D-2)(2I_2-I_1-I_3)\ .\\
\end{split}
\end{equation}
The values of the integrals in the initial point $\Y=1$ 
can be expressed using the values of integrals
with masses equal to one given in section~\ref{resuvalues}:
$I_1(1)=I_2(1)=I_3(1)=I^2(\ref{figself}\text a)$,
$I_4(1)=I(\ref{figself}\text b)$, 
$I_5(1)=-I(\ref{figself}\text b)/3$.  
The singular points of the system are $\Y =1$  and $\Y =0$.
Considering the equation for $I_4$, 
the exponents in these points are
\begin{equation}
\rho_{1,2}^{(\Y=1)}=0, 2  -2\e \ , \qquad 
\rho_{1,2}^{(\Y=0)}=0, 3/2-2\e  \   . 
\end{equation}
In the initial point the integral $I_4(1)$  
and its derivatives are IR finite,
so that the solution must be regular;
therefore the singular solution associated with $\rho_2^{(\Y=1)}$ does not
contribute,
and the second-order equation for $I_4$ needs only $I_4(1)$ 
as initial condition.
In the final point the integral $I_4(0)$ 
is IR finite but $I_4'(0)$ is IR divergent,
so that the singular solution associated with $\rho_2^{(\Y=0)}$ 
must contribute to $I_4(\Y )$.

The integration of the system is divided into two parts:
from $\Y =1$ to $\Y =1/2$ expanding in $\Y =1$,
and from $\Y =1/2$ to the value $\Y =\lambda^2 \ll 1$ expanding in $\Y =0$.
The presence of the cutoff $\lambda$ is needed because the solution
is not regular in the origin.
Perusing the numerical results,
the effect of the cutoff on the coefficient of $\e^{s}$ of the expansion
in $\e$ of the final value turns out to be 
approximately proportional to $\lambda^2\log^{m+s} \lambda$,
where $m$ is some small integer.
A value $\lambda= 10^{-15}$
suffices to obtain the first coefficients with 20 exact digits.
The normalized result is 
\begin{multline}
I_4(0)\Gammae^{-2}=
   -0.5 \e^{-2} -1.25 \e^{-1} -4.6648681336964528729 \\
   -9.595397946879509324 \e
  -26.045799878017610383 \e^2 \\
  -49.501934187562851546 \e^3
  -120.38235865133474218 \e^4 +O(\e^5)\ . 
\end{multline}

Another more complicated example is the integral of Fig.~\ref{figder}b
\begin{equation}
L=\int \dfrac{\dk1 \;\dk2 \;\dk3 }
{(k_1^2+1) \left((k_1-k_2)^2+1\right) \left((k_2-k_3)^2+1\right) k_2^2\;
 k_3^2\; (k_1-k_3)^2}\ .\quad
\end{equation}
As before, we give a square mass $\Y $ to massless denominators.
In this case there are 43 master integrals from 2 to 6 denominators,
of which only 1 with 6 denominators.
The system of differential equations is too long
to be shown here.
The singular points are $\Y =-1,0,1/9,1/4,(3\pm\sqrt{5})/2,1,4,9$;
the points $1/9$, $1/4$, and $(3-\sqrt{5})/2$ on the interval $[0,1]$ 
turn out to be regular points of the solutions.
The effect of the cutoff is different from the previous case,
being proportional to $\lambda\log^{m+s} \lambda$.
A value $\lambda=10^{-30}$
suffices to obtain the first coefficients with 20 exact digits.
The result is 
\begin{multline}\label{resl}
L\Gammae^{-3}=
    2.4041138063191885708 \e^{-1} -3.0270094939876520198 \\
    +22.804522068631748454 \e  -53.102275435449702689 \e^2 \\
    +201.84333994219396694 \e^3 -577.74024368094326834 \e^4 
   +O(\e^5)\ .
\end{multline}
The first two terms agree with the results~\cite{r3.027}; subsequent terms are new.

These calculations were carried out by using the program \SYS;
the calculation from scratch of the integral $L$ 
and all the other master integrals,
including the calculation of the master integrals in the initial point,
required about 1.5 hours on a 133 MHz Pentium PC.

\section{The program \SYS }
\label{calcprog}
\label{program}
Here we report in brief some information concerning the program \SYS~ used to
calculate the values of the integrals of the 
sections~\ref{resu} and~\ref{resumass0}.
Further details will be given in a separate paper.
\begin{itemize}
\item C program, about 23000 lines, 1Mb executable.
\item The program \SYS~ allows one to calculate the value of integrals
      in almost completely automatic way; only needed input is a description
      of the diagram, the constants $a_i$ and $b_i$ of section~\ref{secab}
      which limit the identities generated,
      the number of dimensions $D_0$ about which to expand the integrals,
      and the number of terms of the expansions in $D-D_0$.
\item The program contains a simplified algebraic manipulator,
      used for the solution of the systems of identities.
\item Coefficients of the integrals in the identities can be
      unlimited precision integers, rationals,
      ratios of polynomials in one and two variables
      (for example $D$ and $x$) with integer coefficients.
      At present the values of square masses and products of external momenta
      must be rational numbers. 
\item Efficient management of systems of identities of size up to
      the limit of disk space (tested up to half million of identities).
\item Numerical solution of systems of difference/differential equations
      up to 500 equations.      
\item Numerical variables used in the solution are
      arbitrary precision floating point complex numbers and
      truncated series in $\e$ with this kind of coefficients.
\item Arithmetic libraries which deal with operations on
      integers, polynomials, rationals, floating point numbers and truncated
      series in $\e$ were written on purpose.
\end{itemize}
Two versions of the program exist, one using factorial series
and one using Laplace's transformation.
The Laplace's transformation version is much more complicated
than the factorial series version but
it turns to be faster in many cases:
both systems of difference and Laplace-transformed
differential equations are generated,
then the system of differential equations is solved,
and the functions $U(x)$ are obtained by integrating over $t$ 
the solutions $v(t)$,
also checking that $U(x)$ are solutions of the system of difference equations.
Both versions of program were used for calculating the single-scale integrals
of section \ref{resu}.

\section{Conclusions}\label{Conclusions}
Most part of this paper has been devoted to the description
of a new method of calculation of master integrals, based on numerical
solution of systems of difference equations,
obtained by solving systems of identities obtained by integration-by-parts.
Important features of the method are the applicability to arbitrary diagrams,
inherited from integration-by-parts method,
and the ability to obtain very high precision results (even 100-200 digits)
expanded at will in $\e$,
due to fact that the calculation of integrals is reduced 
to sums of \emph{convergent} factorial or power
series in one variable.

We also have described two important complements to this method: 
an algorithm for the reduction of generic Feynman integrals to master
integrals,
and a procedure for construction and numerical
solution of differential equations is masses and momenta;
among other things, 
at present the latter is needed  
to calculate generic master integrals with massless denominators
or with external momenta `deep' in the non-euclidean region 
(over the deformation thresholds),
at least until a working automated general procedure 
for the calculation of the arbitrary constants of difference equations
in these cases will be found.

The implementation of these methods and algorithms in the program $\SYS$ 
turned out to be essential to test and prove the validity of the approach.

In this first exploratory work,
mostly devoted to test our approach,
we have focused our attention to the calculation of 
single-scale master integrals, in particular 
vacuum and self-energy diagrams  up to three loops and
vertex and box diagrams      up to two loops.
The calculated values of these master integrals are useful, 
as they may appear
expanding Feynman integrals with respect to ratio of different scale 
parameters. 

But the final targets of our approach are 
the calculation of four-loop $g$-$2$ contribution in QED, 
and 
the calculation of multi-scale multi-loop master integrals
especially in cases where there are no hierarchies between scales,
where asymptotic expansions in large masses or momenta seem to be not useful. 
No insurmountable difficulty seems to exist 
for applying our approach to these problems.
Clearly, that means to deal with a larger number of master integrals,
or with more complicated equations,
and that will imply modifications or 
improvements of various parts of the algorithms
implemented in the program $\SYS$ which,
at this stage of development, is far to be optimal.
The experience gained by performing the calculations of this work 
has given
many suggestions 
on the changes which should be made and which
will be discussed in future papers.

\section*{Acknowledgement}
The author wants to thank E. Remiddi and M. Caffo for
useful discussions and encouragement in the very early stage of this work.

\vfill\eject 
\pagenumbering{roman}
\setcounter{page}{1}
\phantom{.}\vspace{5cm}
\section*{Figure Captions}
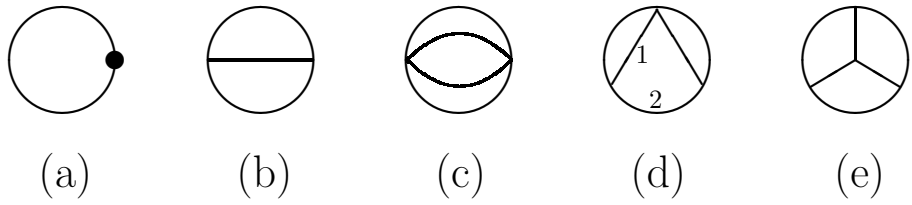
\par\noindent Figure 1: Vacuum diagrams up to three loops.
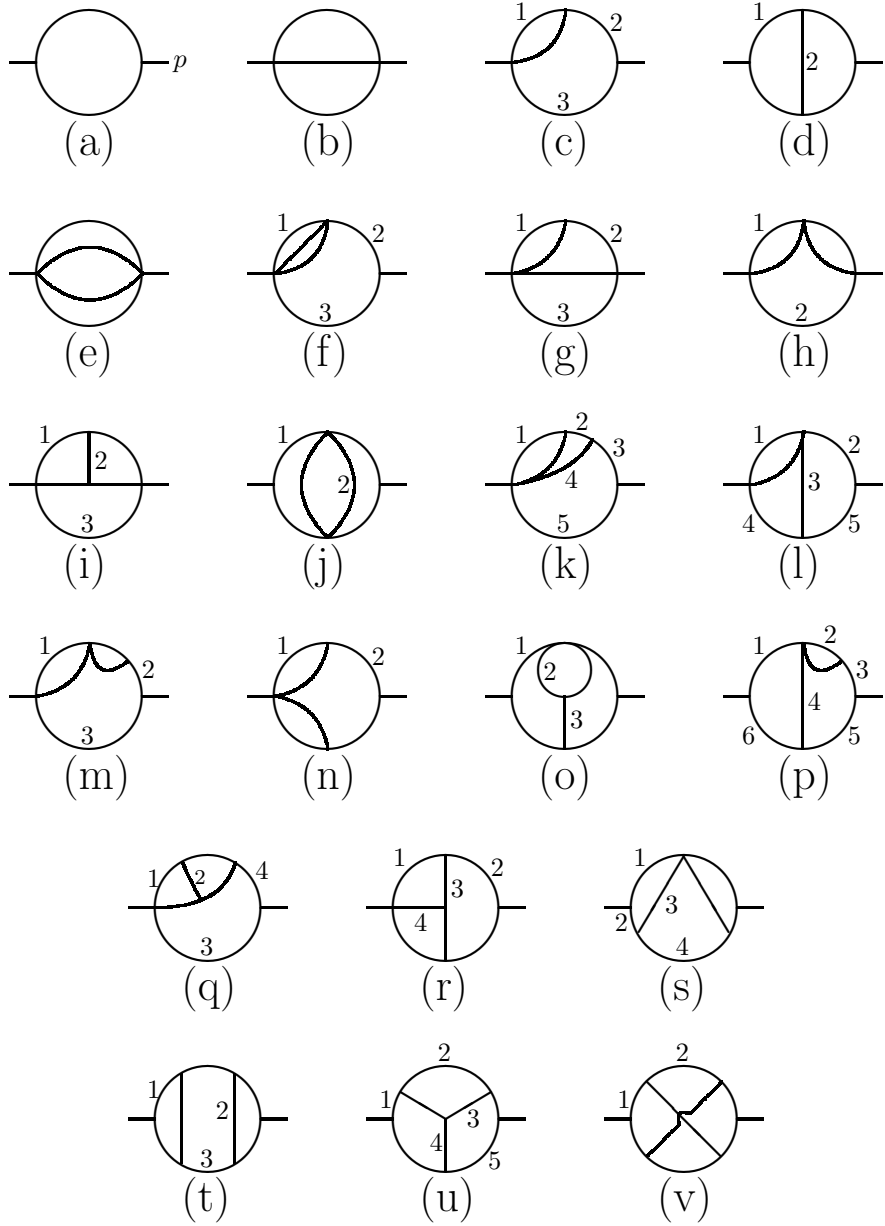
\par\noindent Figure 2: Self-energy diagrams up to three loops.
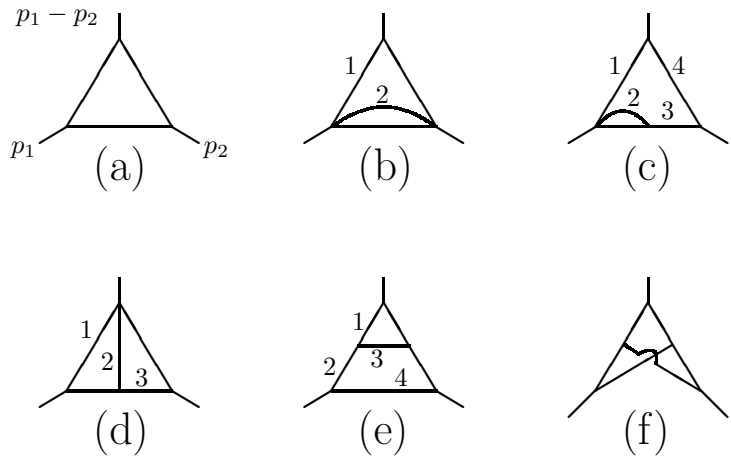
\par\noindent Figure 3: Vertex diagrams up to two loops.
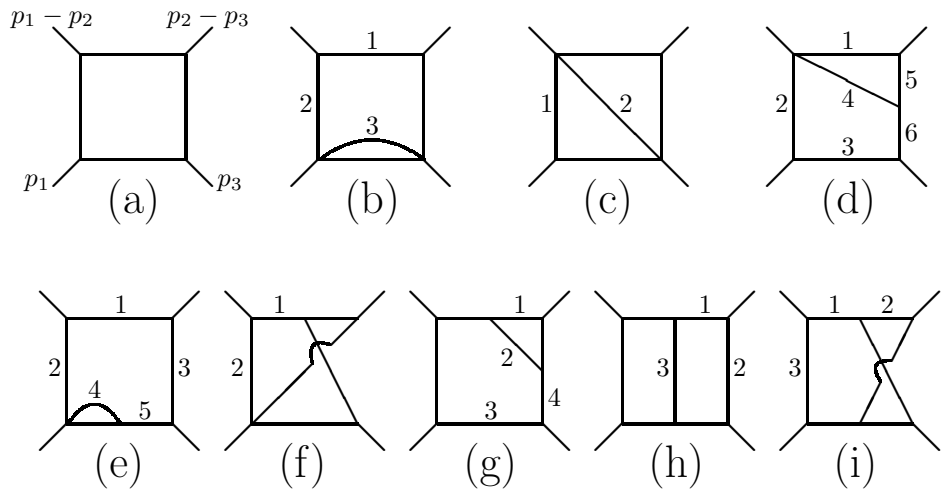
\par\noindent Figure 4: Box diagrams up to two loops.
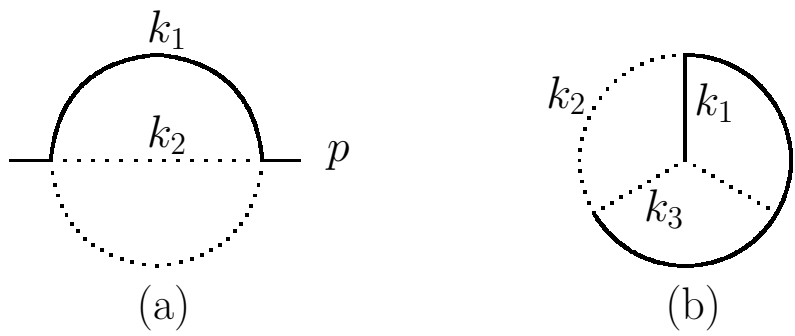
\par\noindent Figure 5: Diagrams with massless lines (dashed).
\phantom{.}\vspace{12cm}\IDENTIFY
\vfill\eject 
\phantom{.}\vspace{5cm}
\section*{Condensed paper title}
{\large Calculation of Feynman integrals by difference equations}
\phantom{.}\vspace{14cm}\IDENTIFY
\vfill\eject 
\phantom{.}\vspace{5cm}
\begin{table}
\notag
\begin{center}
\begin{tabular}{rrrr}
\hline
$x$& $J(x)$ &$\IOMOG_{-}(x)$ & $\INOMOG(x)$   \\ \hline
9  & $0.017857 +0.033442 \e$ &$  0.047713        +0.089160 \e $&
     $ -0.008928   -0.007323 \e$ \\
8  & $0.023809 +0.040621 \e$ &$  0.059006        +0.100337 \e $&
     $ -0.011904   -0.007663 \e$ \\
7  & $0.033333 +0.050203 \e$ &$  0.075609        +0.113249 \e $&
     $ -0.016666   -0.007159 \e$ \\
6  & $0.05     +0.062805 \e$ &$  0.101857        +0.126598 \e $&
     $ -0.025      -0.003929 \e$ \\
5  & $0.083333 +0.076898 \e$ &$  0.148085        +0.133074 \e $&
     $ -0.041666   +0.009010 \e$ \\
4  & $0.166666 +0.070464 \e$ &$  0.245635        +0.090010 \e $&
     $ -0.083333   +0.067207 \e$ \\
3  & $0.5      -0.288607 \e$ &$  0.548843        -0.442122 \e $&
     $ -0.25       +0.534955 \e$ \\
2  & $\e^{-1}  -0.577215\fe$ &$  0.282094\e^{-1} +0.519388 \fe$&
     $ -0.25\e^{-1}+0.144303 \fe$ \\
1  & $-\e^{-1} -0.422784\fe$ &$  0.282094\e^{-1} -1.645293 \fe$&
     $  0.75\e^{-1}+1.067088 \fe$ \\
0  & $0\phantom{.123456\fe}$ &$ -0.564189\e^{-1} -0.238530 \fe$&
     $ -0.5\e^{-1} -0.211392 \fe$ \\
\hline
\end{tabular}
\end{center}
\centerline{Table \ref{tableij}: Values of $J(x)$, $\IOMOG_{-}(x)$ and $\INOMOG(x)$}
\phantom{ }\vspace{3truecm}\phantom{ }
\end{table}
\phantom{.}\vspace{7cm}\IDENTIFY
\vfill\eject 
\phantom{.}\vspace{5cm}
\begin{table}
\notag
\begin{center}
\begin{tabular}{rrc}
\hline
$x_{max}$& terms & finite part of $I(1)$  \\ \hline
$30$  &$   125$ & $-0.3910008887063124$\\
$25$  &$   154$ & $-0.3910149952724784$\\
$20$  &$   217$ & $-0.3910150292106927$\\
$15$  &$   395$ & $-0.3910150291388126$\\
$10$  &$  1470$ & $-0.3910150291357554$\\
$ 9$  &$  2454$ & $-0.3910150291357472$\\
$ 8$  &$  4439$ & $-0.3910150291357503$\\
$ 7$  &$ 13086$ & $-0.3910150291357507$\\
$ 6$  &$ 36210$ & $-0.3910150291357507$\\
\hline
\end{tabular}
\end{center}
\centerline{Table \ref{tablei1}: Dependence of the finite part of $I(1)$ on $x_{max}$.}
\end{table}
\phantom{.}\vspace{10cm}\IDENTIFY
\vfill\eject 
\def\baselinestretch{}
{
\begin{table}
\begin{center}
\vskip -17pt
\begin{tabular}{lcrr|crr}
\hline
\scriptsize{diagram}             & $n_b$ &   $R_C$     & $S$          
& $n_b'$ &  $R_{C'}'$     & $S'$  \\ \hline
\ref{figvac}a     &   1   & $1_1$       &  1           
&   1   & $1_1$        & 1      \\
\ref{figvac}b     &   1   & 2           &  1           
&   1   & 2            & 1      \\
\ref{figvac}c     &  1,2,1& $5_1$       &  4           
&   1   & $2_1$        & 2      \\
\ref{figvac}d1-2  &   1   & 2,5         &  1,1         
&   1   & 2,2          & 1,1    \\
\ref{figvac}e     &   1   & 2           &  1           
&   1   & 2            & 1      \\
\ref{figself}a   &   1   & 2           &  1           
&   1   & $2_1$         & 1      \\
\ref{figself}b   &  1,2,1& $5_1$       &  4           
&   1   & $2_1$         & 2      \\
\ref{figself}c1-3&   1   & 2,5,2       &  1,1,1       
&   1   & 2,2,$2_1$     & 1,1,1  \\
\ref{figself}d1-2&   1   & 2,2         &  1,1         
&   1   & 2,2           & 1,1    \\
\ref{figself}e   & 1,5,5 & $12_4$      &  11          
& 1,5,5 & $4_1$         & 3      \\
\ref{figself}f1-3& 1,2,1 & $5_1$,15,2  &  4,4,1       
&   1   & $2_1$,$5_1$,$2_1$& 2,2,1  \\
\ref{figself}g1-3& 1,2,1 & 2,15,$5_1$  &  1,4,4       
&   1   & 2,5,$2_1$     & 1,2,2  \\
\ref{figself}h1-2&1,2,2,1& $10_1$,$10_1$&  6,6        
&   1   & $2_1$,$2_1$ & 3,3    \\
\ref{figself}i1-3& 1,2,1 & 8,10,$5_1$  &  4,4,4       
&  1,1  & 3,3,$3_1$     & 3,2,3  \\
\ref{figself}j1-2& 1,2,1 & 8,$5_1$     &  4,4         
&  1,1  & $3_1$,$3_1$   & 3,3    \\
\ref{figself}k1-5&   1   & 2,5,5,2,2   &  1,1,1,1,1   
&   1   & 2,2,2,2,$2_1$& 1,1,1,1,1\\
\ref{figself}l1-5& 1,2,1 &$5_1$,8,8,8,10& 4,4,4,4,4   
&  1,1  & $3_1$,$3_1$,3,3,3& 3,3,3,3,2\\
\ref{figself}m1-3&   1   & 2,7,2       &  1,1,1       
&   1   & 2,2,$2_1$     & 1,1,1\\
\ref{figself}n1-2&   1   & 2,5         &  1,1         
&   1   & 2,2           & 1,1\\
\ref{figself}o1-3&   1   & 2,2,5       &  1,1,1       
&   1   & 2,2,3         & 1,1,1 \\
\ref{figself}p1-6&   1   & 2,2,5,2,2,2 &  1,1,1,1,1,1 
&   1   & 2,2,3,2,2,2   & 1,1,1,1,1,1 \\
\ref{figself}q1-4&   1   & 2,2,2,5     &  1,1,1,1     
&   1   & 2,2,$2_1$,3   & 1,1,1,1 \\
\ref{figself}r1-4& 1,2,2 & 6,9,9,6     &  5,5,5,5     
&  1,1  & 3,$4_2$,$4_1$,3& 3,3,3,3 \\
\ref{figself}s1-4& 1,2,2 & 6,9,6,9     &  5,5,5,5     
&  1,1  &$3_1$,$4_2$,3,$4_1$&3,3,3,3 \\
\ref{figself}t1-3&   1   & 2,2,6       &  1,1,1       
&   1   & 2,2,3         &1,1,1 \\
\ref{figself}u1-5&   1   & 6,2,2,6,2   &  1,1,1,1,1   
&   1   & 3,2,2,3,2     &1,1,1,1,1 \\
\ref{figself}v1-2& 1,1,1 & 8,8         &  3,3         
&   1   & 3,3           &2,2 \\
\ref{figvert}a   &   1   & 2           &  1           
&   1   & $2_1$         & 1      \\
\ref{figvert}b1-2& 1,2,1 & 8,$5_1$     &  4,4         
&  1,1  & $3_1$,$3_1$   & 3,3     \\
\ref{figvert}c1-4&   1   & 2,2,5,2     &  1,1,1,1     
&   1   & $2_1$,2,3,$2_1$& 1,1,1,1 \\
\ref{figvert}d1-3& 1,2,2 & 6,6,9       &  5,5,5       
&  1,1  & $3_1$,3,$4_2$  & 3,3,3 \\
\ref{figvert}e1-4&   1   & 2,6,2,2     &  1,1,1,1     
&   1   & 2,3,2,$2_1$   & 1,1,1,1 \\
\ref{figvert}f   & 1,1,1 & 8           &  3           
&   1   & 3             & 2 \\
\ref{figbox}a    &   1   & 2           &  1           
&   1   & $2_1$         & 1  \\
\ref{figbox}b1-3 & 1,2,1 & 8,8,$5_1$   &  4,4,4       
&  1,1  & $4_1$,$4_1$,$3_1$& 3,3,3\\
\ref{figbox}c1-2 &1,4,3  & 12,$9_1$    &  8,8         
& 1,2,1 & $6_1$,$4_1$   & 5,4  \\
\ref{figbox}d1-6 & 1,3,1 &6,10,10,6,9,13& 5,5,5,5,5,5 
&  1,2  & $4_1,$$5_1$,$5_1$,4,$5_2$,$7_4$ & 4,4,4,4,3,5   \\
\ref{figbox}e1-5 &   1   & 2,2,2,2,5   &  1,1,1,1,1   
&   1   & $2_1$,$2_1$,$2_1$,2,3 & 1,1,1,1,1   \\
\ref{figbox}f1-2 &1,3,6,3& (23),(19)   &  (13),(13)   
&1,3,6,1& $16_4$,$13_1$ & 13,13 \\
\ref{figbox}g1-4 &   1   & 2,2,2,6     &  1,1,1,1     
&   1   & 2,2,$2_1$,4   & 1,1,1,1  \\
\ref{figbox}h1-3 & 1,2,2 & 10,6,6      &  5,5,5       
&  1,1  & $5_2$,$3_1$,3 & 3,3,3  \\
\ref{figbox}i1-3 &1,2,3,1& (20),12,10  &  7,7,7       
&  1,1  & $13_4$,5,$3_1$& 7,5,3  \\
\hline
\end{tabular}
\vskip -9pt
\end{center}
\centerline{Table \ref{tablevacself}: Number of master integrals and orders of the equations for the diagrams.}
\phantom{.}\vspace{1cm}\IDENTIFY
\end{table}
}
\def\baselinestretch{\factor}
\vfill\eject 
\begin{figure}
\phantom{.}\vspace{5cm}
\notag
\begin{center}
   \begin{picture}(380,120)(0,0)
   \thicklines
   {
   \put(257,079){\text{1}}   
   \put(262,062){\text{2}}   
\LARGE
   \put(031,032){\text{(a)}}   
   \put(106,032){\text{(b)}}   
   \put(181,032){\text{(c)}}   
   \put(255,032){\text{(d)}}   
   \put(332,032){\text{(e)}}   
   }
   \put(060,080){\circle*{7}}
   \put(040,080){\circle{40}}
   \put(115,080){\circle{40}}
   \put(095,080){\line(1,0){40}}
   \put(190,080){\circle{40}}
   \qbezier(170,080)(190,100)(209.5,080)
   \qbezier(170,080)(190,060)(209.5,080)
   \put(265,080){\circle{40}}
   \put(265,099.2){\line(+3,-5){17}}
   \put(265,099.2){\line(-3,-5){17}}
   \put(340,080){\circle{40}}
   \put(340,080){\line(0,1){20}}
   \put(340,080){\line(+5,-3){17}}
   \put(340,080){\line(-5,-3){17}}
   \end{picture}
\end{center}
\phantom{ }\vspace{10truecm}\phantom{ }
 \centerline{Figure \ref{figvac}}
\phantom{.}\vspace{1cm}\IDENTIFY
 \end{figure}
\begin{figure}
\phantom{.}\vspace{1cm}
\notag
\begin{center}
   \begin{picture}(370,480)(0,0)
   \thicklines
   {
   \put(082,458){\text{$p$}}
   \put(211,476){\text{1}}   
   \put(247,472){\text{2}}   
   \put(227,442){\text{3}}   
   \put(301,476){\text{1}}   
   \put(321,457){\text{2}}   
   \put(121,396){\text{1}}   
   \put(157,392){\text{2}}   
   \put(137,362){\text{3}}   
   \put(211,396){\text{1}}   
   \put(247,392){\text{2}}   
   \put(227,362){\text{3}}   
   \put(301,396){\text{1}}   
   \put(317,362){\text{2}}   
   \put(031,316){\text{1}}   
   \put(052,306){\text{2}}   
   \put(047,282){\text{3}}   
   \put(121,316){\text{1}}   
   \put(144,297){\text{2}}   
   \put(211,316){\text{1}}   
   \put(234,321){\text{2}}   
   \put(248,311){\text{3}}   
   \put(230,299){\text{4}}   
   \put(227,282){\text{5}}   
   \put(301,316){\text{1}}   
   \put(322,298){\text{3}}   
   \put(337,312){\text{2}}   
   \put(337,282){\text{5}}   
   \put(297,282){\text{4}}   
   \put(031,236){\text{1}}   
   \put(070,227){\text{2}}   
   \put(047,202){\text{3}}   
   \put(121,236){\text{1}}   
   \put(157,232){\text{2}}   
   \put(211,236){\text{1}}   
   \put(222,226){\text{2}}   
   \put(232,208){\text{3}}   
   \put(301,236){\text{1}}   
   \put(322,215){\text{4}}   
   \put(337,202){\text{5}}   
   \put(340,227){\text{3}}   
   \put(328,240){\text{2}}   
   \put(297,202){\text{6}}   
   \put(072,148){\text{1}}   
   \put(092,122){\text{3}}   
    {\footnotesize \put(090,149){\text{2}}}   
   \put(113,151){\text{4}}   
   \put(165,156){\text{1}}   
   \put(187,144){\text{3}}   
   \put(202,152){\text{2}}   
   \put(173,131){\text{4}}   
   \put(256,156){\text{1}}   
   \put(249,131){\text{2}}   
   \put(268,138){\text{3}}   
   \put(272,122){\text{4}}   
   \put(072,068){\text{1}}   
   \put(098,060){\text{2}}   
   \put(092,042){\text{3}}   
   \put(160,064){\text{1}}   
   \put(182,082){\text{2}}   
   \put(193,057){\text{3}}   
   \put(179,048){\text{4}}   
   \put(201,041){\text{5}}   
   \put(250,064){\text{1}}   
   \put(272,082){\text{2}}   
\LARGE
   \put(040,425){\text{(a)}}   
   \put(130,425){\text{(b)}}   
   \put(220,425){\text{(c)}}   
   \put(310,425){\text{(d)}}   
   
   \put(040,345){\text{(e)}}   
   \put(130,345){\text{(f)}}   
   \put(220,345){\text{(g)}}   
   \put(310,345){\text{(h)}}   
   
   \put(040,265){\text{(i)}}   
   \put(130,265){\text{(j)}}   
   \put(220,265){\text{(k)}}   
   \put(310,265){\text{(l)}}   
   
   \put(040,185){\text{(m)}}   
   \put(130,185){\text{(n)}}   
   \put(220,185){\text{(o)}}   
   \put(310,185){\text{(p)}}   
   
   \put(085,105){\text{(q)}}   
   \put(175,105){\text{(r)}}   
   \put(265,105){\text{(s)}}   
   
   \put(085,025){\text{(t)}}   
   \put(175,025){\text{(u)}}   
   \put(265,025){\text{(v)}}   
   }

   \put(050,460){\circle{40}}
   \put(030,460){\line(-1,0){10}}
   \put(070,460){\line(+1,0){10}}
   \put(140,460){\circle{40}}
   \put(120,460){\line(1,0){40}}
   \put(120,460){\line(-1,0){10}}
   \put(160,460){\line(+1,0){10}}
   \put(230,460){\circle{40}}
   \put(210,460){\line(-1,0){10}}
   \put(250,460){\line(+1,0){10}}
   \qbezier(210,460)(228,462)(230,480)
   \put(320,460){\circle{40}}
   \put(320,440){\line(0,+1){40}}
   \put(300,460){\line(-1,0){10}}
   \put(340,460){\line(+1,0){10}}
   \put(050,380){\circle{40}}
   \qbezier(030,380)(050,400)(070,380)
   \qbezier(030,380)(050,360)(070,380)
   \put(030,380){\line(-1,0){10}}
   \put(070,380){\line(+1,0){10}}
   \put(140,380){\circle{40}}
   \put(120,380){\line(-1,0){10}}
   \put(160,380){\line(+1,0){10}}
   \qbezier(120,380)(130,390)(140,400)
   \qbezier(120,380)(138,382)(140,400)
   \put(230,380){\circle{40}}
   \put(210,380){\line(-1,0){10}}
   \put(250,380){\line(+1,0){10}}
   \qbezier(210,380)(227,383)(230,400)
   \put(210,380){\line(1,0){40}}
   \put(320,380){\circle{40}}
   \qbezier(300,380)(318,382)(320,400)
   \qbezier(340,380)(322,382)(320,400)
   \put(300,380){\line(-1,0){10}}
   \put(340,380){\line(+1,0){10}}
   \put(050,300){\circle{40}}
   \put(030,300){\line(-1,0){10}}
   \put(070,300){\line(+1,0){10}}
   \put(030,300){\line(+1,0){40}}
   \put(050,300){\line(0,+1){20}}
   \put(140,300){\circle{40}}
   \put(120,300){\line(-1,0){10}}
   \put(160,300){\line(+1,0){10}}
   \qbezier(140,280)(160,300)(140,320)
   \qbezier(140,280)(120,300)(140,320)
   \put(230,300){\circle{40}}
   \put(210,300){\line(-1,0){10}}
   \put(250,300){\line(+1,0){10}}
   \qbezier(210,300)(227,303)(230,320)
   \qbezier(210,300)(233,304)(240,317)
   \put(320,300){\circle{40}}
   \put(300,300){\line(-1,0){10}}
   \put(340,300){\line(+1,0){10}}
   \qbezier(300,300)(317,303)(320,320)
   \put(320,280){\line(0,+1){40}}
   \put(050,220){\circle{40}}
   \put(030,220){\line(-1,0){10}}
   \put(070,220){\line(+1,0){10}}
   \qbezier(030,220)(047,223)(050,240)
   \qbezier(050,240)(052,224)(064,233)
   \put(140,220){\circle{40}}
   \put(120,220){\line(-1,0){10}}
   \put(160,220){\line(+1,0){10}}
   \qbezier(120,220)(137,223)(140,240)
   \qbezier(120,220)(137,217)(140,200)
   \put(230,220){\circle{40}}
   \put(210,220){\line(-1,0){10}}
   \put(250,220){\line(+1,0){10}}
   \put(230,200){\line(0,+1){20}}
   \put(230,230){\circle{20}}
   \put(320,220){\circle{40}}
   \put(300,220){\line(-1,0){10}}
   \put(340,220){\line(+1,0){10}}
   \put(320,200){\line(0,+1){40}}   
   \qbezier(320,240)(322,224)(334,233)
   \put(095,140){\circle{40}}
   \put(075,140){\line(-1,0){10}}
   \put(115,140){\line(+1,0){10}}
   \qbezier(085,157)(088,150)(092,143)
   \qbezier(075,140)(100,140)(105,157)
   \put(185,140){\circle{40}}
   \put(165,140){\line(-1,0){10}}
   \put(205,140){\line(+1,0){10}}
   \put(185,120){\line(0,+1){40}}
   \put(165,140){\line(+1,0){20}}
   \put(275,140){\circle{40}}
   \put(255,140){\line(-1,0){10}}
   \put(295,140){\line(+1,0){10}}
   \put(275,159.2){\line(+3,-5){17}}
   \put(275,159.2){\line(-3,-5){17}}
   \put(095,060){\circle{40}}
   \put(075,060){\line(-1,0){10}}
   \put(115,060){\line(+1,0){10}}
   \put(085,043){\line(0,+1){34}}
   \put(105,043){\line(0,+1){34}}
   \put(185,060){\circle{40}}
   \put(165,060){\line(-1,0){10}}
   \put(205,060){\line(+1,0){10}}
   \put(185,060){\line(0,-1){20}}
   \put(185,060){\line(+5,+3){17}}
   \put(185,060){\line(-5,+3){17}}
   \put(275,060){\circle{40}}
   \put(255,060){\line(-1,0){10}}
   \put(295,060){\line(+1,0){10}}
   \put(261,074){\line(+1,-1){28}}
   \qbezier(261,046)(267,052)(273,058)
   \qbezier(289,074)(283,068)(277,062)
   \qbezier(273,058)(271,064)(277,062)
   \end{picture}
\end{center}
\phantom{ }\vspace{1truecm}\phantom{ }
 \centerline{Figure \ref{figself}}
\phantom{.}\vspace{0.1cm}\IDENTIFY
 \end{figure}
\begin{figure}
\phantom{.}\vspace{5cm}
\notag
\begin{center}
   \begin{picture}(300,200)(0,0)
   \thicklines
   {
   \put(009,130){\text{$p_1$}}
   \put(082,130){\text{$p_2$}}
   \put(011,180){\text{$p_1-p_2$}}
   \put(135,160){\text{1}}   
   \put(147,149){\text{2}}   
   \put(235,160){\text{1}}   
   \put(242,148){\text{2}}   
   \put(255,143){\text{3}}   
   \put(259,160){\text{4}}   
   \put(035,060){\text{1}}   
   \put(043,048){\text{2}}   
   \put(056,042){\text{3}}   
   \put(138,063){\text{1}}   
   \put(127,046){\text{2}}   
   \put(145,049){\text{3}}   
   \put(154,042){\text{4}}   
\LARGE
   \put(040,120){\text{(a)}}   
   \put(140,120){\text{(b)}}   
   \put(240,120){\text{(c)}}   
   \put(040,020){\text{(d)}}   
   \put(140,020){\text{(e)}}   
   \put(240,020){\text{(f)}}   
   }
   \put(030,140){\line(+1,0){40}}
   \put(030,140){\line(+3,+5){20}}
   \put(070,140){\line(-3,+5){20}}
   \put(030,140){\line(-5,-3){10}}
   \put(070,140){\line(+5,-3){10}}
   \put(050,173){\line(0,+1){10}}
   \put(130,140){\line(+1,0){40}}
   \put(130,140){\line(+3,+5){20}}
   \put(170,140){\line(-3,+5){20}}
   \put(130,140){\line(-5,-3){10}}
   \put(170,140){\line(+5,-3){10}}
   \put(150,173){\line(0,+1){10}}
   \qbezier(130,140)(150,155)(169,140)
   \put(230,140){\line(+1,0){40}}
   \put(230,140){\line(+3,+5){20}}
   \put(270,140){\line(-3,+5){20}}
   \put(230,140){\line(-5,-3){10}}
   \put(270,140){\line(+5,-3){10}}
   \put(250,173){\line(0,+1){10}}
   \qbezier(230,140)(240,152)(250,140)
   \put(030,040){\line(+1,0){40}}
   \put(030,040){\line(+3,+5){20}}
   \put(070,040){\line(-3,+5){20}}
   \put(030,040){\line(-5,-3){10}}
   \put(070,040){\line(+5,-3){10}}
   \put(050,073){\line(0,+1){10}}
   \put(050,073){\line(0,-1){33}}
   \put(130,040){\line(+1,0){40}}
   \put(130,040){\line(+3,+5){20}}
   \put(170,040){\line(-3,+5){20}}
   \put(130,040){\line(-5,-3){10}}
   \put(170,040){\line(+5,-3){10}}
   \put(150,073){\line(0,+1){10}}
   \put(140,057){\line(+1,0){20}}
   \put(230,040){\line(+3,+5){20}}
   \put(270,040){\line(-3,+5){20}}
   \put(230,040){\line(-1,-1){10}}
   \put(270,040){\line(+1,-1){10}}
   \put(250,073){\line(0,+1){10}}
   \put(230,040){\line(+5,+3){29}}
   \put(270,040){\line(-5,+3){17}}
   \qbezier(246,054)(243,056)(240.5,057.5)
   \qbezier(246,054)(254,058)(252.5,050.5)
   \end{picture}
\end{center}
\phantom{ }\vspace{7truecm}\phantom{ }
 \centerline{Figure \ref{figvert}}
\phantom{.}\vspace{1cm}\IDENTIFY
 \end{figure}
\begin{figure}
\phantom{.}\vspace{5cm}
\notag
\begin{center}
   \begin{picture}(370,200)(0,0)
   \thicklines
   {
   \put(009,130){\text{$p_1$}}
   \put(082,130){\text{$p_3$}}
   \put(004,192){\text{$p_1-p_2$}}
   \put(063,192){\text{$p_2-p_3$}}
   \put(138,182){\text{1}}   
   \put(113,158){\text{2}}   
   \put(138,150){\text{3}}   
   \put(204,158){\text{1}}   
   \put(234,158){\text{2}}   
   \put(318,182){\text{1}}   
   \put(293,158){\text{2}}   
   \put(318,142){\text{3}}   
   \put(318,160){\text{4}}   
   \put(342,167){\text{5}}
   \put(342,147){\text{6}}
   \put(043,082){\text{1}}   
   \put(018,058){\text{2}}   
   \put(067,058){\text{3}}
   \put(033,050){\text{4}}   
   \put(052,042){\text{5}}   
   \put(103,082){\text{1}}
   \put(087,058){\text{2}}   
   \put(194,082){\text{1}}
   \put(189,062){\text{2}}
   \put(183,042){\text{3}}   
   \put(207,047){\text{4}}
   \put(264,082){\text{1}}
   \put(277,058){\text{2}}
   \put(248,058){\text{3}}
   \put(313,082){\text{1}}
   \put(333,082){\text{2}}
   \put(298,058){\text{3}}   
\LARGE
   \put(040,120){\text{(a)}}   
   \put(130,120){\text{(b)}}   
   \put(220,120){\text{(c)}}   
   \put(310,120){\text{(d)}}   
   \put(035,020){\text{(e)}}   
   \put(105,020){\text{(f)}}   
   \put(175,020){\text{(g)}}   
   \put(245,020){\text{(h)}}   
   \put(315,020){\text{(i)}}   
   }
   \put(030,140){\line(+1,0){40}}
   \put(030,140){\line(0,+1){40}}
   \put(070,140){\line(0,+1){40}}
   \put(030,180){\line(+1,0){40}}
   \put(030,140){\line(-1,-1){10}}
   \put(030,180){\line(-1,+1){10}}
   \put(070,140){\line(+1,-1){10}}
   \put(070,180){\line(+1,+1){10}}
   \put(120,140){\line(+1,0){40}}
   \put(120,140){\line(0,+1){40}}
   \put(160,140){\line(0,+1){40}}
   \put(120,180){\line(+1,0){40}}
   \put(120,140){\line(-1,-1){10}}
   \put(120,180){\line(-1,+1){10}}
   \put(160,140){\line(+1,-1){10}}
   \put(160,180){\line(+1,+1){10}}
   \qbezier(120,140)(140,155)(160,140)
   \put(210,140){\line(+1,0){40}}
   \put(210,140){\line(0,+1){40}}
   \put(250,140){\line(0,+1){40}}
   \put(210,180){\line(+1,0){40}}
   \put(210,140){\line(-1,-1){10}}
   \put(210,180){\line(-1,+1){10}}
   \put(250,140){\line(+1,-1){10}}
   \put(250,180){\line(+1,+1){10}}
   \put(210,180){\line(+1,-1){40}}
   \put(300,140){\line(+1,0){40}}
   \put(300,140){\line(0,+1){40}}
   \put(340,140){\line(0,+1){40}}
   \put(300,180){\line(+1,0){40}}
   \put(300,140){\line(-1,-1){10}}
   \put(300,180){\line(-1,+1){10}}
   \put(340,140){\line(+1,-1){10}}
   \put(340,180){\line(+1,+1){10}}
   \put(300,180){\line(+2,-1){40}}
   \put(025,040){\line(+1,0){40}}
   \put(025,040){\line(0,+1){40}}
   \put(065,040){\line(0,+1){40}}
   \put(025,080){\line(+1,0){40}}
   \put(025,040){\line(-1,-1){10}}
   \put(025,080){\line(-1,+1){10}}
   \put(065,040){\line(+1,-1){10}}
   \put(065,080){\line(+1,+1){10}}
   \qbezier(025,040)(035,055)(045,040)
   \put(095,040){\line(+1,0){40}}
   \put(095,040){\line(0,+1){40}}
   \put(135,040){\line(-1,+2){20}}
   \put(135,080){\line(-1,-1){10}}
   \put(095,040){\line(+1,+1){23}}
   \qbezier(117.5,063)(115,073)(124,070)
   \put(095,080){\line(+1,0){40}}
   \put(095,040){\line(-1,-1){10}}
   \put(095,080){\line(-1,+1){10}}
   \put(135,040){\line(+1,-1){10}}
   \put(135,080){\line(+1,+1){10}}
   \put(165,040){\line(+1,0){40}}
   \put(165,040){\line(0,+1){40}}
   \put(205,040){\line(0,+1){40}}
   \put(165,080){\line(+1,0){40}}
   \put(165,040){\line(-1,-1){10}}
   \put(165,080){\line(-1,+1){10}}
   \put(205,040){\line(+1,-1){10}}
   \put(205,080){\line(+1,+1){10}}
   \put(185,080){\line(+1,-1){20}}
   \put(235,040){\line(+1,0){40}}
   \put(235,040){\line(0,+1){40}}
   \put(275,040){\line(0,+1){40}}
   \put(235,080){\line(+1,0){40}}
   \put(235,040){\line(-1,-1){10}}
   \put(235,080){\line(-1,+1){10}}
   \put(275,040){\line(+1,-1){10}}
   \put(275,080){\line(+1,+1){10}}
   \put(255,080){\line(0,-1){40}}
   \put(305,040){\line(+1,0){40}}
   \put(305,040){\line(0,+1){40}}
   \put(345,040){\line(-1,+2){20}}
   \put(345,080){\line(-1,-2){8}}
   \put(325,040){\line(+1,+2){8}}
   \qbezier(332.5,056)(327,064)(336,064)
   \put(305,080){\line(+1,0){40}}
   \put(305,040){\line(-1,-1){10}}
   \put(305,080){\line(-1,+1){10}}
   \put(345,040){\line(+1,-1){10}}
   \put(345,080){\line(+1,+1){10}}
   \end{picture}
\end{center}
\phantom{ }\vspace{7truecm}\phantom{ }
 \centerline{Figure \ref{figbox}}
\phantom{.}\vspace{1cm}\IDENTIFY
 \end{figure}

\begin{figure}
\phantom{.}\vspace{5cm}
\notag
\begin{center}
   \begin{picture}(370,160)(0,0)
   \linethickness{1pt}
   {
   %\Large
   \LARGE
   \put(145,080){\text{$p$}}   
   \put(077,125){\text{$k_1$}}   
   \put(077,085){\text{$k_2$}}   
   \put(073,020){\text{(a)}}   

   \put(228,100){\text{$k_2$}}   
   \put(265,057){\text{$k_3$}}   
   \put(284,097){\text{$k_1$}}   
   \put(273,020){\text{(b)}}   
   }
   \put(040,080){\line(-1,0){15}}
   \put(120,080){\line(+1,0){15}}
   \qbezier[15](040,080)(080,080)(120,080)
   \qbezier(+120.0,+080.0)(+116.6,+116.6)(+080.0,+120.0)
   \qbezier(+080.0,+120.0)(+043.4,+116.6)(+040.0,+080.0)
   \qbezier[15](+040.0,+080.0)(+043.4,+043.4)(+080.0,+040.0)
   \qbezier[15](+080.0,+040.0)(+116.6,+043.4)(+120.0,+080.0)
   \qbezier[10](+280.0,+120.0)(+257.3,+119.3)(+245.4,+100.0)
   \qbezier[10](+245.4,+100.0)(+234.6,+080.0)(+245.4,+060.0)
   \qbezier(+245.4,+060.0)(+257.3,+040.7)(+280.0,+040.0)
   \qbezier(+280.0,+040.0)(+302.7,+040.7)(+314.6,+060.0)
   \qbezier(+314.6,+060.0)(+325.4,+080.0)(+314.6,+100.0)
   \qbezier(+314.6,+100.0)(+302.7,+119.3)(+280.0,+120.0)
   \qbezier[10](280,080)(297.3,070)(314.6,060)
   \qbezier[10](280,080)(262.7,070)(245.4,060)
   \qbezier(280,080)(280,100)(280,120)
\end{picture}
\end{center}
\phantom{ }\vspace{7truecm}\phantom{ }
 \centerline{Figure \ref{figder}}
\phantom{.}\vspace{1cm}\IDENTIFY
 \end{figure}
\end{document}